\documentclass[fleqn,usenatbib]{mnras}

\usepackage{newtxtext,newtxmath}

\usepackage[T1]{fontenc}
\usepackage{color}
\usepackage{xcolor}
\DeclareRobustCommand{\VAN}[3]{#2}
\let\VANthebibliography\thebibliography
\def\thebibliography{\DeclareRobustCommand{\VAN}[3]{##3}\VANthebibliography}

\usepackage{graphicx}	
\usepackage{amsmath}	

\title[A comprehensive separation of dark matter and baryons in galaxy clusters II]{A comprehensive separation of dark matter and baryonic mass components in galaxy clusters II: an overview of the mass distribution in Abell S1063}

\author[Beauchesne et al.]{
Benjamin Beauchesne,$^{1,2}$\thanks{E-mail: benjamin.e.beauchesne@durham.ac.uk}
Benjamin Cl\'ement,$^{3}$
Marceau Limousin,$^{4}$
Anna Niemiec,$^{5}$
\newauthor Mathilde Jauzac,$^{1,2,6,7}$
Belén Alcalde Pampliega,$^{8,9,10}$
Johan Richard,$^{11}$
Guillaume Mahler,$^{12}$
Jose M. Diego,$^{13}$
\newauthor Pascale Hibon,$^{8}$
Anton M. Koekemoer,$^{14}$
Thomas Connor,$^{15}$
Jean-Paul Kneib $^{3}$,
Andreas L. Faisst$^{16}$
\\
$^{1}$Centre for Extragalactic Astronomy, Department of Physics, Durham University, South Road, Durham DH1 3LE, UK\\
$^{2}$Institute for Computational Cosmology, Department of Physics, Durham University, South Road, Durham DH1 3LE, UK\\
$^{3}$Institute of Physics, Laboratory of Astrophysics, Ecole Polytechnique Fédérale de Lausanne (EPFL), Observatoire de Sauverny, 1290 Versoix, Switzerland\\
$^{4}$Aix Marseille Univ, CNRS, CNES, LAM, Marseille, France\\
$^{5}$Univ. Grenoble Alpes, CNRS, Grenoble INP, LPSC IN2P3, 53, Avenue des Martyrs, 38000 Grenoble, France \\
$^{6}$Astrophysics Research Centre, University of KwaZulu-Natal, Westville Campus, Durban 4041, South Africa\\
$^{7}$School of Mathematics, Statistics \& Computer Science, University of KwaZulu-Natal, Westville Campus, Durban 4041, South Africa\\
$^{8}$ESO Vitacura, Alonso de Córdova 3107,Vitacura, Casilla 19001, Santiago de Chile, Chile\\
$^{9}$Instituto de Estudios Astrofísicos, Facultad de Ingeniería y 455 Ciencias, Universidad Diego Portales, Av. Ejército Libertador 441, Santiago, Chile\\
$^{10}$SKA Observatory, Jodrell Bank, SK11 9FT, UK\\
$^{11}$Univ Lyon, Univ Lyon1, Ens de Lyon, CNRS, Centre de Recherche Astrophysique de Lyon (CRAL) UMR5574, F-69230 Saint-Genis-Laval, France\\
$^{12}$STAR Institute, Quartier Agora - All\'ee du six Ao\^ut, 19c B-4000 Li\`ege, Belgium\\
$^{13}$Instituto de F\'isica de Cantabria (CSIC-UC). Avda. Los Castros s/n. 39005 Santander, Spain \\
$^{14}$Space Telescope Science Institute, 3700 San Martin Dr., Baltimore, MD 21218, USA\\
$^{15}$Center for Astrophysics $\vert$\ Harvard\ \&\ Smithsonian, 60 Garden St., Cambridge, MA 02138, USA\\
$^{16}$Caltech/IPAC, 1200 E. California Blvd. Pasadena, CA 91125, USA
}

\date{Accepted XXX. Received YYY; in original form ZZZ}

\pubyear{2025}

\begin{document}
\label{firstpage}
\pagerange{\pageref{firstpage}--\pageref{lastpage}}
\maketitle

\begin{abstract}
In the first paper of this series, we derived mass constraints on the total mass and the baryonic components of the galaxy cluster Abell S1063. The main focus was to recover stellar masses and kinematics for cluster members, the brightest cluster galaxy (BCG) and the intra-cluster light (ICL). In this second paper, we introduce a multi-probe mass modelling approach that incorporates constraints on both the total mass and the individual baryonic components. We obtain comprehensive mass models of Abell S1063, in which the dark matter distribution is disentangled from the baryonic mass at both cluster and galaxy scales. The best-fitting mass model achieves an RMS of $0.50"$ on the positions of strongly-lensed multiple images. The kinematic profiles of the BCG \& ICL, as well as the X-ray surface brightness of the intra-cluster gas, are accurately reproduced within observational uncertainties. However, a $35~\mathrm{km/s}$ scatter is required for the cluster member line-of-sight dispersions. This method yields the most complex parametric mass model with consistency among almost all available mass constraints. We find a $1\sigma$ agreement between the inferred stellar-to-subhalo mass relation and that predicted by large-scale cosmological simulations. The ICL stellar mass derived from our model is consistent with estimates from stellar population modelling. We present the first multi-probe mass modelling method capable of disentangling the dark matter from the baryonic mass distributions in massive galaxy clusters. Its results, such as the stellar-to-subhalo mass relation or the distribution of each mass component, can be directly compared to hydrodynamical cosmological simulations such as illustrisTNG.
\end{abstract}



\section{Introduction}
In this series of two papers, we aim to develop a comprehensive mass modelling method in which each main cluster mass component is modelled, i.e., where dark matter (DM) is fully separated from baryons. Such a method would generate some of the most sophisticated parametric mass models to date, representing the first approach to achieve consistency across nearly all available mass probes. This new method extends the framework developed by \citet[hereafter B24][]{Beauchesne2024}, which combines constraints from strong lensing together with the X-ray emission from the intra-cluster gas in a self-consistent framework. Thanks to this combined dataset, the intra-cluster gas or intra-cluster medium (ICM), i.e., the main contributor to the baryonic mass budget, can be disentangled from the total mass. Hence, this series of papers now focuses on the remaining baryonic component, i.e. the stellar mass, which is contained in cluster galaxies and the intra-cluster stars. The latter are responsible for emitting the intra-cluster light (ICL). As in the cluster centre, the brightest cluster galaxy (BCG) is difficult to distinguish from the ICL, we thus consider them as a single component, denoted BCG \& ICL.

To test such new method, we are using Abell S1063 (Hereafter AS1063) as a test case. Thanks to an observational dataset of increasing quality over the years, AS1063 has been used as a test case for methodological improvements. It is one of the morphologically simplest clusters with such data. \citet{Bonamigo2018} added a fixed mass component representing the X-ray gas based on \citet{Bonamigo2017} method. \citet{Bergamini2019} and \citet{Granata2022} focus on improving the cluster member modelling by calibrating their scaling laws with galaxy kinematics and morphology. The earlier used a scaling based on the Faber \& Jackson law \citep{Faber&Jackson1976} while the latter made use of the more accurate fundamental plane of elliptical galaxies \citep{Hyde2009}. \citet{Limousin2022} revisited \citet{Bergamini2019} model and tested the addition of a free-form perturbative B-spline component \citep{Beauchesne2021}. B24 incorporated elements from \citet{Granata2022} and \citet{Bonamigo2018} to develop the framework that we intend to extend in this work. That framework allows us to disentangle the DM from the main baryonic component (i.e., intra-cluster gas). It reduces biased on the resulting DM distribution, in particular when the gas distribution does not follow the DM. However, the B24 framework does not account for the baryonic mass in the form of stellar mass. This baryonic contribution is usually small, except locally when a cluster member or the BCG is present. The DM distribution in and around the BCG is of particular interest as cold dark matter (CDM) model and its alternative such as Fuzzy DM\citep[FDM;][]{Hu2000} or self-interacting DM\citep[SIDM;]{Spergel2000} disagree on the expected DM profile slope. CDM prefers a cuspy profile, while the other two favour a cored distribution. Strong lensing models of galaxy clusters tend to favour cored distributions for the total mass profile \citep{Limousin2022}, disentangling the DM distribution up to the BCG would allow us to assess if it is due to the DM profile or an interplay between baryons and DM.

AS1063 has been modelled several other times in the past, with different observational datasets and methods. As part of the Cluster Lensing and Supernova Survey (CLASH; \citet{Postman2012}), the Hubble Frontier Fields (HFF; \citet{lotz2017}), the Beyond Ultra-deep Frontier Fields And Legacy Observations (BUFFALO; \citet{steinhardt2020}) and JWST GLIMPSE program \citep{Atek2025}, AS1063 has been modelled for each new observational dataset. \citet{Monna2014} produced a parametric mass model from the CLASH data with \textsc{GLEE} \citep{Suyu2010,Suyu2012}. In preparation for the upcoming HFF observations, both \citet{Johnson2014,Richard2014} modelled the cluster with \textsc{Lenstool} \citep{jullo2007} using different mass parametrisations. \citet{Caminha2016} made a model with \textsc{Lenstool} based on the HFF data and new spectroscopic measurements from VLT/MUSE and VLT/VIMOS as part of the CLASH-VLT program \citep{Mercurio2021}. Those datasets were also used by \citet{Diego2016} to make a free-form model with \textsc{WSLAP}\citep{diego2005,diego2007}. More recently, with the GLIMPSE data, parametric and free-form models have been presented by \citet{Atek2025} and \citet{Diego2026}, respectively.

\begin{figure*}
    \centering
    \includegraphics[width=\linewidth]{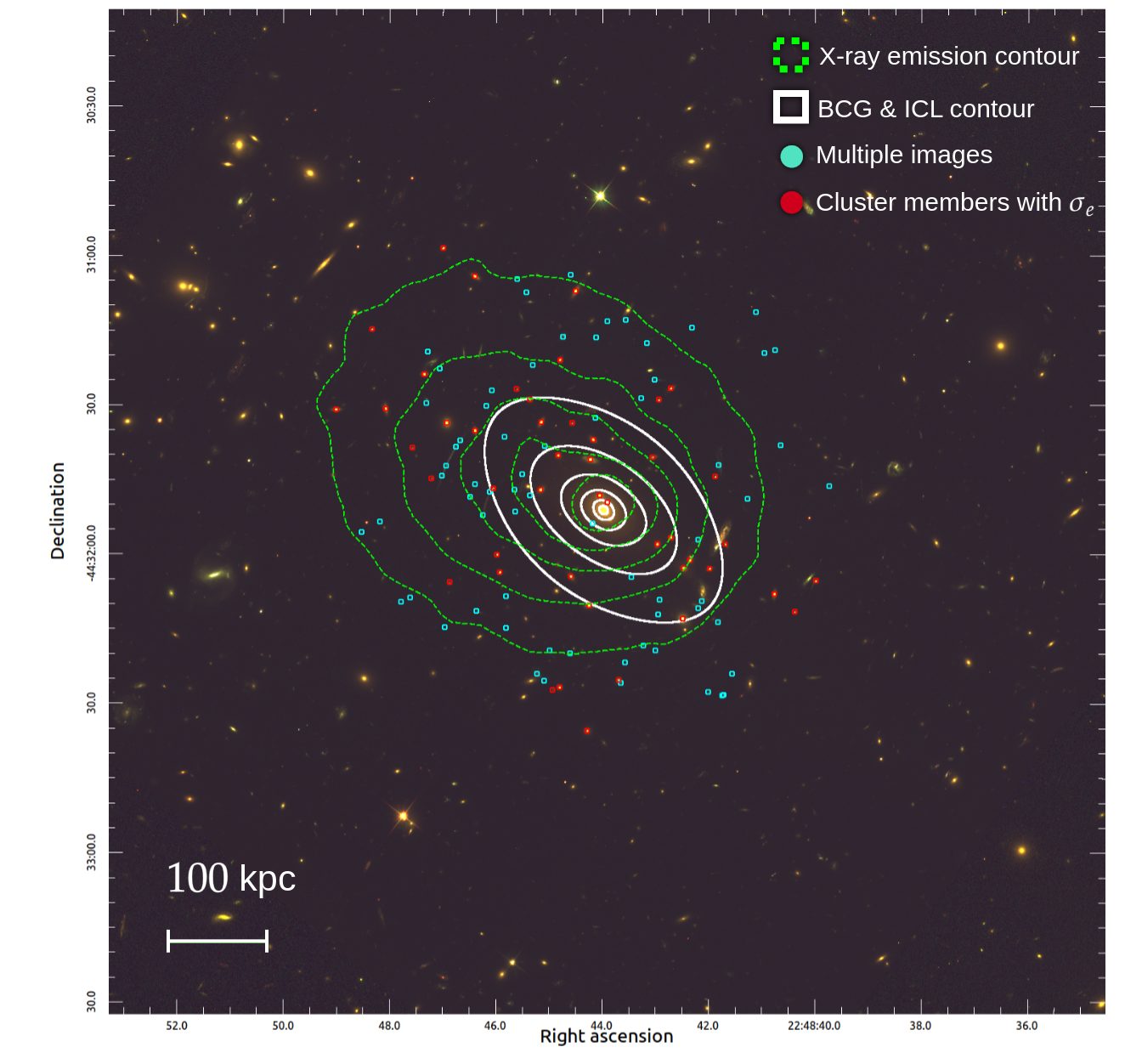}
    \caption{The BUFFALO colour composite image of AS1063 with the following \emph{HST} filters: ${\rm F435W}$ (blue), ${\rm F606W}$ (green) and ${\rm F814W}$ (red). The X-ray surface brightness from the \textsc{Chandra X-ray Observatory} is shown by the green dashed contours. The set of multiple images used in this work is highlighted by the cyan circles. The white contours present the BCG \& ICL light distribution as fitted in B25a. Cluster members for which we measure the velocity dispersion, $\sigma_e$, in B25a are highlighted by red circles.}
    \label{fig:summary_S1063}
\end{figure*}

In \citet{Beauchesne2025a} (hereafter B25a), we detail the mass constraints required to set up our comprehensive mass modelling of AS1063. In section~3, we present the existing set of constraints from strong lensing and X-rays taken from B24. In sections~4 and 5, we focus on the supplementary mass probes, i.e. cluster members and the BCG \& ICL. That set of observations is the same for both components, although we apply different treatments adapted to their properties. We recover the cluster member light distribution by fitting all galaxies in the cluster field to reduce bias from the crowded environment. Thanks to this light model, we are able to subtract the galaxy light model to fit the BCG \& ICL with a multi-Gaussian expansion (MGE). For both components, we fit their spectral energy density (SED) and their stellar kinematics to obtain stellar mass estimates as well as a total mass probe. We recall a part of the mass constraints presented in B25a, in Fig~\ref{fig:summary_S1063}.

In this second paper, we present the mass modelling methodology and detail how we use the supplementary mass constraints from B25a. In Sect.~\ref{sect:mass_model}, we detail the mass model parametrisation of each component, where we use the light distribution from cluster members and the BCG \& ICL as a tracer of their stellar mass distribution. In Sect.~\ref{sect:model_opti}, we define the likelihoods for each mass probe: X-ray surface brightness, strong lensing systems, cluster member kinematics, and the BCG \& ICL component. We conclude this section by describing the tailored optimisation process required by the complex mass model needed to accommodate all these constraints. We then present the mass model results in Sect.\ref{sect:results} and discuss the reliability of the mass estimates from the new component in Sect.\ref{sect:discussion}.

We recall the convention used in B25a, where we adopt a flat $\Lambda$CDM cosmology with $\Omega_\Lambda=0.7$, $\Omega_m=0.3$ and $H_0=70$ km~s$^{-1}$Mpc$^{-1}$. Magnitudes are quoted in the AB system. The uncertainties quoted in that article are the median-centred credible intervals ($\rm CI$) based on the posterior distribution of the considered random variable. These intervals are presented as $n\sigma\, \rm CI$ where $n$ is an integer such that $\rm CI$ contains $100\times \rm erf\left(\frac{n}{\sqrt{2}}\right)$ per cent of the posterior. If we use $\sigma$ to denote a standard deviation, we explicitly notify it in the text.

\section{Mass modelling hypotheses}
\label{sect:mass_model}

In our model, we aim to disentangle the DM from the baryonic components at galaxy and cluster scales. In the following, we define how we model the main cluster components (DM, BCG \& ICL, intra-cluster medium, and cluster members) so that their respective baryonic fractions can be assigned their own mass components.

\subsection{Intra Cluster Medium}

To model the ICM, we follow the same method as in B24, as it already fits our requirement of separating the baryonic mass from the DM. The gas distribution is modelled by a sum of dPIE potentials in a free-form way. All potentials are free to move within the cluster field, and we only fix their cut radii at $1.25$~Mpc, as in B24. Indeed, it is ill-constrained when more than one potential is used. This cut radius limit corresponds to the radius at which the observed number counts are approximately equal to the background value. 

The number of potentials is determined following the goodness-of-fit procedure of B24, whereby potentials are added iteratively until the best-fitting model adequately reproduces the observed X-ray emission according to the B24 criterion. This criterion is an adaptation of the usual goodness-of-fit assessment with $\chi^2~1$ for the X-ray likelihood defined in Sect.~\ref{sect:X-ray-likelihood}. Hence, as in B24, we use three dPIEs to model the hot gas in AS1063.

\subsection{Cluster members}
\label{sect:cluster_member_definition}

The cluster member distribution is one of the less significant mass components of the cluster, although due to the mass concentration of the galaxy, it locally influences the position of multiple images. Hence, they are an important part of a cluster lens mass model, as some systems cannot be reproduced without them. However, due to their large number and the scarcity of lensing constraints, several simplifying assumptions about their parameters are made. These simplifications are usually based on a ``light traces matter'' principle.

In this work, we deviate from the standard modelling of cluster members used in the \textsc{Lenstool} software \citep{Kneib1993,jullo2007} which is based on the Faber \& Jackson law \citep{Faber&Jackson1976}. It scales the dPIE parameters ($r_{\rm core}$, $r_{\rm cut}$ and $\sigma_0$) of each cluster member based on its luminosity $L$ and a reference galaxy ($r^*_{\rm core}$, $r^*_{\rm cut}$, $\sigma^*_0$ and $L^*$). During the parameter inference, only the dPIE parameters of the reference galaxy are optimised. The scaling relations are the following:
\begin{equation}
r_{\rm cut}=r_{\rm cut}^*\left(\frac{L}{L^*}\right)^{1/2} \text{ ; }
r_{\rm core}=r_{\rm core}^*\left(\frac{L}{L^*}\right)^{1/2} \text{ ; }
\sigma_0=\sigma_0^*\left(\frac{L}{L^*}\right)^{1/4}
\end{equation}
the other parameters of the dPIE are usually fixed from the light profile: centre, position angle and ellipticity. Some works prefer circular cluster members rather than the ellipticity of the light distribution. Hence, the DM and baryons of each galaxy are modelled by a single profile.

In this work, we aimed at separating the DM from the baryons in each galaxy using individual mass profiles to be consistent with the cluster-scale model. Since the baryonic distribution of elliptical galaxies is dominated by stars, it can serve as a stellar tracer for kinematic modelling, providing more local mass constraints than lensing. Hence, separating baryons and DM in cluster members not only gives more information to extract from the model, but it also allows the model to use more observational datasets. To make such a framework work, we need relations that link the baryons and DM profiles. So instead of using a relation based on observations such as the Faber \& Jackson law or fundamental plane of an elliptical galaxy, we use the Stellar-to-subhalo mass relation (SsHMR). It has the advantages to be easily comparable to simulation results.

In this new framework, we still use dPIE potentials for their computational efficiency rather than combining a S\'ersic and a power-law profile as it is commonly done in galaxy-scale lensing analyses \citep[e.g.][]{Shajib2021}. However, a dPIE profile can approximate a S\'ersic light profile relatively well for a S\'ersic index between $1$ and $4$ \citep{Dutton2011}. Each galaxy component (baryons and DM) is modelled with a separate dPIE profile, subject to several simplifying assumptions to limit the number of free parameters. Hence, we use the same centre coordinates, ellipticity and position angle for both dPIEs. These parameters are assumed to follow the light distribution. Such approximations are poor for disk galaxies, although the majority of the included cluster members are elliptical galaxies or galaxies in transition to elliptical. We fix these parameters to the best-fitting values of the cluster member light profile with a dPIE presented in B25a (section~4.1). The core and cut radii of the baryonic dPIE ($r_{\rm core,b}$ and $r_{\rm cut,b}$), are fixed to the values of the light profile as well. Only the central velocity dispersion $\sigma_{0,b}$ remains free, which we rescale to match the stellar mass from B25a (section~4.3). Hence, the baryonic component is fixed in the overall model and is determined only from measurements independent of the lensing analysis.

For the DM component, radii, $r_{\rm core,DM}$ and $r_{\rm cut,DM}$, and velocity dispersion, $\sigma_{0,DM}$, are scaled by a spatial and concentration scale, $\alpha_c$ and $\beta_c$, with respect to their baryonic counterpart, ($r_{\rm core,b}$, $r_{\rm cut,b}$ and $\sigma_{0,l}$), such that:
\begin{align}
\label{equ-def-cluster-member}
    &r_{\rm core,DM}=\alpha_c r_{\rm core,b} \\
    &r_{\rm cut,DM}=\alpha_c r_{\rm cut,b} \\
    &\sigma_{0,DM}=\beta_c \sigma_{0,l} 
\end{align}
The spatial scale, $\alpha_c$, is assumed to be the same for all cluster members, and is optimised with our set of heterogeneous constraints. This parameter is expected to exhibit a radial, cluster-centric variation, as it reflects the tidal stripping of DM halos experienced by galaxies falling into the cluster potential well \citep{niemiec2022}. We leave the exploration of such variations for future work.

$\beta$ is different for each cluster member, and is recovered through the modelling of the SsHMR in the model under the form of a double power-law as in \citet{niemiec2022} and \citet{Moster2013}, such that:
\begin{equation}
    M_{\rm tot}=2N\left (\left(\frac{M_*}{M_1}\right)^\delta+\left(\frac{M_*}{M_1}\right)^{-\gamma}\right)M_*
\end{equation}
where $N$, $M_1$, $\delta$ and $\gamma$ are optimised, and define the cluster member distribution model with $\alpha_c$. We could switch roles between $\alpha_c$ and $\beta_c$. In that case, we would optimise $\beta_c$ and set $\alpha_c$ with the SsHMR. However, $\beta_c$ is more likely to be ill-constrained by strong lensing as it mostly affects cluster members inner profiles. It is still not perfect as $\gamma$ and $\alpha_c$ are not well constrained. We impose an upper bound on them, as a large value of $\gamma$ and $\alpha_c$ leads to unrealistic mass for the faintest cluster member, which is not constrained. We lack velocity dispersions for these cluster members, and these unrealistic masses are highly concentrated, which severely reduces the ability to penalise them through strong lensing. More generally, $\alpha_c$ does not seem to be well bounded by the data, and as it increases, it changes the mass balance between the other components, which makes it hard to match the measured velocity dispersion of the BCG within the two-step optimisation process presented in Sect.~\ref{sect:optimisation-process}, for example. We based the upper bound on empirical limits obtained from preliminary models.

In our modelling, the SsHMR is the usual scaling relation used for cluster members modelling with \textsc{Lenstool}. However, instead of using an observational law, we use a parametrisation oriented toward comparison with cosmological simulations. The SsHMR of our model can be directly compared with simulation results, such as those presented in \citet{niemiec2022}. We recover $\beta_c$ from the SsHMR by writing the total mass of the combined dPIE models, $M_{\rm tot}$, such that:
\begin{equation}
    M_{\rm tot}=\left(\alpha_c\beta_c^2\frac{L M_\odot}{L_\odot}+M_*\right)
\end{equation}
where $L$ is the total luminosity of the dPIE fitted to the light distribution, and $M_*$ is the stellar mass associated with the object.

In contrast to B24, we do not inject the cluster member line-of-sight velocity dispersion (LOSVD) in their associated dPIE parameters. We would rather use them as a new set of constraints, as presented in Sect.~\ref{sect:cl_likelihoods}. In doing so, the measurement uncertainties on the LOSVD are more directly accounted for, and their weight naturally balanced against that of other mass probes, such as strong lensing, within the inference process. In B24, these quantities are accounted for with the prior on the galaxy scaling law.

As in B25a (section~4.3), we obtain stellar mass estimates from different SED models used to define three different parametrisations of cluster members. Hence, we gain insights into the influence of SED systematic biases on this type of modelling, particularly on the SsHMR discussed in Sect.~\ref{sect:SsHMR_discuss}.

\subsection{BCG \& ICL}
\label{sect:ICL-model-param}
To model the BCG \& ICL component, we assume a stellar component to represent its baryonic distribution and a DM component for the BCG. The motivation behind the BCG DM component is to allow for different mass slopes at scales that differ from those of the main DM halo.

We model the BCG \& ICL stellar component using an MGE parametrisation. We begin with the MGE model of its light distribution in the ${\rm F160W}$ filter, measured in B25a (section~5.1). Contours of the MGE model are presented in Fig.~\ref{fig:summary_S1063}. We adopt identical parameters for light and mass distributions, except that the Gaussian normalisations now represent the stellar mass-to-light ratio $\Upsilon^{\rm BCG}_{\rm *}$ of the BCG \& ICL component. We designate the estimate of $\Upsilon^{\rm BCG}_{\rm *}$ by \textsc{Lenstool} as $\Upsilon^{\rm BCG}_{\rm *,\,lt}$, in contrast to $\Upsilon^{\rm BCG,\,F160W}_{\rm *}$ that we refer to as $\Upsilon^{\rm BCG}_{\rm *,\,SED}$. We could impose a prior on $\Upsilon^{\rm BCG}_{\rm *,\,lt}$ based on $\Upsilon^{\rm BCG}_{\rm *,\,SED}$ as estimated in B25a (section~5.3). However, we prefer a uniform prior since SED fits can misestimate this ratio when the assumed initial mass function (IMF) is inappropriate for the object. In particular, stellar mass estimates from kinematic modelling of early-type galaxies present significant differences compared to SED estimates \citep[for a review on this topic see][]{Smith2020}. To further accommodate such discrepancies, we define three MGE parametrisations which allow for different degrees of variation of $\Upsilon^{\rm BCG}_{\rm *}$ with the cluster-centric radius. The first parametrisation assumes constant $\Upsilon^{\rm BCG}_{\rm *}$, renormalising the entire light distribution with a single coefficient $\Upsilon^{\rm BCG,\, 1}_{\rm *,\,lt}$. We refer to this parametrisation as ``BCG - ML 1''. The second uses two coefficients: $\Upsilon^{\rm BCG,\, 2}_{\rm *,\,lt\,1}$ for Gaussians with standard deviations smaller than the BCG half-light radius ($R_e=17$~${\rm kpc}$; \citet{tortorelli2018}), and $\Upsilon^{\rm BCG,\, 2}_{\rm *,\,lt\,2}$ for wider Gaussians. This effectively separates the BCG from the ICL contributions. We refer to this parametrisation as ``BCG - ML 2''. To accommodate wider variations of $\Upsilon^{\rm BCG}_{\rm *}$, we make three groups of Gaussians through a K-means clustering method as implemented in \textsc{scikit-learn} \citep{scikit-learn}, each represented by one coefficient, $\Upsilon^{\rm BCG,\, 3}_{\rm *,\,lt\,i}$ where $i\in\{1,2,3\}$. $\Upsilon^{\rm BCG,\, 3}_{\rm *,\,lt\,1}$ represents the same group of Gaussians as $\Upsilon^{\rm BCG,\, 2}_{\rm *,\,lt\,1}$, while $\Upsilon^{\rm BCG,\, 3}_{\rm *,\,lt\,2}$ and $\Upsilon^{\rm BCG,\, 3}_{\rm *,\,lt\,3}$ allow for more variation of $\Upsilon^{\rm BCG}_{\rm *}$ within the ICL. We refer to this last parametrisation as ``BCG - ML 3''.

Since \textsc{Lenstool} lacks native MGE potential implementation, we incorporate them as fixed external maps of lensing quantities (computed in B25a, section~5.1) scaled by normalisation factors. When we split the MGE parametrisation into groups, we apply the same procedure to each group to compute the lensing quantities as if they were separate MGE potentials. We include the map of the lensing quantities by interpolating them with degree $3$ B-spline surfaces, allowing us to use the spline implementation of \textsc{Lenstool} \citep{Beauchesne2021}. In contrast to \citet{Beauchesne2021}, we do not impose boundary conditions, and there is a different B-spline surface for each quantity: lensing potential, deflection angle, convergence and shear. We simply use the implementation to compute the value of the B-spline surface. This approach generalises to any light distribution parametrisation, accommodating complex cases such as SMACS\,J0723.3-7323 \citep{Mahler2023}. We chose an MGE parametrisation specifically to enable stellar kinematic analysis through Jeans Anisotropic Modelling (JAM), which requires MGE-parametrised mass and light profiles.

To model the DM component of the BCG, we use a dPIE profile. Regarding its parameters, we fix its centre to the BCG centroid and leave the other parameters free to vary. We allow the position angle and ellipticity to vary, as their values may differ from those of the light distribution. We also allow for possible offsets between the DM and light distributions, as parametrised by the smooth cluster-scale DM halo that we define in the following section. We note that this DM component was required in addition to the cluster-scale DM to obtain a sufficiently high velocity dispersion near the BCG and match the observation.

\subsection{Dark matter}

For the DM cluster-scale smooth component, we use a modelling approach similar to other \textsc{Lenstool} models to represent the bulk of the cluster mass. This component is often referred to as the cluster DM component by neglecting baryons contained in the ICM and the ICL. In our case, since we are modelling the baryons with other profiles, it only accounts for the DM mass.

We follow the approach of B24 by using two dPIEs haloes, one representing the main cluster DM halo, and a second one representing the North-East (NE) halo. Regarding the model parameters, we use the same assumptions as B24 for the NE halo position and radii, but we leave its position angle and ellipticity free to vary. We allow all of these parameters to vary for the main clump, and we constrain its central position to a square of $10$~${\rm arcsec}$ centred around the BCG position. Similarly to B24, we fix the cut radius to $3$~${\rm Mpc}$.

These two additional dPIEs complete the cluster mass model. Overall, the model includes $588$ different profiles. Cluster members dominate this count, with each galaxy contributing two profiles (one for baryons, one for DM). Despite the large profile count, most free parameters characterise cluster-scale components, such as the main DM halo and ICM, which make fewer assumptions than the cluster member population.

\section{Likelihoods}

\label{sect:model_opti}

\begin{figure*}
    \centering
    \includegraphics[width=0.8\linewidth]{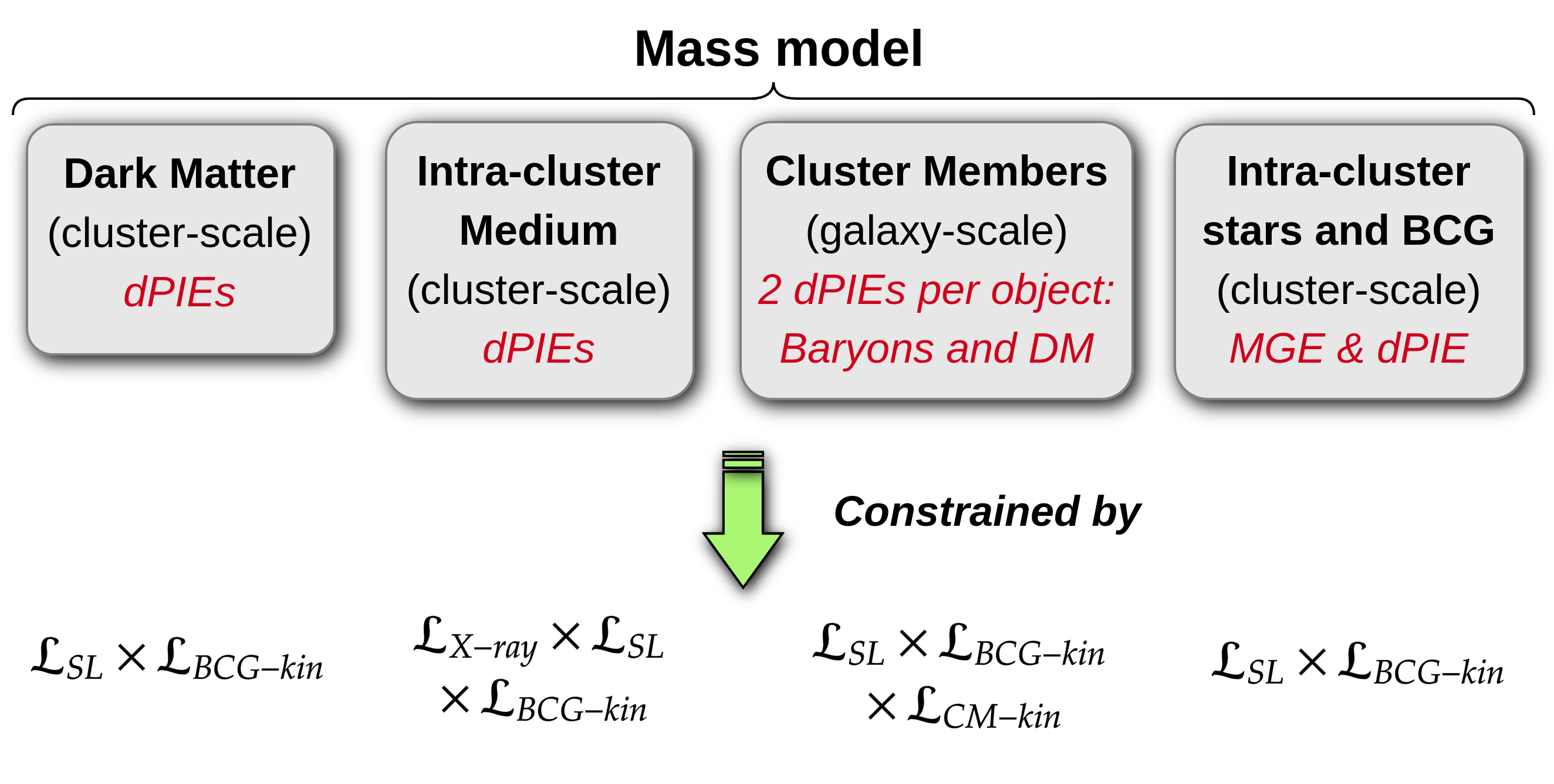}
    \caption{Overview of each model component. Each one of them is associated with a set of analytical profiles and the likelihood that they are affecting.}
    \label{fig:Model_overview}
\end{figure*}
Thanks to our comprehensive model presented in Sect.~\ref{sect:mass_model}, we can define likelihoods for our set of heterogeneous constraints: lensing, cluster members, $\sigma_e$, X-ray surface brightness and BCG \& ICL, $V_{\rm rms}$. We recall the set of constraints in Fig.~\ref{fig:summary_S1063}, with X-ray surface brightness contours, multiple image systems and cluster members with measured kinematics. Each of these likelihoods constrains different parts of the model, as shown in Fig.~\ref{fig:Model_overview}, where each model component is associated with a different set of likelihoods. We have strong lensing ($\mathcal{L}_{SL}$), X-rays ($\mathcal{L}_{ X-ray}$), cluster member kinematics ($\mathcal{L}_{CL- kin}$) and the BCG \& ICL kinematics ($\mathcal{L}_{BCG- kin}$) likelihoods. Only the first one, $\mathcal{L}_{SL}$, constrains the entire mass distribution, as it is sensitive to the cluster's total mass. At the exception of $\mathcal{L}_{BCG-kin}$, which also depends on the total mass in the cluster core, the last two, $\mathcal{L}_{X-ray}$ and $\mathcal{L}_{CL-kin}$, only constrain one model component each. We define each of these likelihoods in the following sections. In Sect.~\ref{sect:optimisation-process}, we detail our parameter inference procedure.

\subsection{X-ray surface brightness}
\label{sect:X-ray-likelihood}
To allow our model to account for the ICM X-ray emission as a constraint, we use the mass distribution of the dPIEs associated with the gas to predict the cluster X-ray surface brightness. We use the same transformation as in B24 to convert the gas mass distribution to its associated electron density, and then its surface brightness. We refer the reader to B24 and the references within for a detailed procedure description. 

To define the likelihood and to take into account the limitations of our modelling choices, we incorporate an intrinsic error on the X-ray photon count, $\sigma_X$, and we use a likelihood based on the Negative Binomial distribution ($\mathbb{NB}(r=\mu_i^2/\sigma_{\rm X}^2,p=\mu_i/(\mu_i+\sigma_{\rm X}^2))$ with $\mu_i$ the expected photon count in a bin). The full likelihood is then defined as follows:

\begin{align}
    \mathcal{L}_{X-ray}=\prod_i \frac{\Gamma\left(k_i+\frac{\mu_i^2}{\sigma_{\rm X}^2}\right)}{k_i!\Gamma\left(\frac{\mu_i^2}{\sigma_{\rm X}^2}\right)}\left(\frac{\mu_i}{\mu_i+\sigma_{\rm X}^2}\right)^\frac{\mu_i^2}{\sigma_{\rm X}^2}\left(\frac{\sigma_{\rm X}^2}{\mu_i+\sigma_{\rm X}^2}\right)^{k_i}
\end{align}

where $\mu_i$ is the expected photon counts from the model in each pixel of the map, $k_i$ is the observed number of photon counts in a bin, and $\Gamma$ represents the well-known gamma function.

In contrast to B24, we do not use a constant value of $\sigma_{\rm X}^2$ for the whole field of AS1063. We use an intrinsic error proportional to the expected value of the surface brightness model, such that $\sigma_{\rm X}=a\mu_i$, where $a$ is the error parameter to optimise. It is better suited to the motivation behind adding this error. The inclusion of $\sigma_{\rm X}^2$ was motivated in B24 by the inability of the model to fully reproduce gas mass variations on scales below that of the cluster. The gas sloshing pattern identified in B24 is an example of such variations. Hence, we expect such phenomena to have an amplitude correlated to the gas density as reflected by the expression of $\sigma_{\rm X}$. If a constant error is still needed, it would likely indicate a misestimation of the background sky.

\subsection{Multiple images positions}

For the strong lensing constraints, we also follow the same definition as in B24, particularly in choosing the source position of each multiply-imaged system. Indeed, we do not rely on a weighted barycenter of the positions of each image in the source plane but rather use it as a starting point that we optimise to find the best-available source position for a given system. We assume that the uncertainties on the positions of multiply-imaged systems are well represented by independent Gaussian distributions, which gives the following likelihood definition:

\begin{equation}
    \mathcal{L}_{SL}=\prod^{N_{\rm sys}}_j \frac{1}{\prod^{N_{\rm im,j}}_i \sigma_{ij}\sqrt{2\pi}} \exp\left(-\chi^2_j/2\right)
\end{equation}
with $N_{\rm sys}$, the total number of multiply-imaged systems, $N_{\rm im,j}$, the number of images in the $j^{th}$ system, and $\sigma_{ij}$, the observational error on the $i^{th}$ images of the $j^{th}$ system. $\chi^2_j$ is the $\chi^2$ statistics associated with the $j^{th}$ system and reads as follows:
\begin{equation}
    \chi^2_j=\sum^{N_{\rm im,j}}_i \frac{\left|\left|\vec{\theta}^{\rm obs}_{i,j}-\vec{\theta}^{\rm pred}_{i,j}\right|\right|^2}{\sigma_{ij}^2}
\end{equation}

A positional uncertainty of $0.55$~${\rm arcsec}$ is adopted to account for both the observational uncertainties in image position identification and the systematic limitations of the model. This value is determined by a maximum-likelihood estimate, calibrated such that the best-fitting model achieves a reduced $\chi^2$ of approximately $1$.

\subsection{Cluster member kinematics}
\label{sect:cl_likelihoods}
Thanks to our detailed cluster members modelling, we can access both their light and mass distributions in terms of dPIEs. We use a similar approach to B24 to model the LOSVD of the cluster members at the half-light radius, $\sigma_e$, from the light and mass distributions. The main difference is that we do not try to obtain $\sigma_0$ from $\sigma_e$, but $\sigma_e$ as a function of dPIEs parameters.

Following \citet{Beauchesne2024,Bergamini2019}, the LOSVD, $\sigma_{\rm LOS}$, is expressed as follows:
\begin{equation}
    \sigma^2_{\rm LOS}(R)=\frac{2G}{I(R)}\int^{\infty}_{R} \nu(r) M_{\rm 3D}(r) \mathcal{F}(r)\, r^{2\beta_{\rm aniso}-2}{\rm d}r
\end{equation}
where $I$ and $\nu$ are the surface density and density of the luminosity distribution, respectively. $M_{\rm 3D}$ is the total mass enclosed in a sphere of radius $r$. It leaves us with two terms, $\mathcal{F}$ and $\beta_{\rm aniso}$, which are linked to the anisotropy of stars' orbits. For simplicity, we assume that these orbits are isotropic, which gives the following formulas \citep{Cappellari2008}:
\begin{equation}
    \mathcal{F}(r)=\sqrt{r^2-R^2} \text{   and   } \beta_{\rm  aniso}=0
\end{equation}
To match the actual measurements of $\sigma_{\rm ap}$ and their extraction apertures, $\sigma_{\rm LOS}$ is averaged in circular apertures of radius $R'$. We obtain the following expression:
\begin{equation}
   \label{Eq:sigma_formula_compact}
   \sigma^2_{\rm ap}(R')=\frac{2\pi}{L(R')}\int_0^{R'} \sigma^2_{\rm LOS}(R) I(R) R {\rm d}R
\end{equation}
where $L(R')$ is the luminosity enclosed in a circle of radius $R'$. In our case, we have three dPIEs, one for the light $(\sigma_{\rm 0,l},r_{\rm core,l},r_{\rm cut,l})$, one for the baryons $(\sigma_{\rm 0, b},r_{\rm core, b},r_{\rm cut, b})$ and one for DM $(\sigma_{\rm 0, DM},r_{\rm core, DM},r_{\rm cut, DM})$ masses. We assume that the mass and light of the considered cluster members are the only components that significantly affect the LOSVD. Thus, we obtain the following expressions for the different terms presented in the previous equations:

\begin{align}
    \nu\left(r\right)=&\frac{\sigma_{\rm 0,l}^2(r_{\rm core,l}+r_{\rm cut,l})}{2 \Upsilon_{\odot} \pi G r_{\rm core,l}^2 r_{\rm cut,l}}\frac{1}{\left(1+\left(\frac{r}{r_{\rm core,l}}\right)^2\right)\left(1+\left(\frac{r}{r_{\rm cut,l}}\right)^2\right)}\label{Eq:dPIE_formula_sigma-1}\\
    I\left(r\right)=&\frac{2\sigma_{\rm 0,l}^2 r_{\rm cut,l}}{G\left(r_{\rm cut,l}-r_{\rm core,l}\right)}{\Upsilon_{\odot}}\left(\frac{1}{\sqrt{r^2+r_{\rm core,l}^2}}-\frac{1}{\sqrt{r^2+r_{\rm cut,l}^2}}\right)\label{Eq:dPIE_formula_sigma-2}\\
    M_{\rm 3D}\left(r\right)=&\sum_{i\in\{b,DM\}} \Biggl(\frac{2\sigma_{\rm 0,i}^2 r_{\rm cut,i}}{G\left(r_{\rm cut,i}-r_{\rm core,i}\right)}\times\\
    &\left(r_{\rm cut,i} \arctan\left(\frac{r}{r_{\rm cut,i}}\right)-r_{\rm core,i} \arctan\left(\frac{r}{r_{\rm core,i}}\right)\right)\Biggr)\label{Eq:dPIE_formula_sigma-3}\\
    L\left(r\right)=&\frac{2\sigma_{\rm 0,l}^2 r_{\rm cut,l}}{G\left(r_{\rm cut,l}-r_{\rm core,l}\right)}{\Upsilon_{\odot}}\times\\
                    &\left(\sqrt{r^2+r_{\rm core,l}^2}+r_{\rm cut,l}-\sqrt{r^2+r_{\rm cut,l}^2}-r_{\rm core,l}\right)\label{Eq:dPIE_formula_sigma-4}
\end{align}

In our case, we extract the LOSVD in a radius equal to the half-light radius of the considered cluster member, which in the case of the dPIE is \citep{Eliasdottir2007}:
\begin{equation}
    R_e=\frac{3}{4}\sqrt{r^2_{\rm core,l}+\frac{10}{3}r_{\rm core,l}r_{\rm cut,l}+r^2_{\rm cut,l}}
\end{equation}
We have now settled everything to compute $\sigma_e=\sigma_{\rm ap}(R_e)$, and define the likelihood associated with cluster member kinematics, $\mathcal{L}_{\rm CM-kin}$. We assume that each $\sigma_e$ measurement is well-represented by a Gaussian distribution, which gives the following definition:
\begin{equation}
    \mathcal{L}_{CM-kin}=\prod^{N_{\rm gal}}_j \frac{1}{\delta_{\sigma_{e,i}}\sqrt{2\pi}} \exp\left(-\frac{\left(\sigma_{e,i}-\sigma_{\rm e,model,i}\right)^2}{2\delta^2_{\sigma_{e,i}}}\right)
\end{equation}
where $N_{\rm gal}$ is the number of cluster members with a measured LOSVD, and $\delta_{\sigma_{e,i}}$ is the observational error on that measurement. $\sigma_{e,i}$ and $\sigma_{\rm e,model,i}$ are the values of the measured and model-predicted LOSVD, respectively.

As our model parameterisation includes several hypotheses to simplify the model, we cannot reproduce each LOSVD within its error bars. Similarly to the lensing constraints, we add an intrinsic error, such that the associated reduced $\chi^2$ is approximately $1$ for the mode of the posterior. We estimate the systematic error to be $35$~${\rm km/s}$ based on a preliminary maximum-likelihood estimate.

\subsection{BCG \& ICL kinematics}
\label{sect:BCG_ICL_kin_def}
To model the kinematics of the BCG \& ICL, we rely on JAM to obtain a 2D model and the \textsc{JamPy} package \citep{Cappellari2008}. We already have an axisymmetric MGE representation of the light distribution as measured in B25a (section~4.1). Thus, we need to define a similar representation for the mass so we can compute the JAM kinematic model. To be computationally efficient, we base the mass model on the 1D mass profile of the galaxy cluster. We compute the average mass density in spherical shells, which we fit with a 1D MGE representation. We include all profiles that define cluster-scale distributions, such as the smooth DM, the gas and the BCG \& ICL components, in the mass density. In preliminary tests, we find that adding some cluster members is necessary to account for all the mass required by the stellar kinematics. Hence, we include each cluster member within the kinematics extraction area. This selection is correct under the assumption of a single mass plane, although in reality, we may include cluster members that would not be in the spherical shells used to estimate the 1D mass profile. We build the 2D MGE mass by assuming the same position angle as the light distribution. This is required by the JAM method. To estimate an ellipticity as close as possible to the underlying mass distribution, we uniformly sample points in the combined area of all extraction regions of the kinematics. We compute the second moment of the mass distribution in the $x$ and $y$ axes of the \textsc{lenstool} reference frame and estimate the ellipticity with the same formulae as \textsc{Sextractor}\footnote{\url{https://sextractor.readthedocs.io/en/latest/Position.html}}. In preliminary tests, we tried to use the 2D mass density and obtain the 2D MGE representation directly. However, the requirement of having the same position angle led to a poor fit of the mass by the 2D MGE.

Given a representation of light and mass as MGEs, we need to specify the remaining parameters of the JAM method to properly define the kinematic model. These parameters are the inclination with respect to the LOS, and the anisotropy of the stars' orbits. We marginalise over the inclination by uniformly sampling it within the range that yields a viable kinematic model. We define the minimal inclination, $\theta_{\rm min}$, with the following equation from the JAM code parameter restriction:
\begin{equation}
    q_{\rm obs,min}=\sqrt{\cos^2(\theta_{\rm min})+q_{\rm min}\sin^2(\theta_{\rm min})}
\end{equation}
where $q_{\rm obs,min}$ is the minimal axis ratio among the Gaussians of both MGE representations (i.e. light and mass), and $q_{\rm min}$ is the minimal axis ratio of the deprojected Gaussians in 3D. As in the JAM code, we fix $q_{\rm min}$ to $0.05$. $\theta_{\rm max}=90$~degree represents the edge-on case. In the JAM code, there are multiple choices for the anisotropy of the stars' orbits: cylindrical alignment of the velocity ellipsoid \citep{Cappellari2008}, spherical alignment \citep{Cappellari2020} and custom functions such as a logistic \citep{simon2024} or a constant function. In the preliminary test, we found no significant difference between the two alignment types when integrating the kinematic model across the observational bins. We try to use a logistic function, as in \citet{simon2024}, to constrain the anisotropy using the data. We used a gradient descent method to optimise the anisotropy in a constant-mass model, but the parameters remained poorly constrained and stayed near the starting points. Hence, we use the cylindrical alignment, as it is less computationally costly than the spherical one. Once the JAM 2D model is defined in the area where $V_{\rm rms}$ is measured, we integrate it over the elliptical annuli used for extraction using a trapezoidal method.

In the observation bins, we then define the BCG \& ICL kinematics likelihood $\mathcal{L}_{\rm BCG-kin}$ as a Gaussian likelihood on the $V_{\rm rms}$ measurements such that:
\begin{equation}
    \mathcal{L}_{BCG-kin}=\prod^{N_{\rm bin}}_j \frac{1}{\delta_{V_{\rm rms,i}}\sqrt{2\pi}} \exp\left(-\frac{\left(V_{\rm rms,i}-V^{\rm model}_{\rm rms,i}\right)^2}{2\delta^2_{V_{\rm rms,i}}}\right)
\end{equation}
where $N_{\rm bin}$ is the number of extraction bins where $V_{\rm rms,i}$ has been measured with the observational error $\delta_{V_{\rm rms,i}}$. $V^{\rm model}_{\rm rms,i}$ is the model predicted LOS velocity moment.

\subsection{Optimisation process}
\label{sect:optimisation-process}
\begin{figure*}
    \centering
    \includegraphics[width=.8\linewidth]{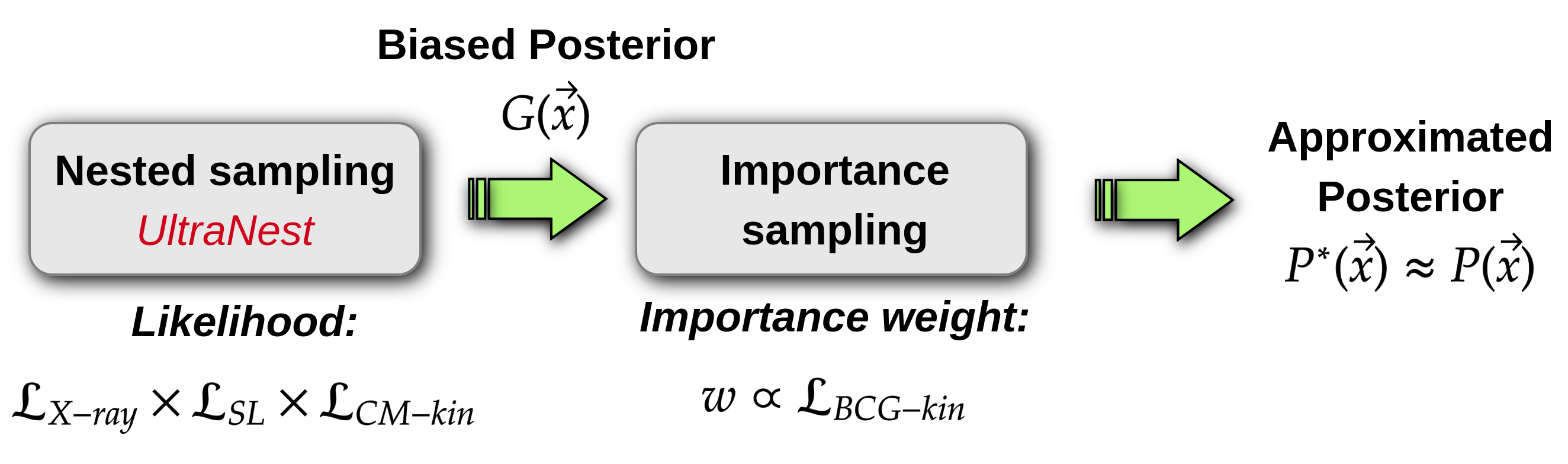}
    \caption{Diagram of the inference workflow in two steps. A biased posterior distribution is obtained on the combined likelihood excluding the BCG \& ICL kinematics. In the second step, we resample the biased posterior through importance sampling to obtain the final posterior estimate.}
    \label{fig:inference-workflow}
\end{figure*}

In the case of our modelling, the computational cost of the combined likelihood is too high to use regular sampling methods for the posterior, such as Markov Chain Monte Carlo or Nested sampling. However, this cost begins to be too expensive only when we consider the kinematic model of the BCG \& ICL. Indeed, this likelihood constrains only a subset of the model parameters, yet its inclusion increases the computational cost of each likelihood evaluation, leading to a significant slowdown.

The BCG \& ICL kinematics are particularly interesting as they complete the strong lensing constraints in the cluster core. Hence, we do not want to exclude them. We converge on the optimisation process presented in Fig.~\ref{fig:inference-workflow}, which uses an approximation to obtain the posterior of the total likelihood. We start the inference scheme by running a regular sampling method to obtain the posterior of the incomplete likelihood $\mathcal{L}_{CM-kin}\times\mathcal{L}_{SL}\times\mathcal{L}_{X-ray}$. In our case, we rely on the nested sampling Monte Carlo algorithm, \textsc{MLFriends} \citep{Buchner2014,Buchner2019} implemented in the Python package \textsc{Ultranest}\footnote{\url{https://johannesbuchner.github.io/UltraNest/}} \citep{Buchner2021}. We then consider this posterior as a biased distribution in the context of importance sampling. We aim to obtain an approximation to the posterior defined by our total likelihood, which serves as the target distribution for the importance sampling. In that case, we can determine the importance weight, $w$, as follows:
\begin{align}
    w&\propto\frac{\mathcal{L}_{CM-kin}\times\mathcal{L}_{SL}\times\mathcal{L}_{X-ray}\times\mathcal{L}_{BCG-kin}}{\mathcal{L}_{CM-kin}\times\mathcal{L}_{SL}\times\mathcal{L}_{X-ray}}\frac{{\rm P}(\theta)}{{\rm {\rm P}(\theta)}}\\
               &\propto\mathcal{L}_{BCG-kin}
\end{align}
where ${\rm P}(\theta)$ represents the priors of the whole model parameters. As we can see, the importance weights vastly simplify in our case, such that they are only proportional to $\mathcal{L}_{\rm BCG-kin}$. We then use the systematic resampling algorithm presented in \citet{Hol2006} to obtain our approximated posterior. 

We note that such a procedure works only if the biased and target posteriors are not too different, such that the resampling keeps as many samples as possible from the biased posterior. We address this issue in our case in Sect.~\ref{sect:BCG-ICL-kin-res}, where we present the resulting posterior on the kinematic model of the BCG \& ICL.

Due to the complexity of the current model, particularly the number of dPIE potentials, we have to speed up the first inference step further. Due to limitations in multi-threading/processing in the Bayesian inference code presented above, the first step would take approximately a month on $60$ CPU cores spread across several computing nodes. Hence, we use a particle swarm optimiser \citep{Bonyadi2017} to obtain a maximum likelihood estimate and a sample of models. From this set of points, we model the likelihood using a multivariate normal distribution, fixing the mean to the best fit and scaling the log-likelihood so that the true and surrogate likelihoods have the same value at the best-fitting point. We sample from the surrogate likelihood posterior and estimate a 1D posterior for each parameter using a Gaussian kernel density estimator (KDE). We ensure that the cumulative distribution function (cdf) of the Gaussian KDE encompasses the whole prior by constraining the cdf to be $0$ and $1$ at the lower and higher bounds of the prior. We use these 1D Gaussian KDE estimations as priors for the \textsc{Ultranest} run.

Thanks to this process, we can perform the first step of the inference in $4$ to $5$ days, including the particle swarm optimiser, the estimation of the approximated prior and the nested sampling run with the true likelihood. In particular, this approach can bias the estimation of the Bayesian evidence (i.e. marginal likelihood) of the nested sampling run. However, it is not usable in our case as we miss $\mathcal{L}_{\rm BCG-kin}$ from its computation. We rely on nested sampling for its robustness in finding the mode of the posterior distribution and not the marginal likelihood estimate. To ensure that this process does not bias the posterior, we provide in Appendix~\ref{app:model_parameters} the density plots of the approximated posterior from the surrogate, the posterior from the nested sampling run and the approximated posterior obtained via the final importance sampling step.

A possible issue with the importance sampling step is the undersampling of the true posterior. Such a case happens when a small number of samples have a non-negligible weight. In particular, the inclination relative to the plane of the sky influences the resulting mass models. In the case of large ellipticities (i.e. $A/B\gtrapprox2.0$ with $A$ and $B$ the major and minor axis sizes), different inclinations can have a $V_{\rm rms}$ difference of more than $20$~${\rm km/s}$. Such differences are enough for a set of parameters from the mass model to be located outside of the final posterior. A solution to avoid undersampling is to increase the sampling of the inclination relative to the mass models. Hence, we are sampling $3$ different inclinations for each mass model.

\section{Results}
\label{sect:results}
\subsection{Model discrimination}

\begin{table}
\caption{Widely applicable information criterion (WAIC) for the different BCG \& ICL parametrisations and estimation of the cluster member stellar masses with their associated root mean squared error (RMS) on the multiple image positions. We highlight in bold the two models favoured by the WAIC.}             
\label{table:WAIC_model}      
\centering                          
\begin{tabular}{c c c c}        
\hline\hline                 
BCG  & SED cluster & WAIC& RMS\\
model & member & $\rm mean\pm std$& $(")$\\
\hline 
BCG - ML 1 & \textsc{LePhare} & $-33538.94\pm0.65$& $0.50"$ \\
BCG - ML 1 & Delayed& $-33548.34\pm1.47$& $0.54"$\\
BCG - ML 1 & Dbl. power-law& $-33536.99\pm1.05$& $0.50"$\\
\textbf{BCG - ML 2} & \textbf{lePhare} & $\mathbf{-33523.48\pm0.94}$& $\mathbf{0.53"}$\\
\textbf{BCG - ML 2} & \textbf{Delayed}& $\mathbf{-33526.09\pm1.16}$& $\mathbf{0.53"}$\\
BCG - ML 2 & Dbl. power-law& $-33536.19\pm1.35$ & $0.53"$\\
BCG - ML 3 & \textsc{LePhare} & $-33552.96\pm1.75$& $0.53"$\\
BCG - ML 3 & Delayed& $-33554.16\pm1.87$& $0.54"$\\
BCG - ML 3 & Dbl. power-law& $-33544.20\pm1.48$& $0.53"$\\

\hline\hline   
\end{tabular}
\end{table}

As we defined three parametrisations for the BCG \& ICL mass components (Sect.~\ref{sect:ICL-model-param}) and three estimates of cluster member stellar masses (B25a, section~4.3), we produce $9$ models combining each possibility. To distinguish between them, we rely on the Widely Applicable Information Criterion (WAIC) \citep{Watanabe2010}, which uses the posterior distribution to compute the criterion value. It is quite similar to the Bayesian Information Criterion \citep[BIC][]{Schwarz1978} or Akaike Information Criterion \citep[AIC][]{Akaike1973}, and is a readaptation of such criteria in a Bayesian inference context. It replaces the maximum-likelihood estimate with its average over the posterior distribution, and switches to an effective number of parameters instead. The number of effective parameters is also estimated from the parameters posterior distribution.

Table~\ref{table:WAIC_model} presents the WAIC estimate performed for each model through the implementation in the \textsc{ArviZ} library \citep{Kumar2019}. We estimate the mean and standard deviation of the WAIC by randomly selecting $1000$ models from the posterior, and repeating the operation $1000$ times. Focusing on the BCG \& ICL parametrisation, ``BCG - ML 2'' models are the best-performing models when averaging over the cluster member SED estimates. They are followed by the ``BCG - ML 1'' models and then the ``BCG - ML 3'' models. Except for the model with a double power-law SFH and ``BCG - ML 2'' parametrisation, there is an improvement at a fixed specific SED model from a model with a ``BCG - ML 3'' to a ``BCG - ML 1'' and from ``BCG - ML 1'' to ``BCG - ML 2'' parametrisation. Hence, according to the WAIC and our set of constraints, the best parametrisation of the BCG \& ICL component is to assume different stellar-mass-to-light ratios for the BCG and the ICL. Providing further freedom to the varying $\Upsilon^{\rm BCG}_*$ does not improve the models, and the last coefficient, $\Upsilon^{\rm BCG,\, 3}_{\rm *,\,lt\,3}$, is not constrained. 

Regarding cluster member stellar masses, there is no clear preference for one SED model that would perform better for each BCG \& ICL parametrisation. The SED model from \textsc{LePhare} is the best, on average, among all BCG \& ICL parametrisation. Although models with \textsc{LePhare} do not outperform the other models with the same BCG \& ICL parametrisation. For ``BCG - ML 2'' parametrisation, it is equally favoured by the WAIC as the model with a delayed SFH, while for ``BCG - ML 1'' and ``BCG - ML 3'' parametrisations, it is the double power-law SFH model that has the best WAIC. In contrast to the BCG \& ICL parametrisation, which can provide a significant increase in the WAIC for most SED models, these cluster member parametrisations have a smaller effect on the overall models and constraints reproduction. Another way to distinguish between these SED models would be to test their prediction of $\Upsilon^{\rm BCG}_{\rm *}$ after selecting the best BCG \& ICL parametrisation. This aspect is discussed in Sect.~\ref{sect:BCG_upsilon}, together with the biases on $\Upsilon^{\rm BCG}_{\rm *}$. Focusing on models with a ``BCG - ML 2'' parametrisation, the Delayed SFH and \textsc{LePhare} SED models are the two best models according to the WAIC. The uncertainties on their WAICs are too large to allow us to discriminate between them. To simplify the discussion and focus on a single model, we present the mass distribution results on the \textsc{LePhare} SFH model in the following sections, as it has the best-fitting models among all. We provide the posteriors and $1\sigma\, \rm CI$ for all models with a ``BCG - ML 2'' parametrisations in Appendix~\ref{app:model_parameters}.

As shown in Table~\ref{table:WAIC_model}, the root mean squared error (RMS) on the multiple image positions is similar among all models. We observe the same behaviour for X-ray and the cluster-member likelihood values, which are of similar order across all models. The most impactful likelihood in the WAIC results is the BCG \& ICL ones, which account for most of the combined likelihood difference between each model. As this likelihood particularly constrains the BCG \& ICL component, it explains why we can discriminate between its parametrisations but not the cluster member one.

\subsection{Reproduction of the constraints}
\label{sect:constraint_repro}
In this section, we present how the selected mass model (i.e. \textsc{LePhare} SFH and ``BCG - ML 2'') reproduces the constraints on each dataset. We start with lensing and X-rays in Sect.~\ref{sect:lensing_cons} and \ref{sect:X-ray-cons-recons}, respectively. We finish with the two new datasets, i.e. the BCG \& ICL and cluster member stellar kinematics, in Sect.~\ref{sect:BCG-ICL-kin-res} and \ref{sect:cl-cons}, respectively.

\subsubsection{Lensing}
\label{sect:lensing_cons}

We quantify the reproduction of the lensing constraints with the root mean square (RMS) error on the position of the multiply-imaged constraint between the observation and model prediction. Our best model, which uses the ``BCG - ML 2'' parametrisation and the SED model with a \textsc{LePhare} SFH, has achieved a RMS of $0.53$~${\rm arcsec}$. This is close to the RMS achieved by the other tested models. In comparison, our worst model according to the WAIC (i.e. ``BCG - ML 3'' parametrisation and SED model with a delayed SFH) has a RMS of $0.54$~${\rm arcsec}$. Hence, lensing constraints are not the dominant dataset to distinguish between the models presented here, as they all perform equivalently. As we use the same set of images as B24, it is the closest comparison we can make with pre-existing models. In that case, the B24 model without a B-spline perturbation achieves an RMS of $0.60$~${\rm arcsec}$, which highlights that we improve the reproduction of the lensing constraints with our new treatment of the BCG \& ICL and cluster members. Our present model has a higher RMS than the B24 model with a B-spline perturbation, although both are close, with RMS values of $0.44$~${\rm arcsec}$. We did not try to include such a component here as the B-spline perturbation as defined in \citet{Beauchesne2021} are only defined on projected quantities in the plane of the sky, although the BCG \& ICL stellar kinematics require the definition of a deprojection in 3D space. 

Interestingly, AS1063 has a multiply imaged system with a central image close to the BCG \citep{Balestra2013}, which is more sensitive to the BCG \& ICL baryons than other systems. On that specific model, the system RMS is of $0.40$~${\rm arcsec}$, slightly lower than the overall RMS and the central image has an error of $0.22$~${\rm arcsec}$. In comparison, B24 model had a RMS of $0.50$~${\rm arcsec}$ for that system and $0.51$~${\rm arcsec}$ for the central image. This system is a good argument in favour of our modelling approach.

\subsubsection{X-rays}
\label{sect:X-ray-cons-recons}
The mass model of the intra-cluster gas is almost identical to the model obtained in B24. The two parametrisations of the likelihood intrinsic error lead to similar likelihood values according to their variations. Mock distribution of $10^5$ likelihoods based on the best-fitting model yields a standard deviation of $87$, which is way larger than the likelihood difference between this work and B24, with a $\mathcal{L}_{\rm X-ray}$ of $-33208$ and $-33189$, respectively. The log-likelihood distribution among the pixels is also close, though with the new error parametrisation, it is slightly wider, with a few pixels with worse and better likelihoods. Hence, we can estimate that the addition of the other likelihoods has not affected the estimate of the gas distribution as presented in B24.

\subsubsection{Cluster member kinematics}
\label{sect:cl-cons}
\begin{figure*}
    \begin{minipage}{0.48\linewidth}
    \centering
    \includegraphics[width=\linewidth]{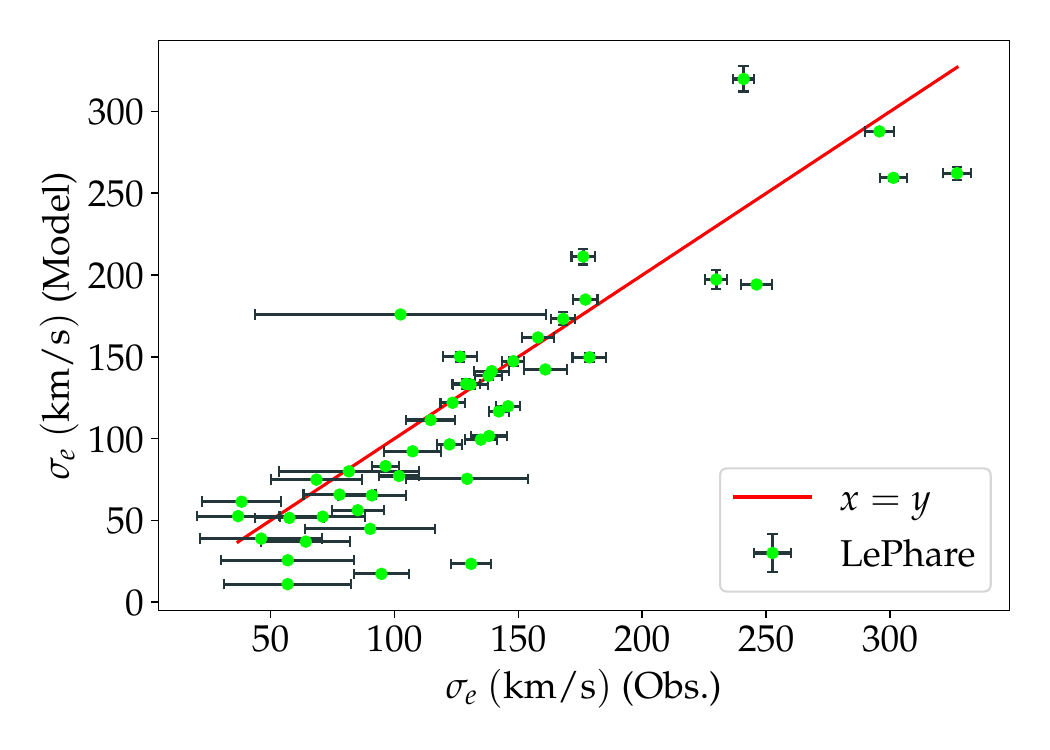}
    \includegraphics[width=\linewidth]{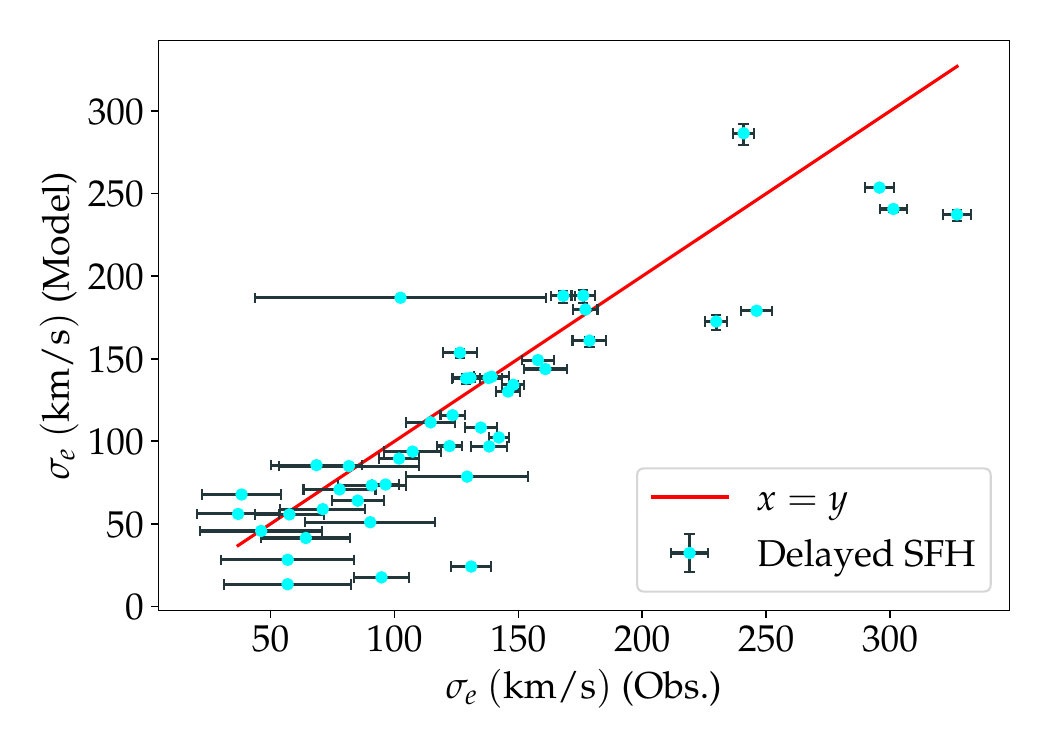}
    \end{minipage}
    \begin{minipage}{0.48\linewidth}
    \centering
    \includegraphics[width=\linewidth]{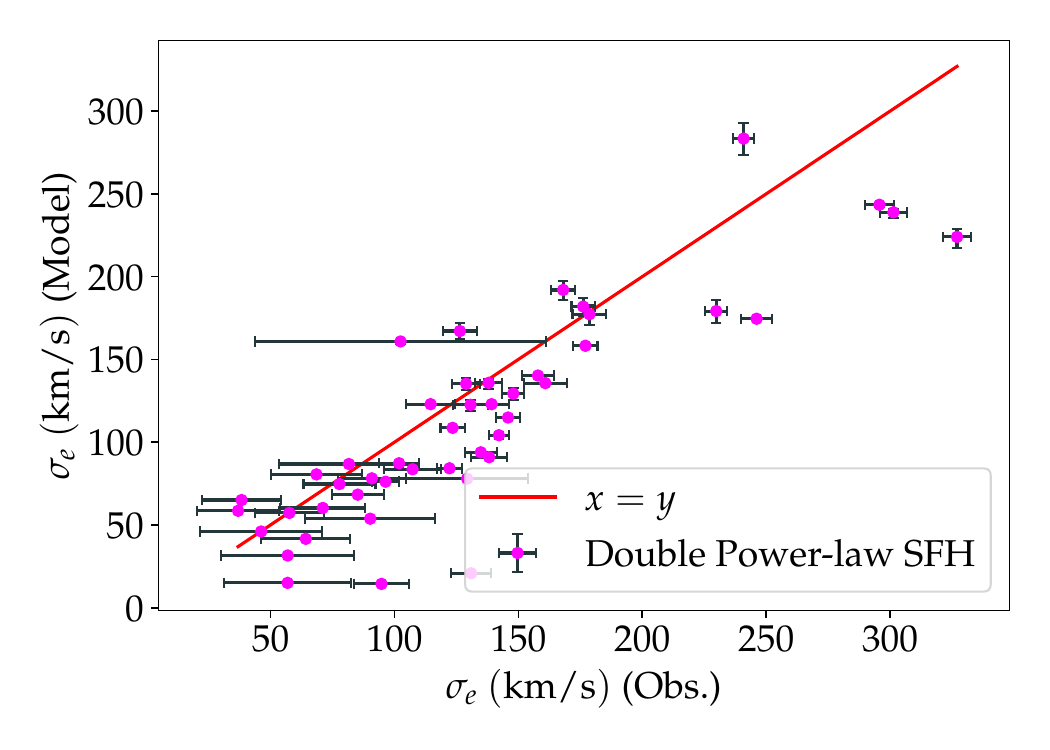}
    \includegraphics[width=\linewidth]{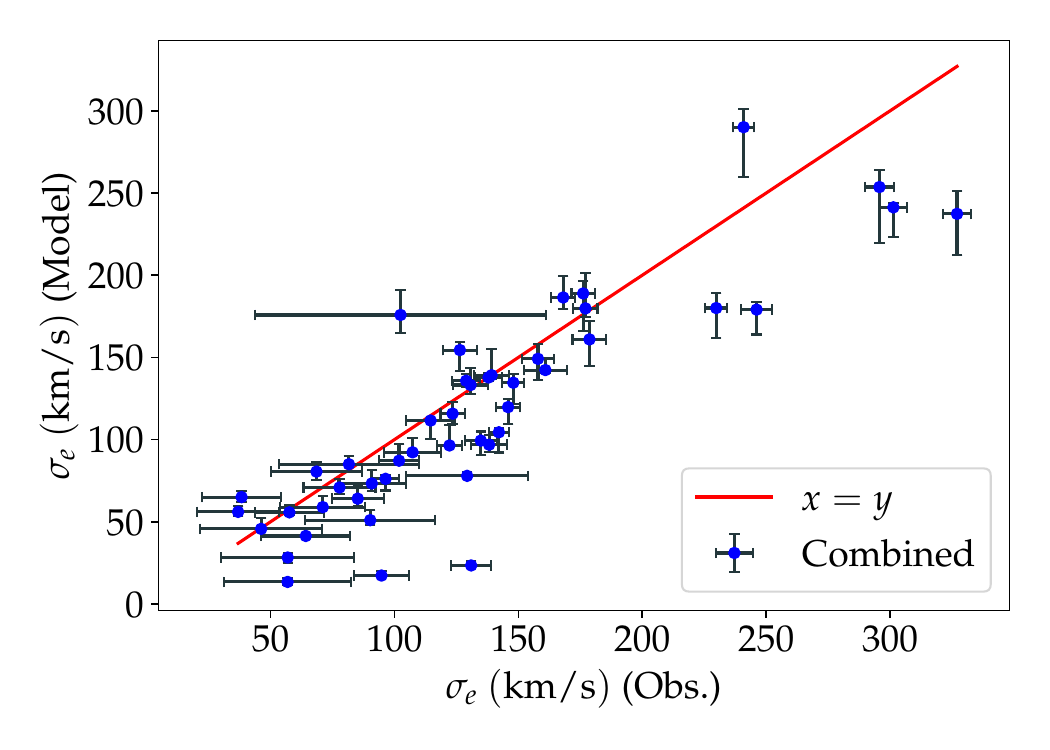}
    \end{minipage}
    
    \caption{Model predicted $\sigma_e$ as a function of the observed one, for each SED model. The uncertainties for the observed $\sigma_e$ are the measurement standard deviations. For the model predicted $\sigma_e$, the uncertainties show the $1\sigma\, \rm CI$ from the mass model posterior. The \textit{top left}, \textit{top right} and \textit{bottom left} panels present the results from the mass model with \textsc{LePhare}, the double power-law SFH and the delayed SFH SED models, respectively. The \textit{bottom-right} panel combines the predictions from the three mass models.}
    \label{fig:kinematic_cl_model_vs_obs}
\end{figure*}

Fig.~\ref{fig:kinematic_cl_model_vs_obs} presents the model-predicted velocity dispersion $\sigma_e$ as a function of the observational measurements. To appreciate the variation due to the SED model, we provide the results for each model with the ``BCG - ML 2'' parametrisation and their combined results. Each mass model presents similar results in terms of scatter and uncertainties, though some galaxies show significant differences depending on the SED model. This result is reflected by the reduced $\chi^2$ of the best-fitting model of each mass modelling that varies from $0.92$ to $1.03$ between these three models with an assumed scatter of $35$~${\rm km/s}$. 

According to the bottom right panel of Fig.~\ref{fig:kinematic_cl_model_vs_obs}, the difference between the $\sigma_e$ estimates relying on different SED models is larger than the statistical uncertainty of the mass models. It is unclear how much the discrepancies are due to the cluster member stellar mass or the interplay between the different components of the mass model. Our results highlight the importance of the selected SED models, as the mass-model uncertainty is too low to blend them. These discrepancies in $\sigma_e$ estimates appear to correlate with increasing values, as points with $\sigma_e>200$~${\rm km/s}$ present larger error bars. To assess such a correlation, we analyse this dataset with a Spearman correlation \citep{Spearman1904} based on $\sigma_e$ and $\delta_{\rm e,sys}/\sigma_e$, with $\delta_{\rm e,sys}$ being half of the width of the $1\sigma\, \rm CI$. We report a slight anti-correlation of $-0.12$ with a large p-value of $0.42$, suggesting no correlation. The same test as with $\sigma_e$ and $\delta_{\rm e,sys}$ gives a correlation coefficient of $0.86$, with p-values inferior to $10^{-13}$. Hence, the amplitude of the systematic error appears to correlate with the accessible mass range of a cluster member within our parametrisation.


\subsubsection{BCG \& ICL kinematics}
\label{sect:BCG-ICL-kin-res}

\begin{figure}
    \centering
    \includegraphics[width=\linewidth]{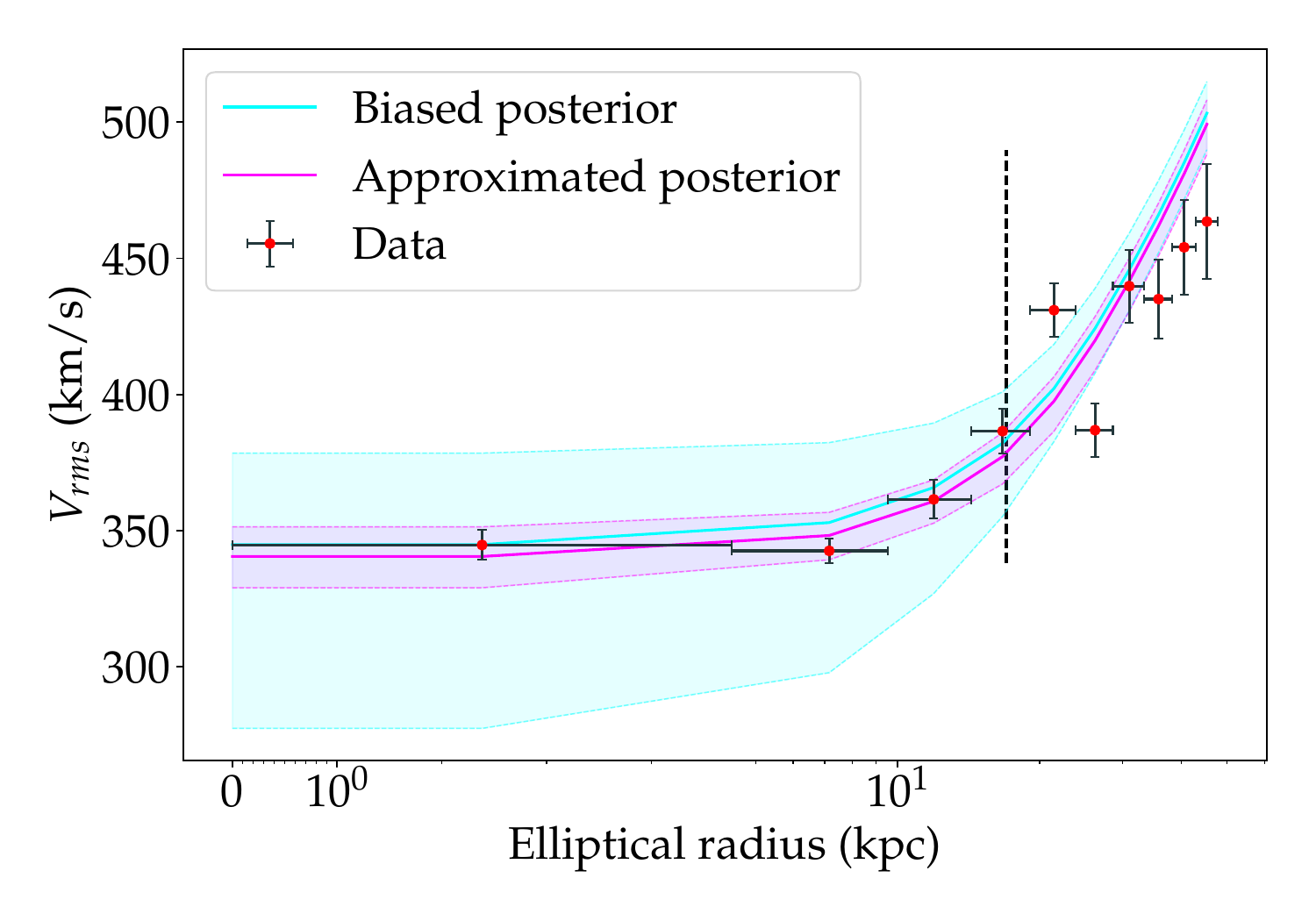}
    \caption{Stellar kinematic data and model of the BCG \& ICL component as a function of the elliptical radius in ${\rm kpc}$. The cyan distribution represents the biased posterior before the importance sampling step, while the magenta one is the approximated posterior after the resampling procedure. The scattered points represent the observational data points. The elliptical radius is defined with the same ellipticity as used for the fit of the $V_{\rm rms}$. The shaded area represents the $3\sigma$ uncertainties of each posterior, while the plain lines highlight the median. The errors of the data points represent the standard deviation of the $V_{\rm rms}$ fit. The dashed black vertical lines show the BCG half-light radius as measured by \citet{tortorelli2018} of $17$~${\rm kpc}$}
    \label{fig:kinematic_ICL}
\end{figure}

Fig.~\ref{fig:kinematic_ICL} presents the kinematic model of the BCG \& ICL component through its $V_{\rm rms}$. As shown by the cyan distribution, mass models are already in good agreement with the observed stellar kinematics even when not fitted to those data. This agreement is a necessary condition to validate the importance sampling procedure, but it also increases confidence in the mass model parameterisation. Indeed, as the inner mass model is driven by prior assumptions on the mass profile, it could have deviated from the true mass without direct constraints. Thanks to the importance sampling procedure, we can significantly reduce the width of the kinematic posterior, particularly in the BCG and its vicinity, as highlighted by the dashed black line representing the BCG half-light radius.

When looking more closely at the reduced $\chi^2$ of each model in the resampled posterior, we obtain a value of $3.60^{+0.23}_{-0.18}$ for its $1\sigma\, \rm CI$ with a minimum of $3.24$, while the best-fitting model among all datasets reaches a reduced $\chi^2$ of $4.02$. More precisely, the median distance to the data points from the model is below $2$ standard deviations for every point, except the fifth and sixth ones, which are at $3.36$ and $3.38$, respectively. As these two points deviate from the trend adopted by the other data points, their uncertainties are likely not accounting for systematic biases. If we exclude them, we obtain $2.72^{+0.45}_{-0.35}$ for the reduced $\chi^2$ $1\sigma\, \rm CI$ with a minimum at $2.02$. If our kinematic model posterior agrees with the observed data, it does not include a model that agrees with the measurement uncertainties. It may highlight a limitation of our kinematic modelling or an undersampling of the posterior due to our two-step optimisation procedure.

\subsection{Mass distribution and profiles}

\begin{figure*}
    \centering
    \includegraphics[width=\linewidth]{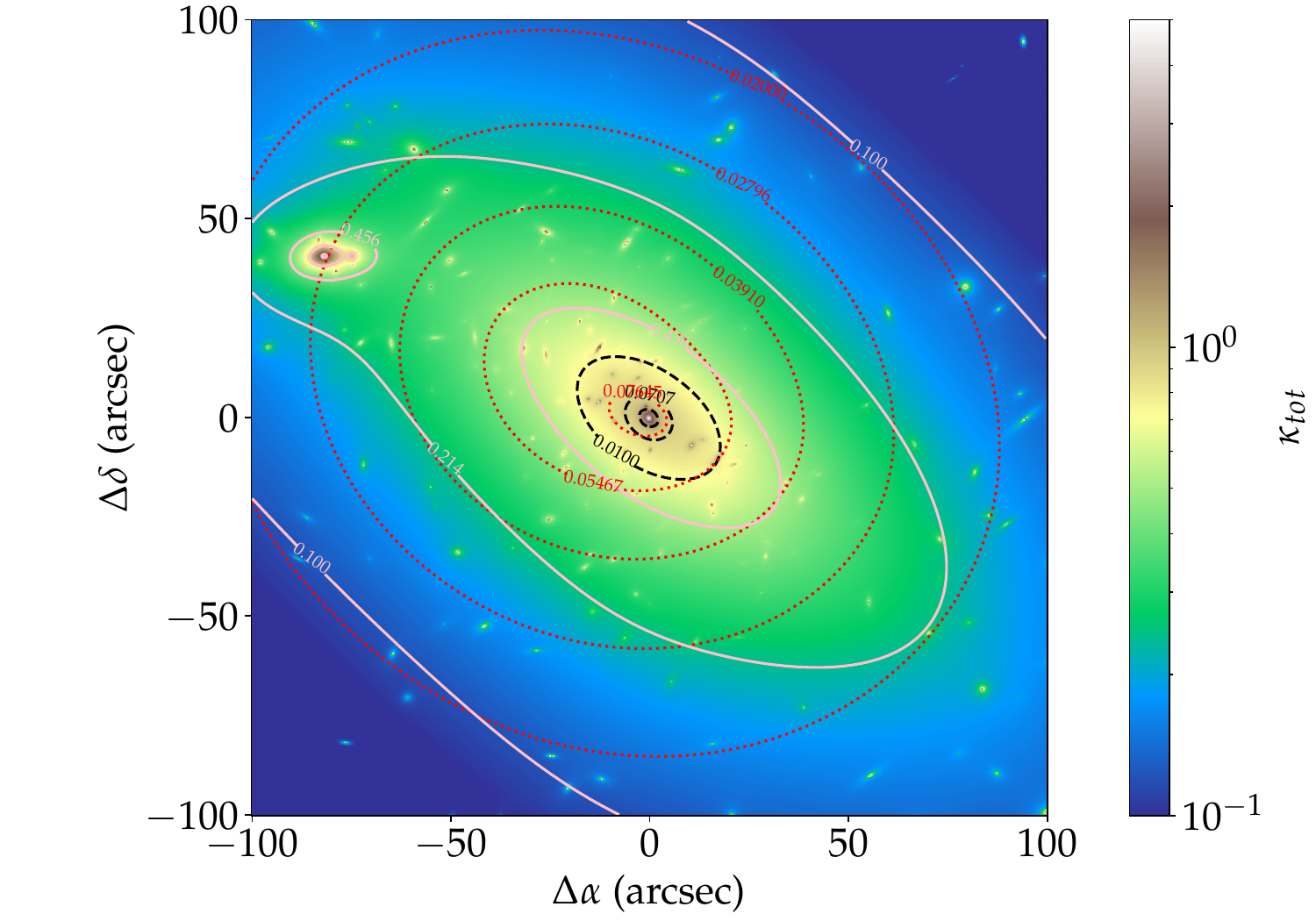}
    \caption{2D normalised surface mass density (convergence) associated with the total mass distribution in the same field of view as the RGB image presented in Fig.~\ref{fig:summary_S1063}. Each cluster-scale mass component is highlighted by contours. The BCG \& ICL baryons are highlighted in dashed black, while the intra-cluster gas and DM are in dotted red and solid pink contours, respectively.}
    \label{fig:mass_whole}
\end{figure*}

Fig.~\ref{fig:mass_whole} presents the normalised surface mass density or convergence of the mass model. Each cluster-scale mass component is highlighted by dashed contours. Thanks to our highly detailed mass modelling, we can disentangle each mass component. Similarly to B24, we observe a discrepancy between the intra-cluster gas and DM. The gas distribution presents a smaller ellipticity and an asymmetry toward the NE halo. It highlights the importance of considering the gas, even in slowly perturbed clusters. Regarding the BCG \& ICL baryons, their contribution is negligible compared to the gas and DM, except in the vicinity of the BCG. However, within a few ${\rm kpc}$ of the cluster centre, the ellipticity of the BCG \& ICL highest surface mass density contour closely matches that of the DM. Moving inward, the two innermost contours of the BCG \& ICL show progressively lower ellipticities, indicating a systematic deviation from the DM ellipticity toward the BCG one.

Overall, the mass distribution is similar to that of B24 and other lensing studies on AS1063 \citep{Caminha2016,Bergamini2019,Granata2022,Limousin2022}, which are mostly unimodal with an asymmetry toward the NE halo. As found in B24, it is likely the remnant of a small cluster/group of galaxies infalling in AS1063. Such an event induces gas sloshing, as confirmed by a specific pattern in the gas properties.

\begin{figure*}
    \begin{minipage}{.49\linewidth}
    \centering
    \includegraphics[width=\linewidth]{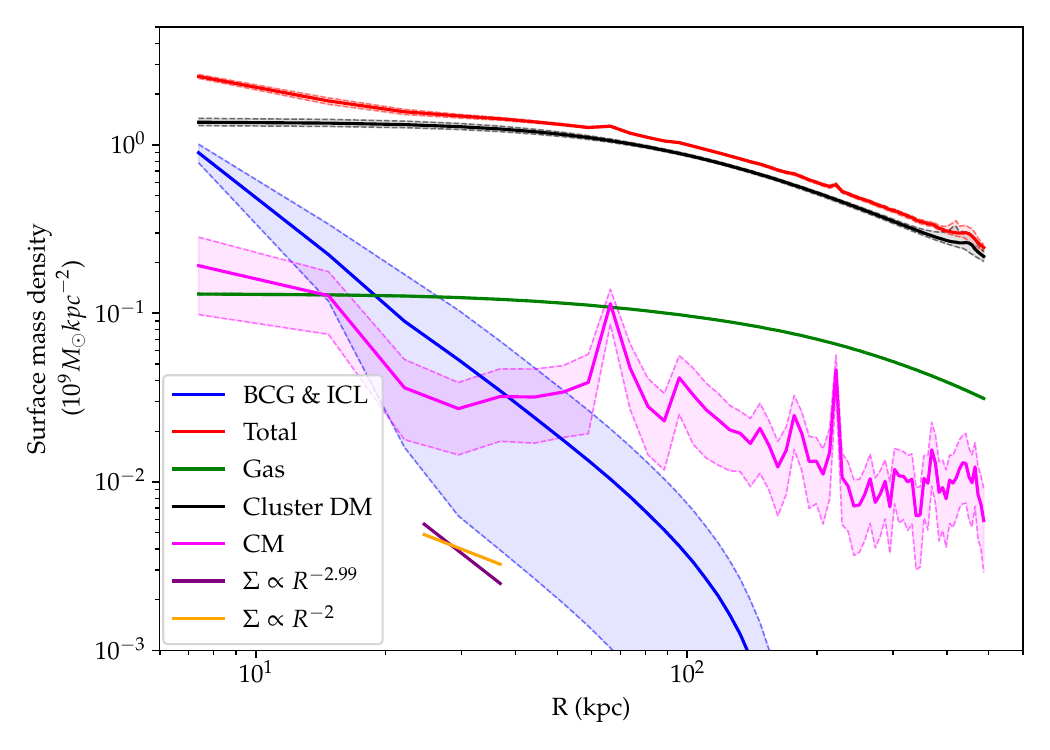} 
    \end{minipage}
    \begin{minipage}{.49\linewidth}
    \centering
    \includegraphics[width=\linewidth]{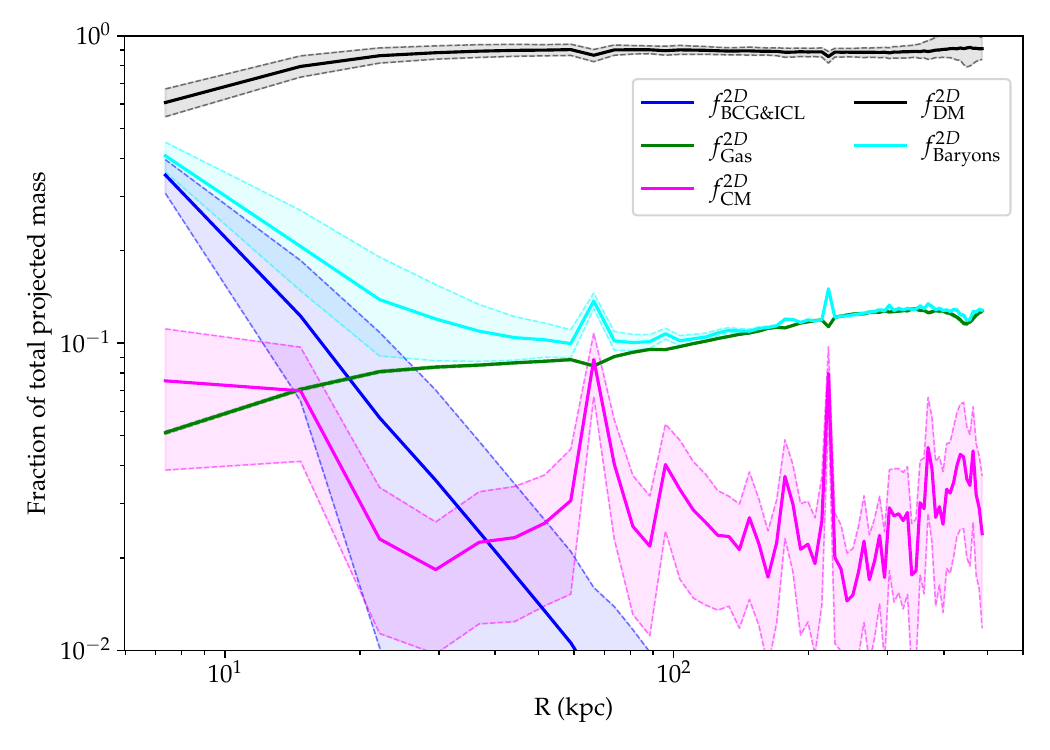}
    \end{minipage}
    \caption{\textit{Left panel:} Surface mass density of the whole cluster (red) and each of its components: the cluster-scale DM component (black), the BCG \& ICL baryons (blue), the gas (green) and the cluster members (magenta). The plain lines represent the median of each distribution among the model posterior, while the shaded area shows its $3\sigma\, \rm CI$. As a qualitative comparison to BCG \& ICL profile slope, we present two power-law profiles with a slope of $-2$ and $-2.99$ (best fitting solution on the BCG \& ICL profile), respectively. \textit{Right panel:} Projected fraction of each cluster mass component relative to the total mass. It presents the mass profiles of each component divided by the total mass density profile. We present BCG \& ICL baryons (blue), the gas (green) and the cluster members (magenta) with the same profile as the left panel. For the DM component (black) and baryonic (cyan) mass profiles, we present the overall DM including the cluster-scale and cluster member counterpart, while the baryons include the stellar mass from the BCG \& ICL, the cluster member as well as the X-ray gas.}
    \label{fig:mass_profile}
\end{figure*}

Fig.~\ref{fig:mass_profile} presents the surface mass density profile of each mass component with the total mass in the left panel, and the component profiles relative to the total mass in the right panel. We present the DM distribution through the cluster-scale DM component in the left panel and the overall DM, including the cluster member DM, in the right panel. The cluster member distribution includes both DM and baryons. Thanks to our detailed modelling, we have access to the baryonic fraction and the contribution to each baryonic component. We can see that they reach a maximum of $\approx40$ per cent within the BCG, where the baryons are dominated by the BCG \& ICL stellar masses. Beyond a radius of $20$~${\rm kpc}$, the ICM becomes the dominant baryonic component, having reached the contribution of the BCG \& ICL at that radius, each accounting for $8$ per cent of the overall mass. The bulk of the baryonic mass in the form of ICM gas represents $\approx10$ per cent of the total mass up to the end of the strong lensing region at $\approx250$~${\rm kpc}$. The cluster member baryons contribute only locally to the overall baryonic content. Although their total mass can represent as much as the gas distribution when the density of the galaxy is large enough. In particular, there are three cluster members around the BCG which contribute up to $8$ per cent of the overall mass locally.

Our mass profile also shows lower statistical uncertainties than the B24 results. Within the strong lensing area ($R<250$~${\rm kpc}$), our total mass profile presents a $1\sigma\, \rm CI$ of $1.45^{+0.54}_{-0.42}$ per cent of error among the considered radial bins. It represents a $21$ per cent reduction compared to the same measurement from the B24 model, which has a $1\sigma\, \rm CI$ of $1.82^{+1.06}_{-0.58}$. Similar error reductions are observed within the BCG and its close surroundings for $R<30$~${\rm kpc}$. We obtain $3.08^{+0.88}_{-0.52}$ and $4.06^{+3.87}_{-1.39}$ for our model and B24's one, respectively. In addition to the significant error reduction of the median, the width of the $1\sigma\, \rm CI$ has been divided by a factor $\approx4$, highlighting the benefit of the BCG stellar kinematics constraints. Thanks to these reduced uncertainties, we obtain an error of $2.15^{+1.16}_{-0.41}$ and $4.68^{+0.24}_{-0.34}$ per cent for $R<250$~${\rm kpc}$ and $R<30$~${\rm kpc}$, respectively, for the DM-only profiles.

\begin{figure*}
    \begin{minipage}{.49\linewidth}
    \centering
    \includegraphics[width=\linewidth]{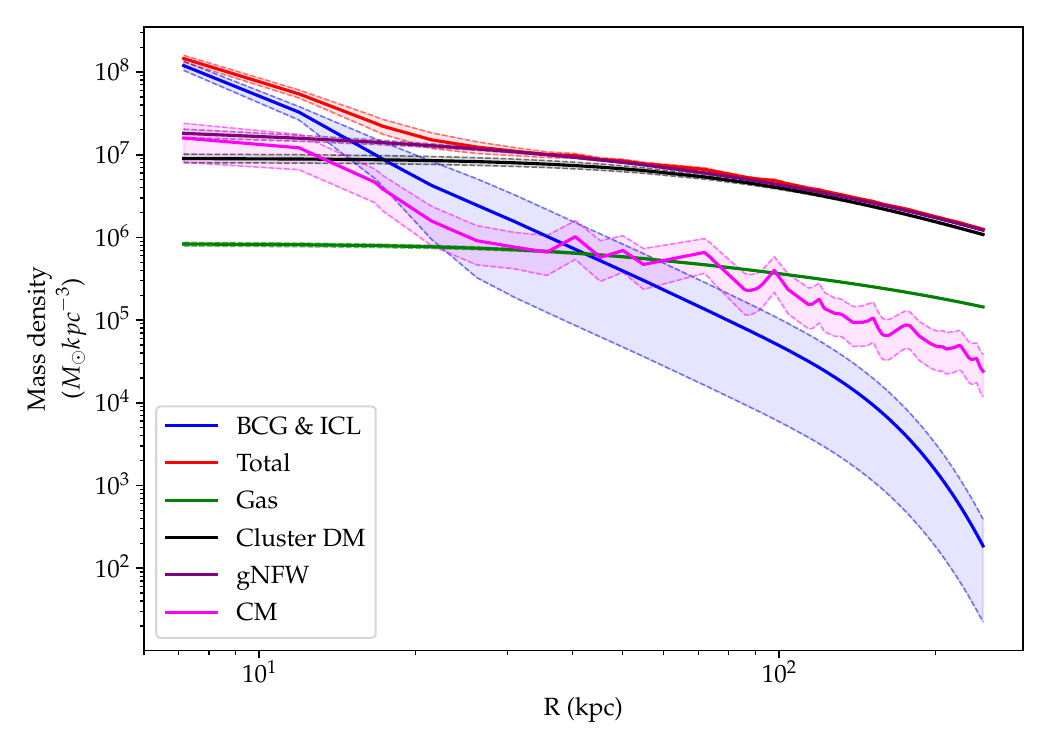} 
    \end{minipage}
    \begin{minipage}{.49\linewidth}
    \centering
    \includegraphics[width=\linewidth]{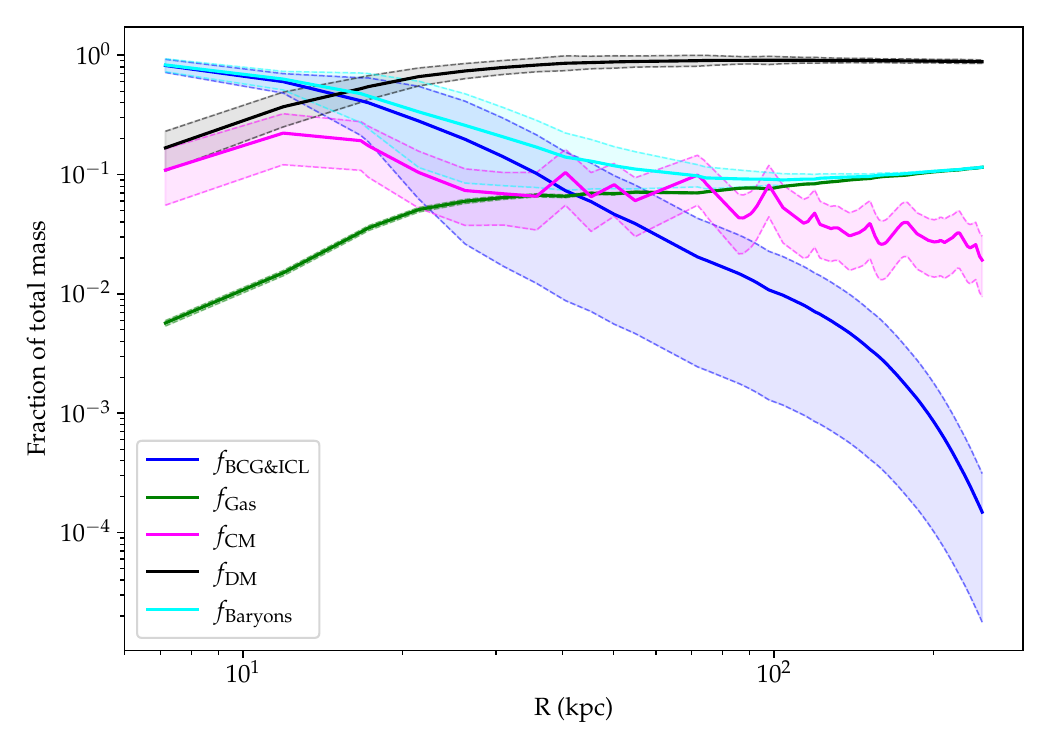}
    \end{minipage}
    \caption{\textit{Left panel:} Mass density of the whole cluster (red) and each of its components: the cluster-scale DM component (black), the BCG \& ICL baryons (blue), the gas (green) and the cluster member (magenta). A gNFW fit to the total mass distribution is presented as a comparison to model the mass slope. Plain lines represent the median of each distribution among the model posterior, while the shaded area shows its $3\sigma\, \rm CI$. \textit{Right panel:} Fraction of each cluster mass component relative to the total mass. For the BCG \& ICL (blue) and the gas (green), we present the same profile as in the left panel. For the DM (black) and baryonic (cyan) fraction, we account for the DM and baryons in each component similarly to the right panel of Fig.~\ref{fig:mass_profile}. The colour scheme and meaning of the plain line and shaded area are the same as in the left panel. The combined baryonic profile is presented in cyan.}
    \label{fig:mass_profile_3D}
\end{figure*}

To complete the picture of the mass profiles, we present the mass density of each cluster component with the total mass density in the left panel of Fig.~\ref{fig:mass_profile_3D}. Their relative fraction with respect to the total mass is shown in the right panel of Fig.~\ref{fig:mass_profile_3D}. We add a fit by a generalised Navarro-Frenck-White profile \citep[gNFW;][]{Zhao1996} of the total mass density as a qualitative comparison to the model mass slope. In contrast to the projected profile presented in Fig.~\ref{fig:mass_profile}, there is a clear domination of the baryonic mass of the BCG up to around $20$~${\rm kpc}$, which is approximately the BCG half-light radius (i.e. $17$~${\rm kpc}$). The ICM becomes the main baryonic contributor at radii larger than the projected counterparts, i.e. around $50$~${\rm kpc}$. Considering the total mass, the gNFW profile can provide a good fit to the total mass from $20$~${\rm kpc}$ and up to the end of the constrained area. Hence, it is unable to account for the BCG mass distribution when it dominates in the inner core, which can be put into perspective with the results of \citet{Newman2013b}, who added a dPIE to account for the BCG in addition to the gNFW profile. In our case, we obtain a slope $\beta=0.14^{+0.03}_{-0.03}$, significantly lower than $1$, where $1$ is the slope of a NFW profile. Similarly, \citet{Cerny2025} presented a sample of cluster mass models where the BCG kinematics was used along with strong lensing to constrain the model parameters. In contrast to this work, in their models the mass components are all modelled with dPIEs, and they rely on a scaling for the cluster member based on the Faber \& Jackson law \citep{Faber&Jackson1976}. We reproduce their slope measurement $\gamma=-\frac{{\rm d} \log_{10}(\rho)}{{\rm d} \log_{10}(r)}$ over $5$ to $50$~${\rm kpc}$ for the DM distribution ($\gamma= 0.182^{+0.008}_{-0.009}$) and the total mass with the BCG \& ICL subtracted ($\gamma=0.68^{+0.040}_{-0.049}$). We obtain a similar discrepancy as \citet{Cerny2025} results, which is mostly due to the cluster members distribution, as the cluster scale DM and gas distribution combined leads to $\gamma=0.186^{+0.008}_{-0.007}$. Hence, our new cluster-member scheme produces results similar to those of the regular Faber \& Jackson law scaling. However, this result, as our mass model assumes that all cluster members lie in the same plane in 3D, likely overestimates the influence of cluster members near the BCG, where their projected density is the highest. In particular, their contribution is larger than the cluster-scale and BCG DM component in the inner $15$~${\rm kpc}$ toward the BCG, which likely highlights the de-projection effect.

Our results highlight the importance of providing an accurate census of the baryonic components to robustly recover the DM profile down to the few ${\rm kpc}$ of the inner core. In particular, the DM profile's shape in that area is sensitive to alternatives to the cold DM (CDM) paradigm, such as self-interacting or Fuzzy DM models that feature a cored distribution rather than the cuspy CDM profile. The parameters of these DM models are correlated with the shape and size of these cores, making our modelling well-suited to robustly measure them \citep{Robertson2021}.

Another advantage of our self-consistent approach, which relies on disentangling each cluster component, is its ability to assess which baryonic component is the best tracer of DM. It is particularly useful in the context of clusters or groups that do not present lensing features, but deep observations can map some of their baryonic components and then recover an estimated DM profile. For example, the previous analysis by \citet{Montes2019} provided evidence that the ICL is a good tracer of the DM profile slope.

However, we leave analyses such as measuring DM properties or assessing which baryonic components are the best DM tracers to future work. Before concluding, we must address the reliability of our stellar mass estimates, as inaccurate BCG \& ICL stellar content could introduce significant bias into the analysis presented above.

\section{Discussion}
\label{sect:discussion}
We choose to focus the discussion on the estimate of the new baryonic components added to the self-consistent multi-probe approach developed in B24. We start in Sect.~\ref{sect:BCG_upsilon} by analysing our estimate of the BCG \& ICL stellar mass in light of the SED estimates provided in B25a (section~5.3). We follow up in Sect.~\ref{sect:SMBH_discuss} and \ref{sect:Kurtosis_discuss}, where we discuss the probable biases of BCG \& ICL stellar mass estimates with the neglected BCG super massive black hole (SMBH) and assumption of the BCG \& ICL stellar kinematics. We complete the discussion on the BCG \& ICL component with its degeneracy with the DM mass profile at Sect.~\ref{sect:ML_DM_degeneracy}. We then discuss our estimate of the SsHMR and its dependence on the initial stellar mass estimated from cluster members in Sect.~\ref{sect:SsHMR_discuss} and \ref{sect:IMF_CL}.

\subsection{BCG \& ICL stellar masses estimate}
\label{sect:BCG_upsilon}
\begin{figure*}
    \begin{minipage}{0.33\linewidth}
    \centering
    
    \includegraphics[width=\linewidth]{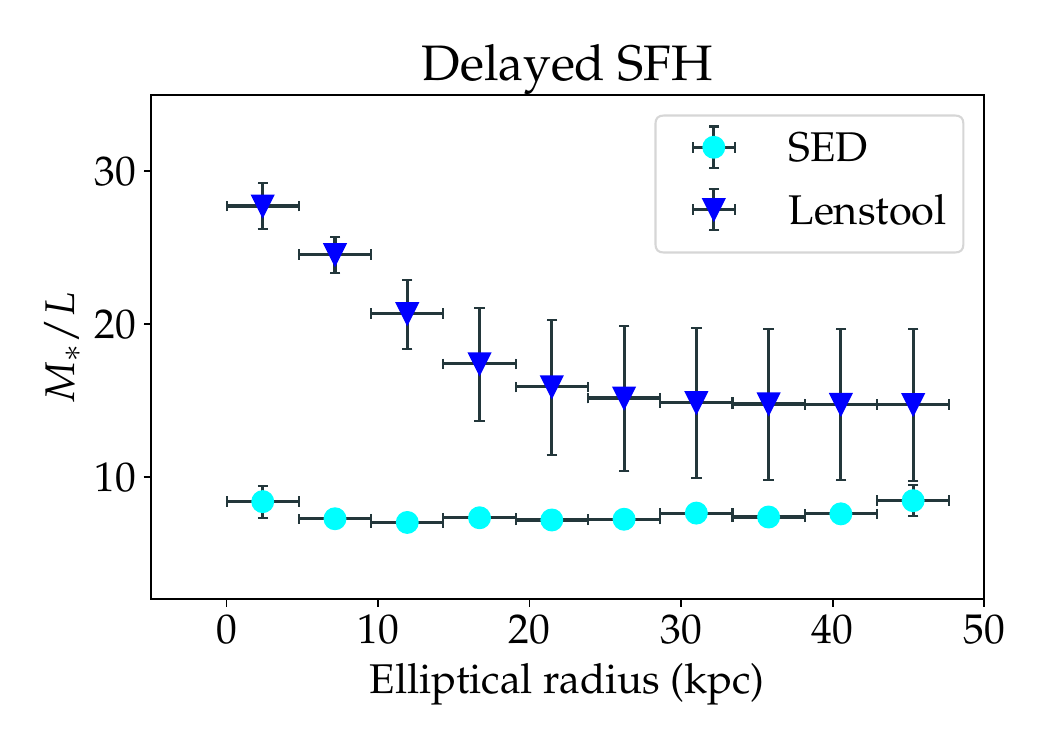}
    \end{minipage}
    \begin{minipage}{0.33\linewidth}
    \centering

    \includegraphics[width=\linewidth]{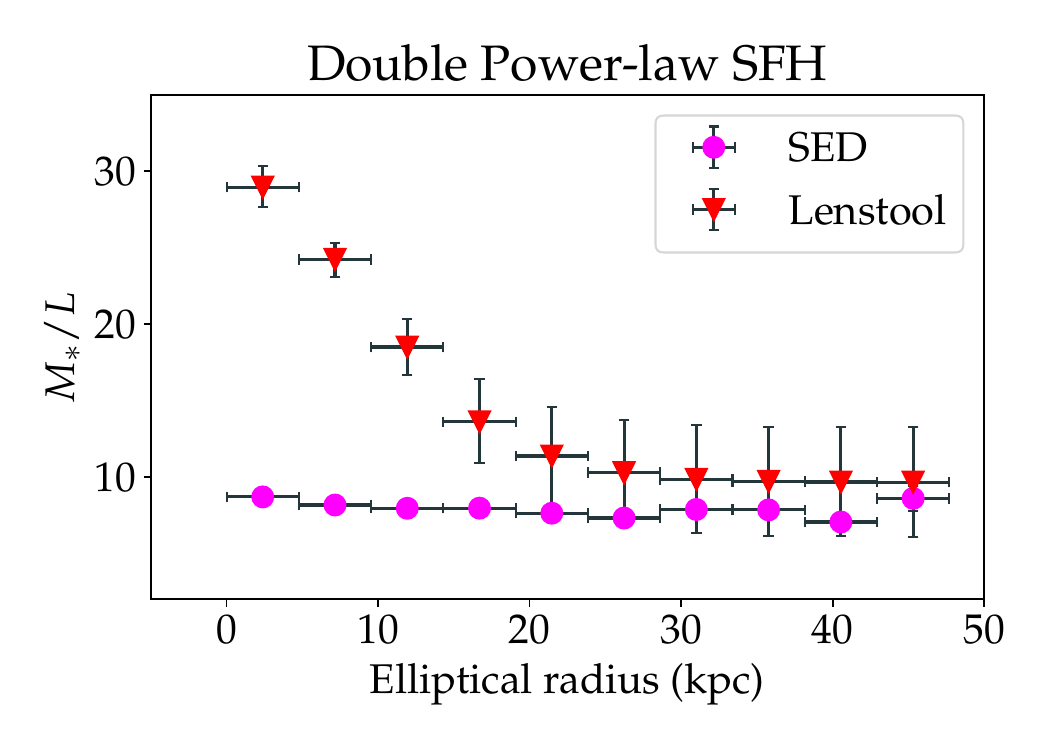}
    \end{minipage}
    \begin{minipage}{0.33\linewidth}
    \centering

    \includegraphics[width=\linewidth]{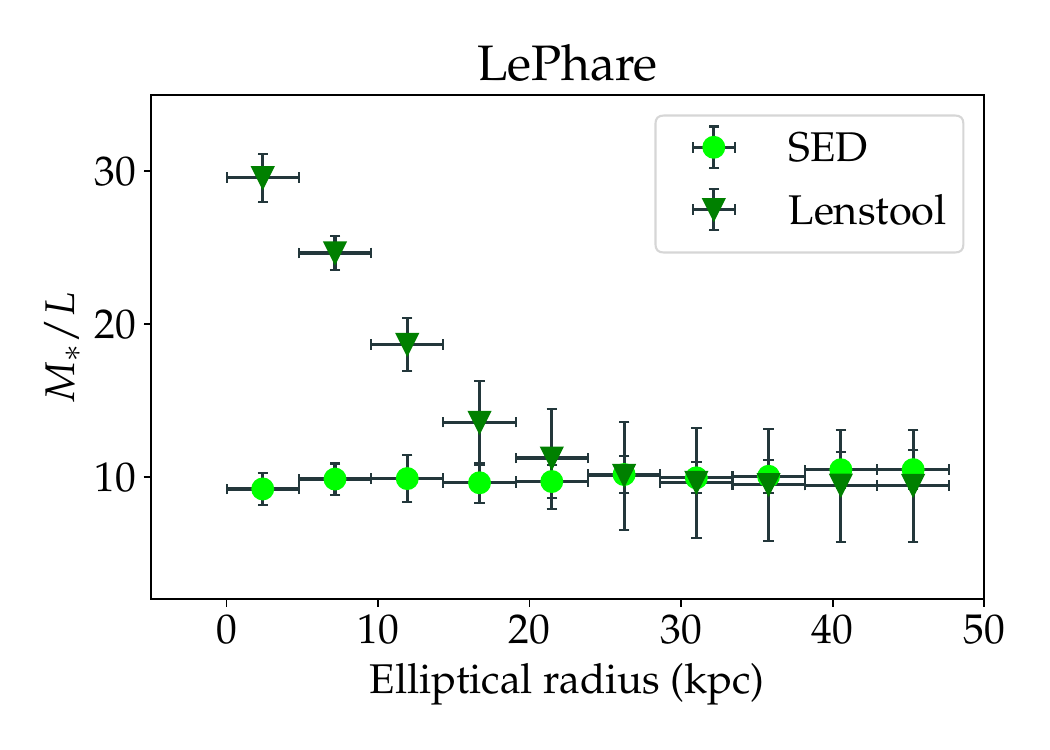}
    \end{minipage}
    \caption{Stellar-mass-to-light ratio of the BCG \& ICL component, $\Upsilon^{\rm BCG}_{\rm *, lt}$ for the three models with a ``BCG - ML 2'' parametrisation with their equivalent estimated through SED fitting, $\Upsilon^{\rm BCG}_{\rm *, SED}$ as a function of the elliptical radii defined in B25a (section~5.2). Error bars represent the $1\sigma\, \rm CI$ for both $\Upsilon^{\rm BCG}_{\rm *}$ estimates, while errors on the radius represent the width of the elliptical bins in which it is averaged. Each model uses a different SED model. They are presented in the following order from \textit{left to right}: Delayed SFH, double power-law SFH and \textsc{LePhare}. The colour schemes highlight $\Upsilon^{\rm BCG}_{\rm *, lt}$ (blue, red and green) with darker colours than $\Upsilon^{\rm BCG}_{\rm *, SED}$ (cyan, magenta and lime). Similar colours regroup models that are associated with the same SED model.}
    \label{fig:ML_profile}
\end{figure*}

Fig.~\ref{fig:ML_profile} presents the estimates of the stellar-mass-to-light ratio of the BCG \& ICL component. It compares the estimates from each mass model with a ``BCG - ML 2'' parametrisation with the SED fit results provided in B25a (section~5.3). For each mass model, there is an apparent discrepancy between the estimate from \textsc{Lenstool} and the SED fit within the BCG (i.e. $R_e=17$~${\rm kpc}$). In contrast, at larger radii, both estimates agree within their $1\sigma\, \rm CI$ for the \textsc{LePhare} SED model. The models with a double power-law and a delayed SFH present a less good agreement with the SED estimate, although $\Upsilon^{\rm BCG}_{\rm *, SED}$ is included in the $3\sigma\, \rm CI$ of $\Upsilon^{\rm BCG}_{\rm *, lt}$. These results are in line with the stellar-mass-to-light ratio mismatch observed in early-type galaxies (ETG) between SED and stellar kinematics estimate (see \citet{Smith2020} for a review). These mismatches are observed when the kinematic analysis relies on a constant value for the BCG, $\Upsilon_*$ \cite{Posacki2015,Lu2024}. We are in a similar case as our varying $\Upsilon^{\rm BCG}_{\rm *, lt}$ parametrisation has only one coefficient associated with the BCG. \citet{Oldham2018a} and \citet{Collett2018} find evidence for a steep gradient of $\Upsilon_*$ within ETGs associated with an IMF variation with radius from a \citet{Salpeter1955} IMF to a Milky-way like IMF that would produce such a mismatch. As reported by \citet{Mehrgan2024}, such gradients are visible for very small radii, $R \lesssim 1$~${\rm kpc}$, which cannot be probed with our datasets. In addition to this picture, our results suggest that the ICL in AS1063 follows a Milky-way like IMF. Hence, the BCG and the ICL would present a similar IMF at their common border. 

These discrepancies in the estimate of the stellar-mass-to-light ratio imply a larger stellar mass for the BCG \& ICL component than estimated in B25a (section~5.3). We obtain $1.70 \pm 0.32$, $1.41 \pm 0.24$ and $1.41 \pm 0.24\times10^{12}\,M_\odot$ for the model with a delayed SFH, double power-law SFH and \textsc{LePhare} SED models, respectively. These masses are estimated within the same elliptical bins as presented in B25a (section~5.2). These new estimates are from $57$ to $149$ per cent higher than their SED counterparts, which are mostly due to the underestimation of the BCG stellar mass. The models with a delayed SFH and a double power-law SFH, which exhibit discrepancies larger than $100$ per cent, also yield higher estimates of the stellar mass within the ICL.

To further compare our results and in particular the $\Upsilon_*$ estimate mismatches, we define the following mismatch parameter $\alpha$:
\begin{equation}
    \alpha=\frac{\Upsilon^{\rm BCG}_{\rm *, lt}}{\Upsilon^{\rm BCG}_{\rm *, SED}}
\end{equation}
We estimate the posterior distribution of $\alpha$ by propagating the distribution of $\Upsilon^{\rm BCG}_{\rm *, lt}$ from mass models with a single evaluation of $\Upsilon^{\rm BCG}_{\rm *, SED}$ represented by its median. We do not include the uncertainty of $\Upsilon^{\rm BCG}_{\rm *, SED}$ as it is negligible for two SED models, and we do not have the full posterior shape for every SED fit, in particular the one made with \textsc{LePhare}. We compare $\alpha$ with the results from \citet{Lu2024}, who analysed the IMF variation for the MaNGA DynPop project. This project aimed to observe $10000$ early-type galaxies (ETGs) to perform dynamical and stellar population synthesis analyses. In particular, they estimate $\alpha$ as a function of the galaxy velocity dispersion, $\sigma_e$, for a \citet{Salpeter1955} IMF. We convert their results to a Milky-way like IMF following the conversion factor provided in \citet{Wright2017}, and we use a value of $341$~${\rm km/s}$ for $\sigma_e$, which has been measured in the most central bins of our BCG \& ICL stellar kinematics measurement.

\begin{figure*}
    \centering
    \begin{minipage}{0.33\linewidth}
    \centering
    \includegraphics[width=\linewidth]{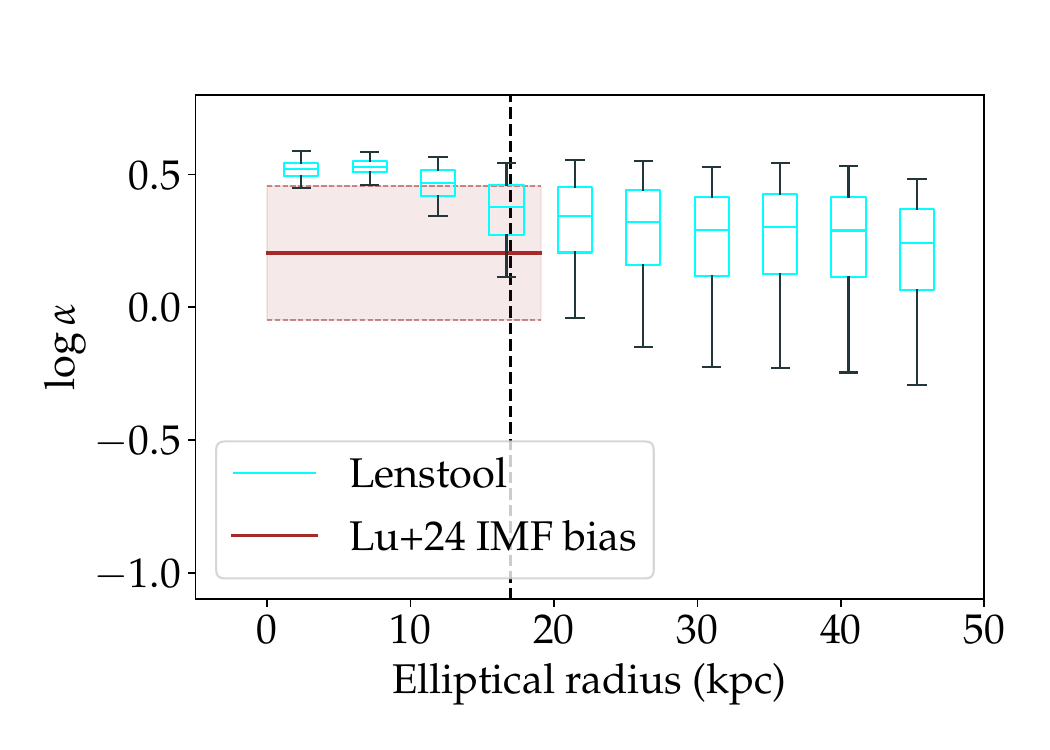}
    \end{minipage}
    \begin{minipage}{0.33\linewidth}
    \centering
    \includegraphics[width=\linewidth]{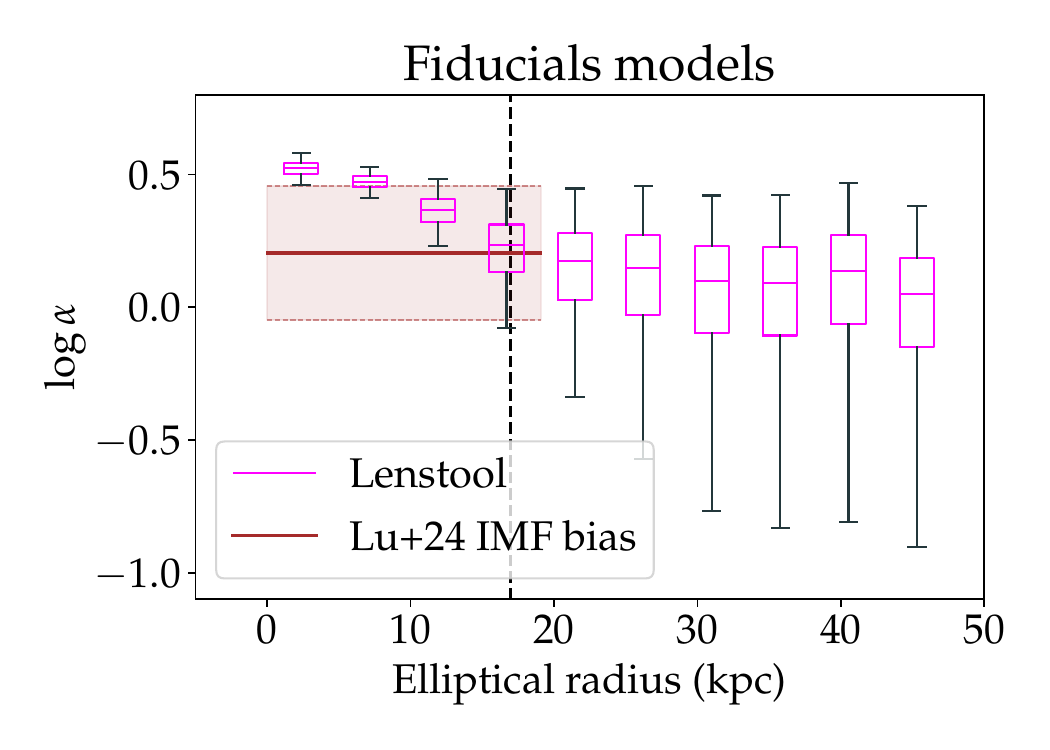}
    \end{minipage}
    \begin{minipage}{0.33\linewidth}
    \centering
    \includegraphics[width=\linewidth]{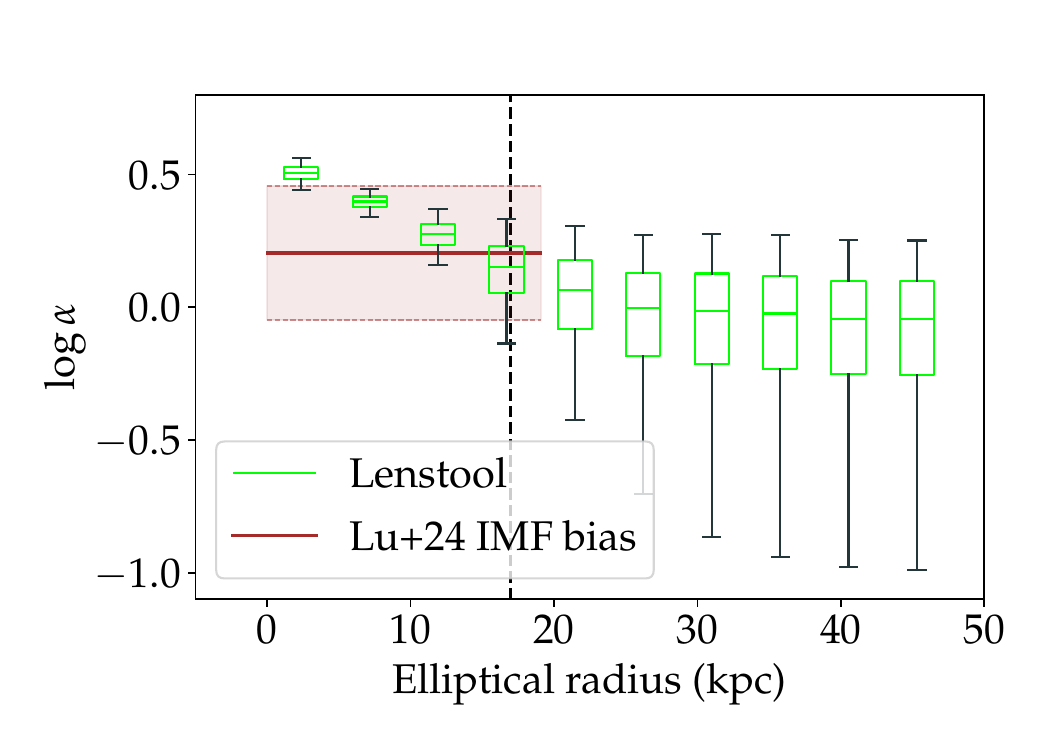}
    \end{minipage}
    
    \begin{minipage}{0.33\linewidth}
    \centering   
    \includegraphics[width=\linewidth]{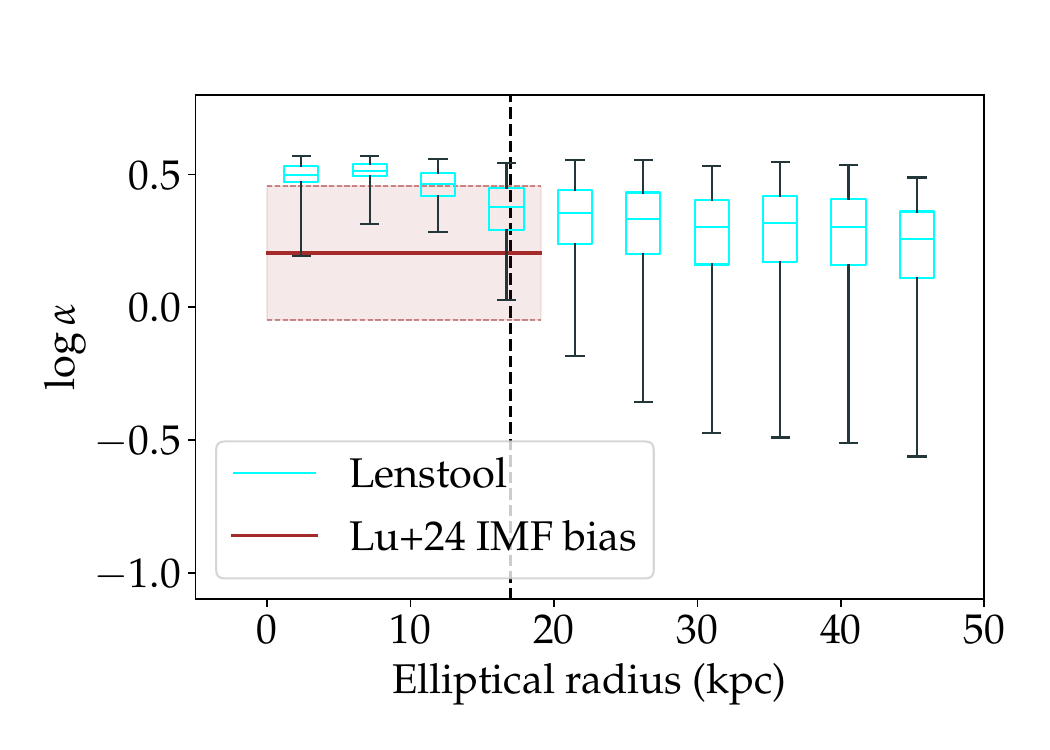}
    \end{minipage}
    \begin{minipage}{0.33\linewidth}
    \centering
    \includegraphics[width=\linewidth]{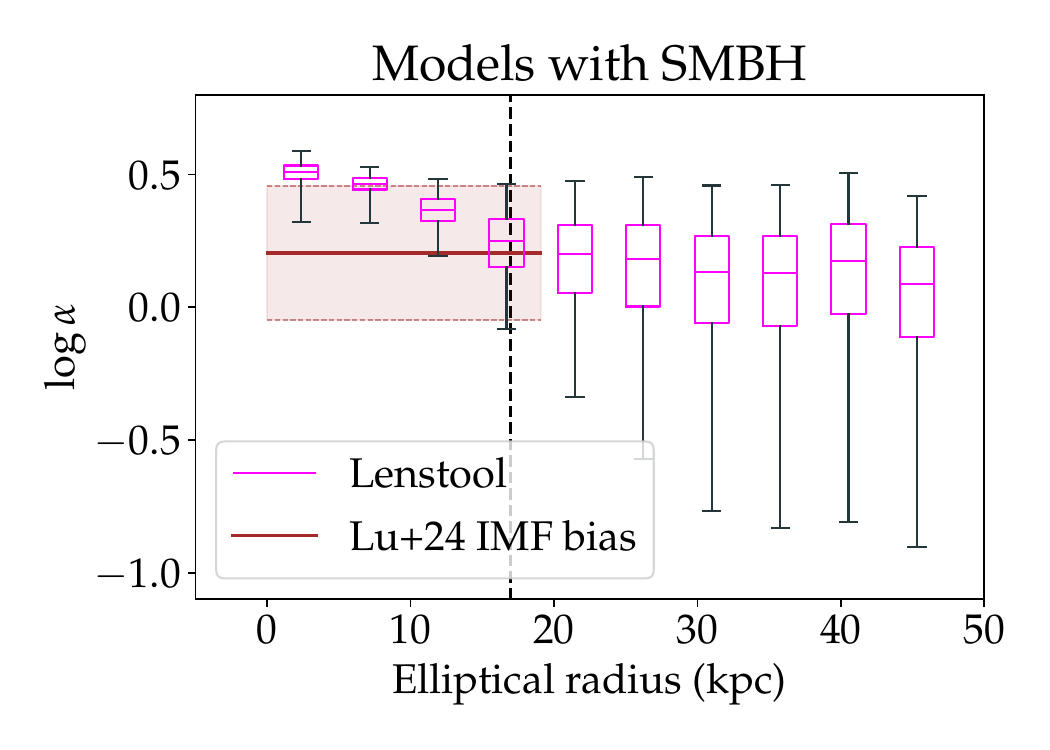}
    \end{minipage}
    \begin{minipage}{0.33\linewidth}
    \centering
    \includegraphics[width=\linewidth]{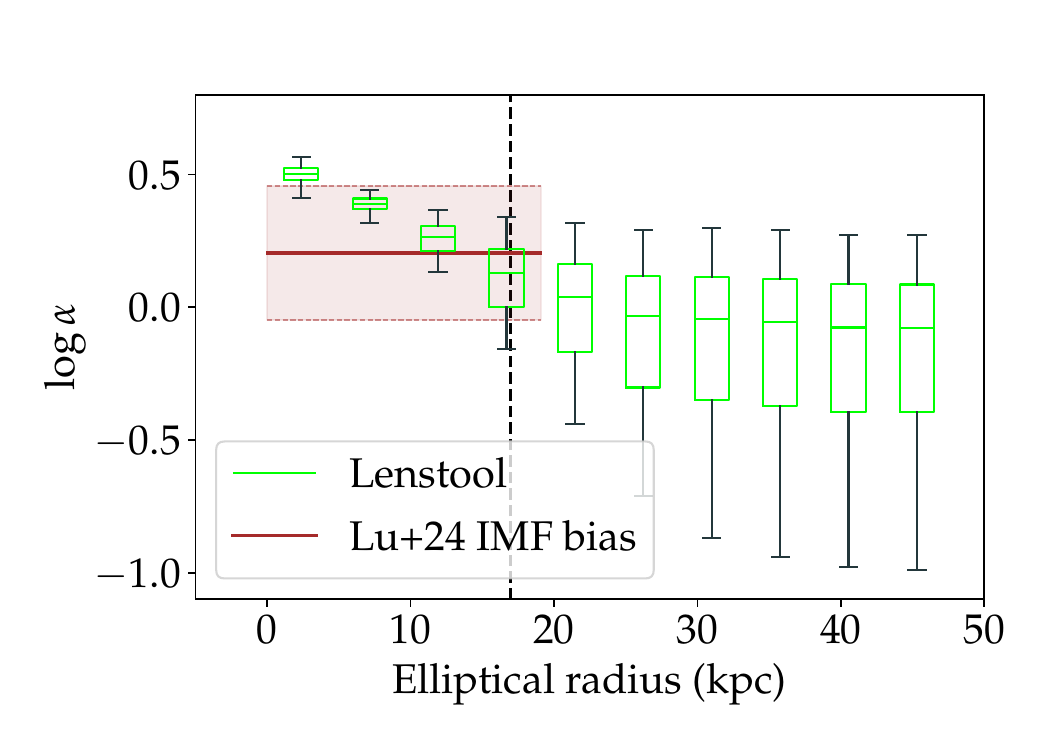}
    \end{minipage}
    
    \begin{minipage}{0.33\linewidth}
    \centering
    \includegraphics[width=\linewidth]{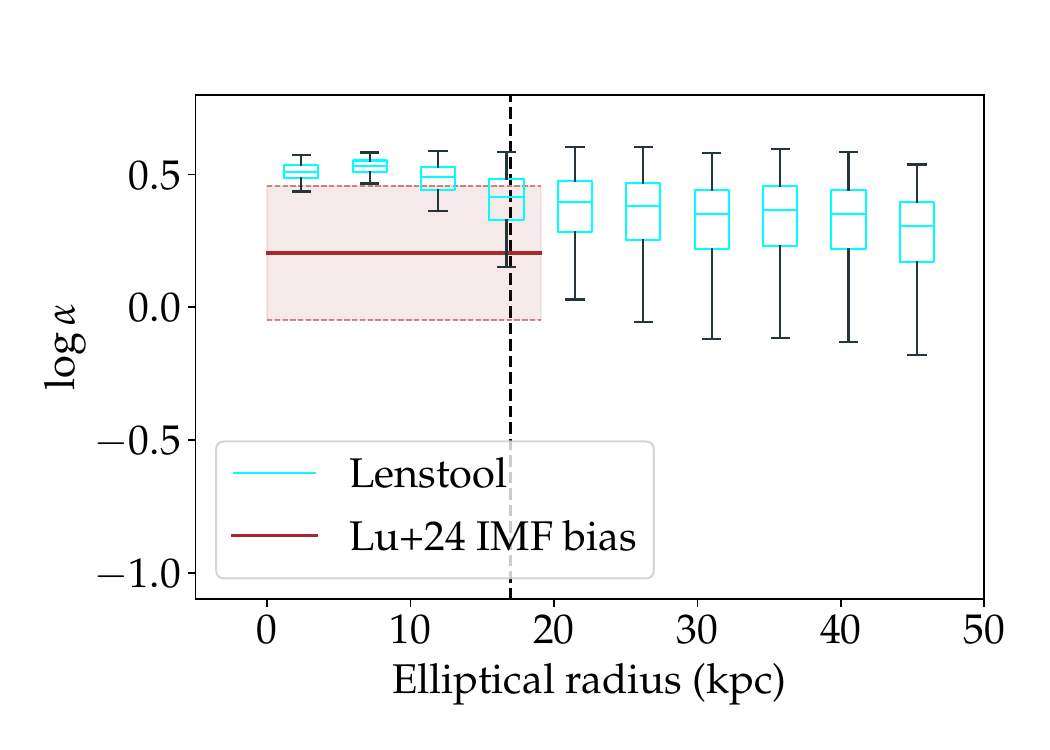}
    \end{minipage}
    \begin{minipage}{0.33\linewidth}
    \centering
    \includegraphics[width=\linewidth]{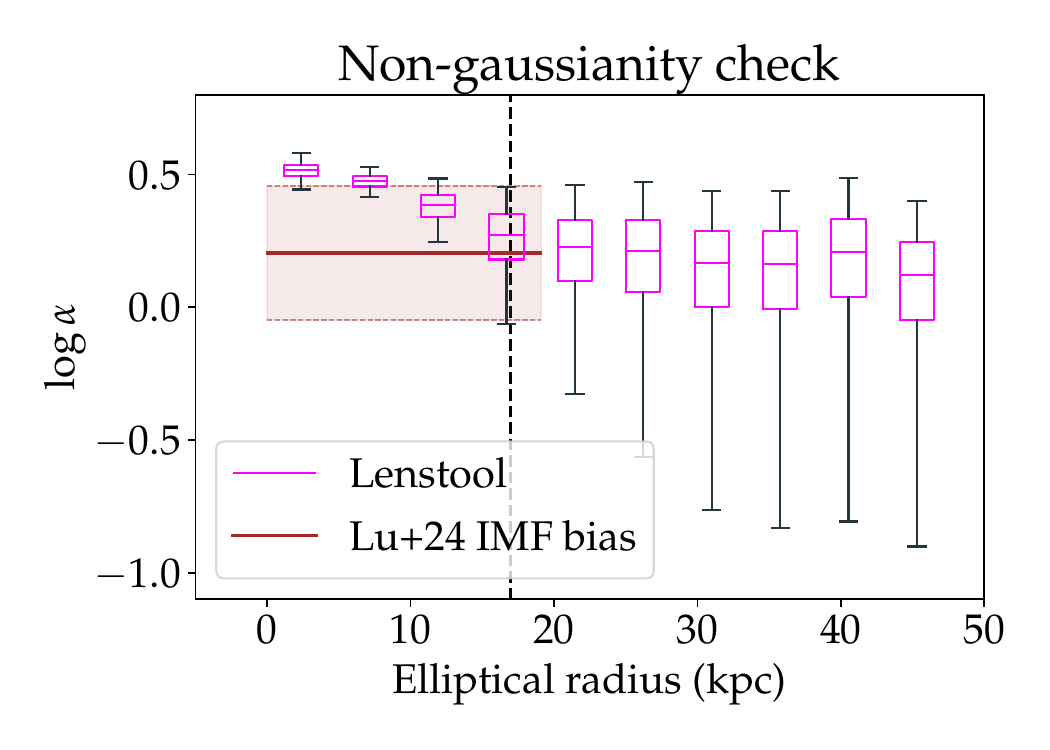}
    \end{minipage}
    \begin{minipage}{0.33\linewidth}
    \centering
    \includegraphics[width=\linewidth]{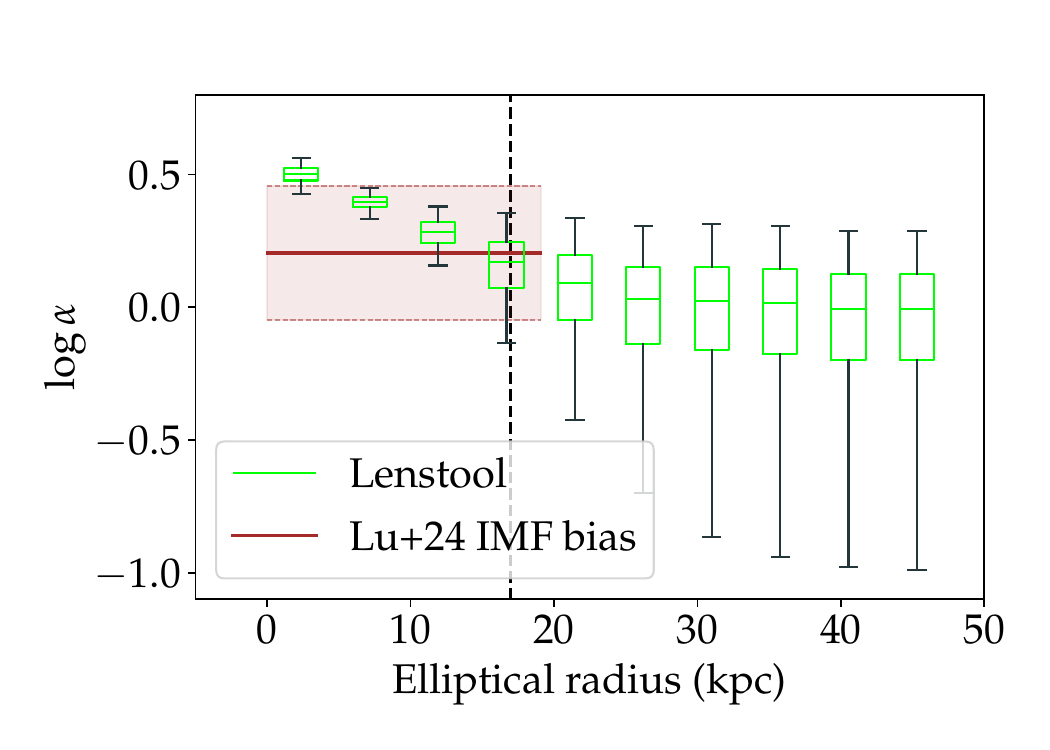}
    \end{minipage}

    \caption{Stellar-mass-to-light ratio mismatch parameter $\alpha$ as a function of the elliptical radius defined in B25a (section~5.2). In each plot, we highlight the BCG size using its half-light radius, represented by the dashed black line. The shaded brown area represents the expected mismatch due to IMF variation in early-type galaxies as reported by \citet{Lu2024}. The width of this area represents $3$ times the standard deviations from \citet{Lu2024} fit, while the plain line represents their best-fitting solution. The $1\sigma\, \rm CI$ and $3\sigma\, \rm CI$ of the posterior distribution of $\alpha$ are represented by the boxes and the associated error bars, respectively. The median is highlighted by the horizontal bars within each box. Each column represents a model with a ``BCG - ML 2'' parametrisation that used different SED models for the cluster member components. They are ordered from \textit{left to right}: \textsc{Bagpipes} Delayed SFH, \textsc{Bagpipes} double power-law SFH and \textsc{LePhare}. The colour scheme of the boxes represents the associated SED models. Each row represents a different test. \textit{Top row:} Mass models with the methodology outlined in this work. \textit{Middle row:} Mass models with a central SMBH. \textit{Bottom row:} Mass models where the kinematic models of the BCG \& ICL component are compared against the observed kinematics, in which we account for the deviation from Gaussian stellar velocity distributions.}
    \label{fig:alpha_profile}
    
\end{figure*}

The top row of Fig.\ref{fig:alpha_profile} presents the estimation of $\alpha$ compared to the results of \citet{Lu2024}. The $3\sigma\, \rm CI$ of each model agrees with the expected mismatch from \citet{Lu2024} for the data points within the BCG. However, the agreement gets weaker towards the cluster centre. Indeed, the $1\sigma\, \rm CI$ of only the two/three last data points of the BCG agree with the expected mismatch. It may indicate an overestimate of $\Upsilon^{\rm BCG}_{\rm *, lt}$ at the first data point if the BCG is not an outlier of the \citet{Lu2024} relations. This could be due to the presence of a SMBH with a mass above the $M_\bullet-\sigma_e$ relation as we discuss in Sect.~\ref{sect:SMBH_discuss}. The assumption behind the BCG \& ICL stellar kinematic models may also have an influence. A JAM model assumes a single ellipticity and a Gaussian line-of-sight velocity distribution. A lower ellipticity would increase the $V_{\rm rms}$ of the kinematic model at a fixed mass, which may reduce the $\Upsilon^{\rm BCG}_{\rm *, lt}$ required at the first data point. However, a scheme with varying ellipticity should be employed to avoid an increase of $V_{\rm rms}$ within the ICL. Similarly, a misalignment with the light tracer and mass distribution that is forbidden by the axisymmetric requirement of the JAM method would result in deviation from a Gaussian distribution for the line-of-sight velocity distribution. A proper assessment of such bias requires a 3D modelling method, such as Schwarzschild's orbit-superposition \citep{Bovy2015}, but is beyond the scope of this work. In Sect.~\ref{sect:Kurtosis_discuss}, we try to correct the effect of possible deviation from a Gaussian line-of-sight velocity distribution by refitting these velocity distributions without the extra-moments, $h_3$ and $h_4$.

\subsection{BCG SMBH}
\label{sect:SMBH_discuss}
In our modelling, we neglect the influence of a SMBH which mass can be degenerated with the BCG stellar mass. Hence, to assess the reliability of $\Upsilon^{\rm BCG}_{\rm *, lt}$ estimates to that bias, we add a point mass profile with a fixed position at the BCG centre in the lensing mass model. We modify our stellar kinematic model by adding a SMBH with its mass contained in the point mass profile of the lensing model. We apply these modelling changes to each ``BCG - ML 2'' model and reproduce the whole optimisation procedure presented in Sect.~\ref{sect:model_opti}. 

We apply this procedure for two SMBH priors, one assuming the mass from the $M_\bullet-\sigma_e$ relation compiled in \citet{vandenBosch2016} and another wider prior to test the deviation from this relation. Using the relation of \citet{vandenBosch2016}, we obtain $M_\bullet=3.66^{+0.62}_{-0.53}\times10^{9}\,M_\odot$ using the uncertainties from their relation fit and our best-fitting velocity dispersion in the first bin (i.e. $\sigma_{BCG}=341$~${\rm km/s}$). The associated prior is a Gaussian distribution with the $M_\bullet$ median as the mean and half of the distance between the two quantiles as a standard deviation. For the prior over the wider range, we use a particularly wide log-uniform prior from $10^{7}$ to $10^{11}\,M_\odot$. 

When the SMBH is assumed from the $M_\bullet-\sigma_e$ relation, we observe mostly negligible changes to the model parameters and the reproduction of the constraints. The posterior on $\Upsilon^{\rm BCG}_{\rm *, lt}$ is mostly the same with a slight decrease of the $1\sigma\, \rm CI$ width within the ICL. It shows that neglecting the BCG SMBH, which would follow the $M_\bullet-\sigma_e$ relation, does not significantly affect the estimate of $\Upsilon^{\rm BCG}_{\rm *, lt}$.

With the larger prior on the SMBH mass, we obtain better reproduction of the stellar kinematics only with the model featuring a double power-law SFH, as in the previous case. However, there are larger changes in the model parameter, and in particular with $\Upsilon^{\rm BCG}_{\rm *, lt}$. These changes appear in the width of the $3\sigma\, \rm CI$, while the median and $1\sigma\, \rm CI$ tend to remain the same as in the model without a SMBH. As our constraints on the SMBH are weak, the modification of $\Upsilon^{\rm BCG}_{\rm *, lt}$ can be overestimated. The constraints are too weak to set a higher bound on the SMBH mass across all three models. The middle row of Fig.~\ref{fig:alpha_profile} presents the mismatch parameter, $\alpha$, for these three models. The data points within the BCG highlight the modification of the $3\sigma\, \rm CI$, which overlap more favourably with the expected mismatch from \citet{Lu2024}. It indicates that a SMBH with masses above the $M_\bullet-\sigma_e$ relation can lead to an overestimation of $\Upsilon^{\rm BCG}_{\rm *, lt}$ within the BCG. However, our dataset is insufficient to assess the presence of such a SMBH. 

Another aspect of the SMBH addition is the slight modification of $\Upsilon^{\rm BCG}_{\rm *, lt}$ within the ICL, as the $3\sigma\, \rm CI$ of the double power-law SFH and \textsc{LePhare} SED models are closer to each other than previously. In the case of the double power-law SFH, $\Upsilon^{\rm BCG}_{\rm *, lt}$ are in better agreement with $\Upsilon^{\rm BCG}_{\rm *, SED}$ than before, as the $1\sigma\, \rm CI$ is now including $\log \alpha=0$. In contrast, the model with a delayed SFH, presents a narrower $3\sigma\, \rm CI$ which barely includes the horizontal line at $\log \alpha=0$. Hence, it points toward a more complex correlation with other model components rather than a degeneracy of the mass between the BCG \& ICL and SMBH.

\subsection{Non-gaussianity of the BCG stellar velocity distribution}
\label{sect:Kurtosis_discuss}
In B25a (section~5.2), we fit the line-of-sight velocity distribution assuming a Gaussian modified by a Gauss-Hermite expansion. In our parametrisation, only the coefficients for the third and fourth Hermite polynomials are non-zero (i.e. parameters $h_3$ and $h_4$). To account for the $V_{\rm rms}$ differences if the velocity distribution is purely Gaussian, we fit a Gaussian distribution to the velocity distribution obtained in B25a. It mostly affects points in the ICL, as they exhibit larger $h_4$ values of around $0.05$. Hence, we observe an opposite pattern from \citet{Gavazzi2005}, as the deviation from Gaussianity appears at a larger radius than closer to the cluster centre. 

We reproduce the importance sampling procedure with those modified $V_{\rm rms}$, and the impact on the mismatch parameter is presented in the bottom row of Fig.~\ref{fig:alpha_profile}. It does not change the agreement with the \citep{Lu2024} expected mismatch for the BCG, although models tend to prefer slightly higher $\Upsilon^{\rm BCG}_{\rm *, lt}$ for the ICL. In particular, the model with \textsc{LePhare} has a narrower $1\sigma\, \rm CI$ as its lower tails lean towards higher $\Upsilon^{\rm BCG}_{\rm *, lt}$. Hence, the deviation from a Gaussian stellar velocity distribution does not impact $\Upsilon^{\rm BCG}_{\rm *, lt}$ in the BCG according to this test. However, it leaves other deviations from the JAM method, such as varying ellipticity, as potential biases for future work.


\subsection{Degeneracy between $M_*/L$ and the DM profiles}
\label{sect:ML_DM_degeneracy}
An additional potential source of the observed discrepancies in the stellar mass-to-light ratio is the assumed DM profile. In the case of the BCG, it is represented by two dPIEs, one for the BCG DM and the other for the main DM halo. As shown by the parameter posterior in Appendix~\ref{app:model_parameters}, the $r_{\rm core}$ of these two profiles converge to values larger than $100$~${\rm kpc}$ (i.e. $20$~${\rm arcsec}$), yielding a cored profile around the BCG. As our mass probes are constraining the total mass, the BCG \& ICL stellar mass may account for some DM to accommodate the limitation of dPIEs. 

To test such degeneracies, we use the mass profiles presented in Fig.~\ref{fig:mass_profile_3D}, and we swap the DM profiles for other parametrisations such as a gNFW and a power-law profile and allow for variations of the BCG \& ICL stellar-mass-to-light ratio. The modelled total density, $\rho^{\rm mod}_{\rm tot}$, as the following form:
\begin{equation}
    \rho^{\rm mod}_{\rm tot}(r)=\rho_{\rm Gas}(r)+a\rho_{\rm BCG}(r)+\rho^{\rm mod}_{\rm DM}(r)
\end{equation}
where $\rho_{\rm Gas}$ and $\rho_{\rm BCG}$ are the gas and BCG \& ICL stellar mass density from the mass model. We do not account for the cluster member in that test, aside from the BCG. $a$ represent the variation of the stellar-mass-to-light ratio of BCG \& ICL while $\rho^{\rm mod}_{\rm DM}$ is the DM density from an analytical profile. We are fitting $\rho^{\rm mod}_{\rm tot}$ on the total density of the model without the cluster members in log space, and we assume an error of $0.1$~$\rm dex$. We deliberately choose an error larger than the model uncertainty to account for systematic error due to the mass model D. We perform that fitting for each ``BCG - ML 2'' model with three different parametrisations of the DM profile. Two of these parametrisations are a gNFW and a power-law profile. For the third parametrisation, we use a similar formula as for the BCG \& ICL stellar mass density but with the DM, which we denote as ``fiducial''. This last parametrisation allows us to measure the effect of the fitting procedure, in particular, the uncertainty expected from the assumed error.

\begin{figure*}
    \centering
    \begin{minipage}{0.33\linewidth}
    \centering
    \includegraphics[width=\linewidth]{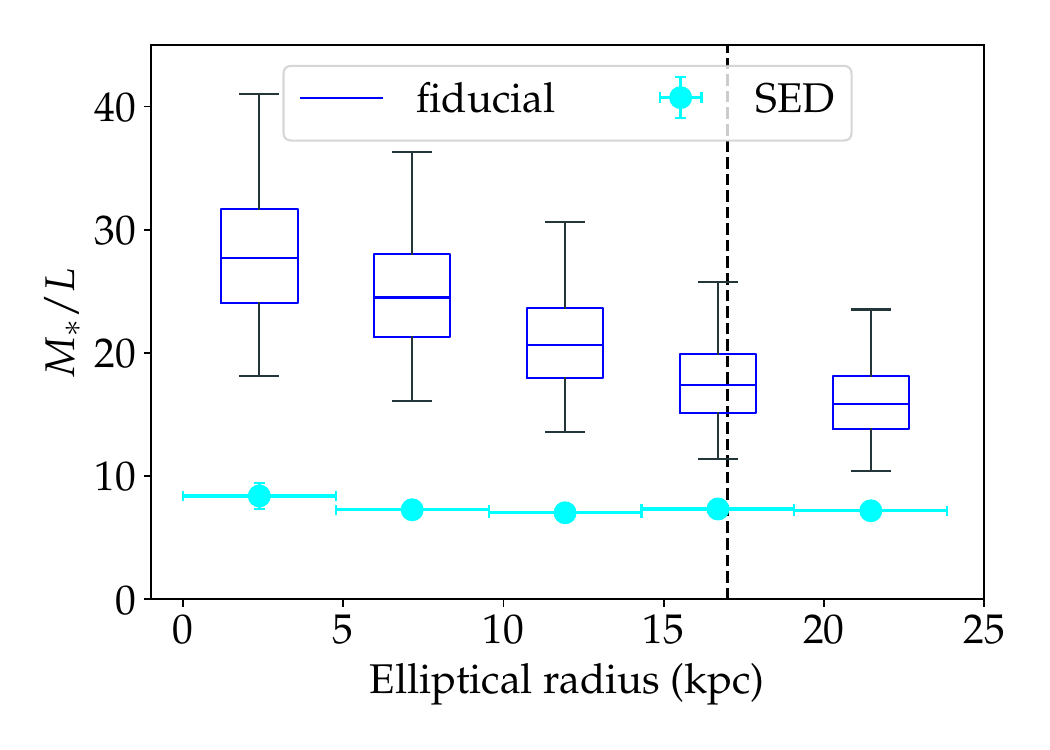}
    \end{minipage}
    \begin{minipage}{0.33\linewidth}
    \centering
    \includegraphics[width=\linewidth]{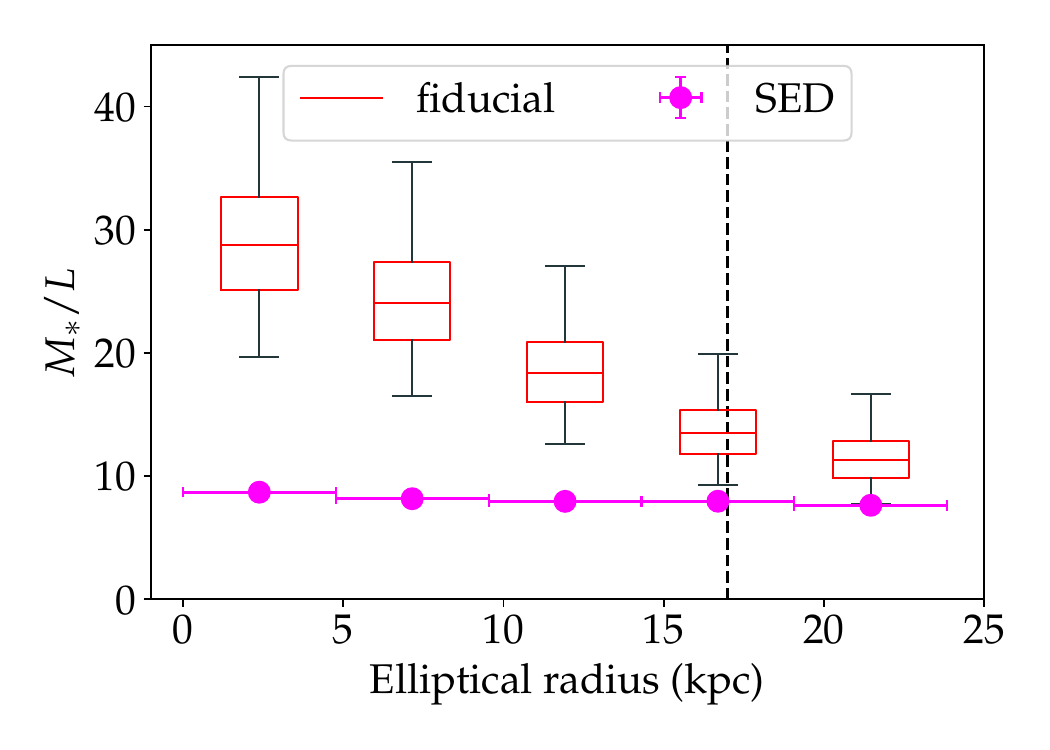}
    \end{minipage}
    \begin{minipage}{0.33\linewidth}
    \centering
    \includegraphics[width=\linewidth]{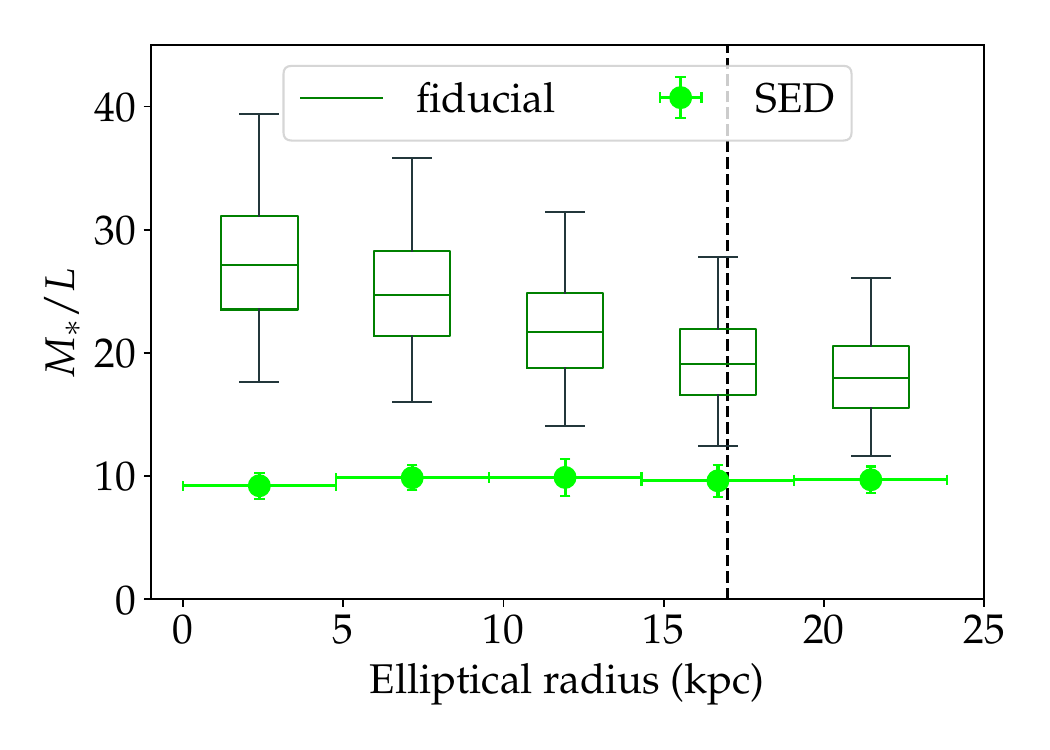}
    \end{minipage}
    
    \begin{minipage}{0.33\linewidth}
    \centering   
    \includegraphics[width=\linewidth]{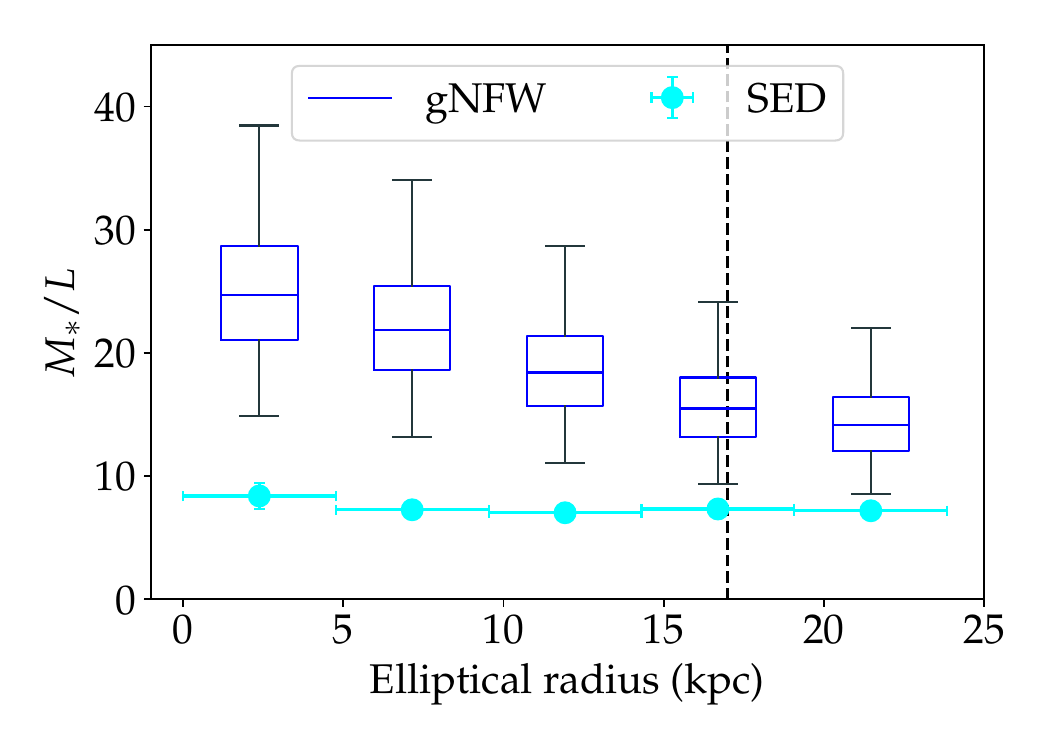}
    \end{minipage}
    \begin{minipage}{0.33\linewidth}
    \centering
    \includegraphics[width=\linewidth]{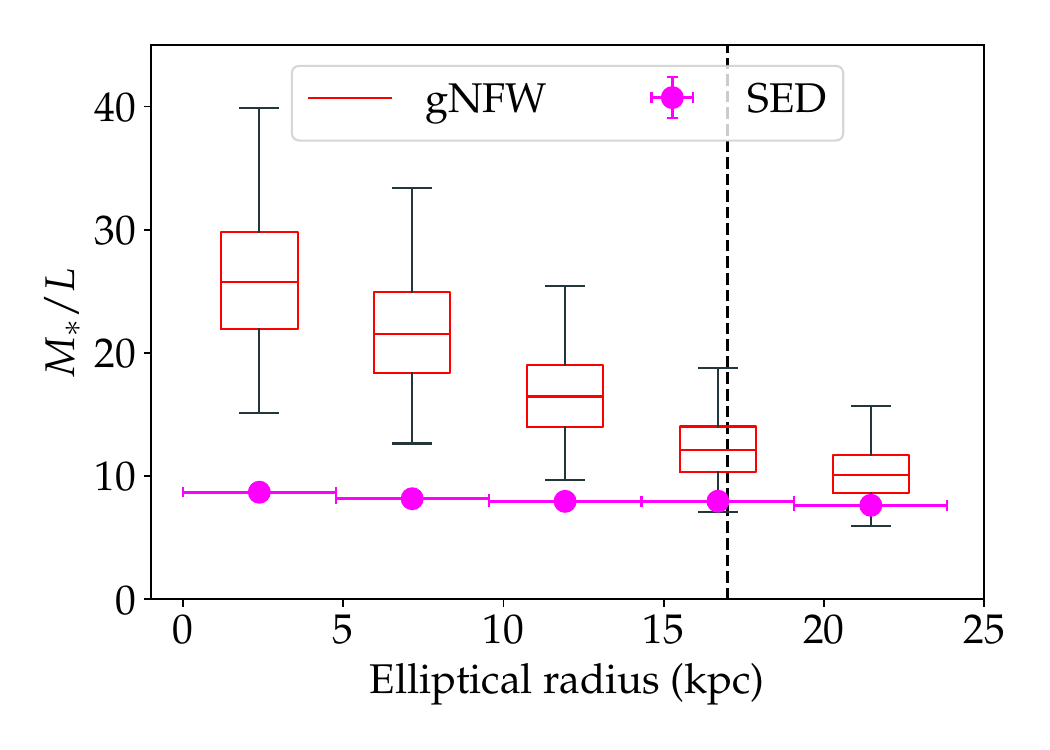}
    \end{minipage}
    \begin{minipage}{0.33\linewidth}
    \centering
    \includegraphics[width=\linewidth]{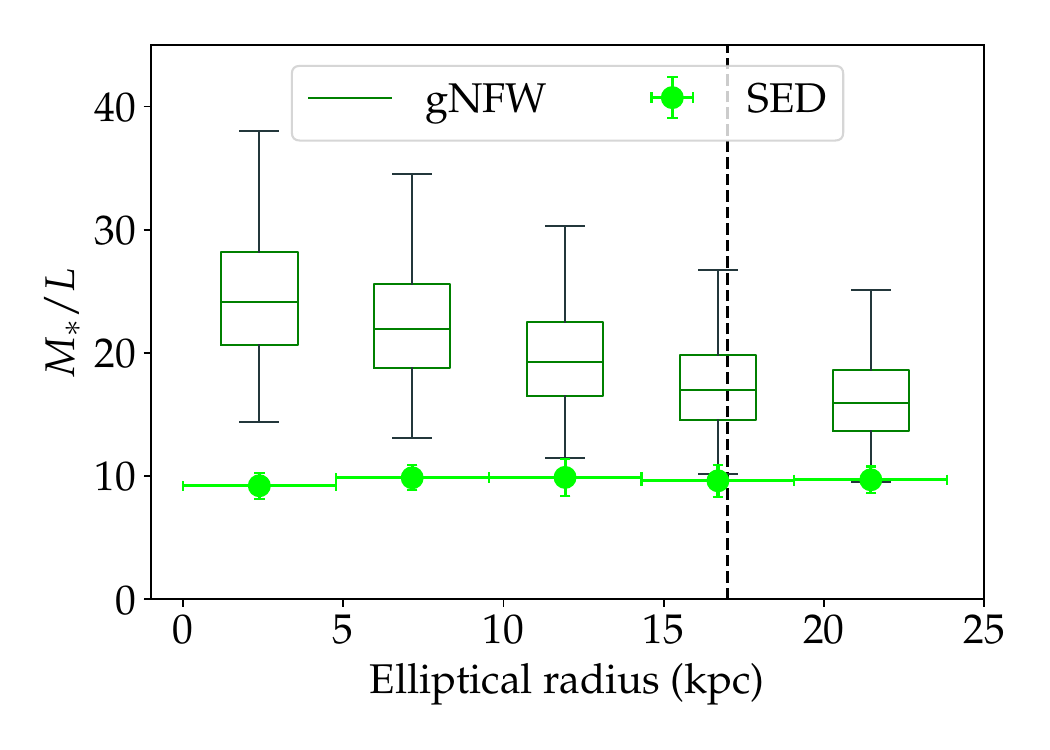}
    \end{minipage}
    
    \begin{minipage}{0.33\linewidth}
    \centering
    \includegraphics[width=\linewidth]{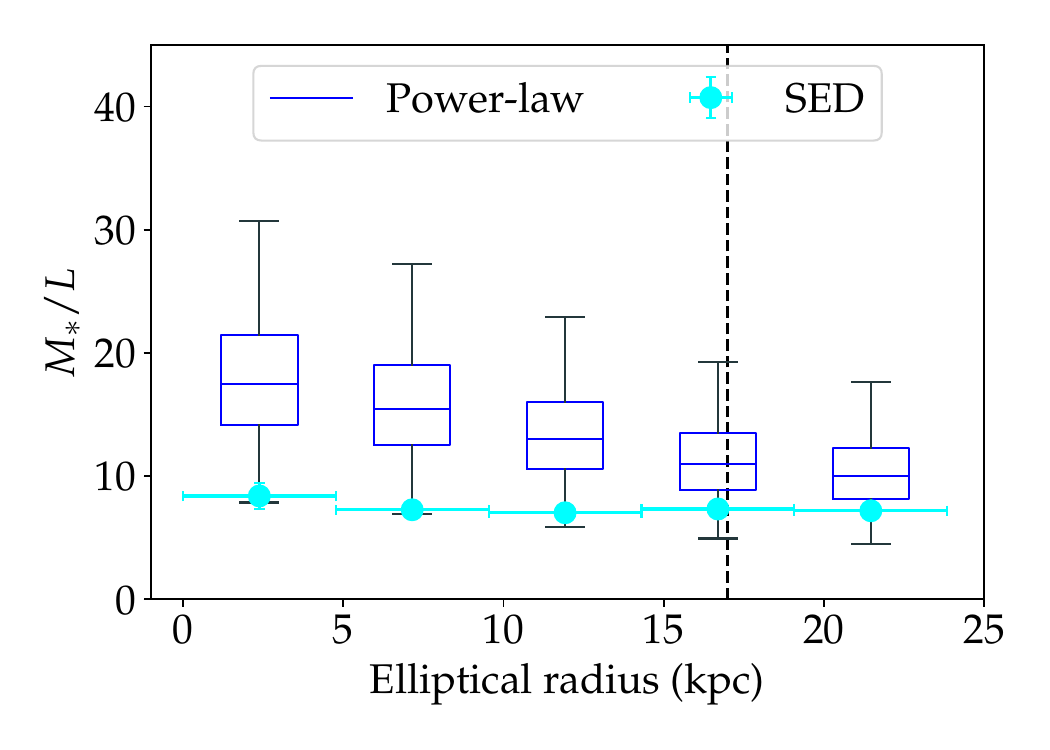}
    \end{minipage}
    \begin{minipage}{0.33\linewidth}
    \centering
    \includegraphics[width=\linewidth]{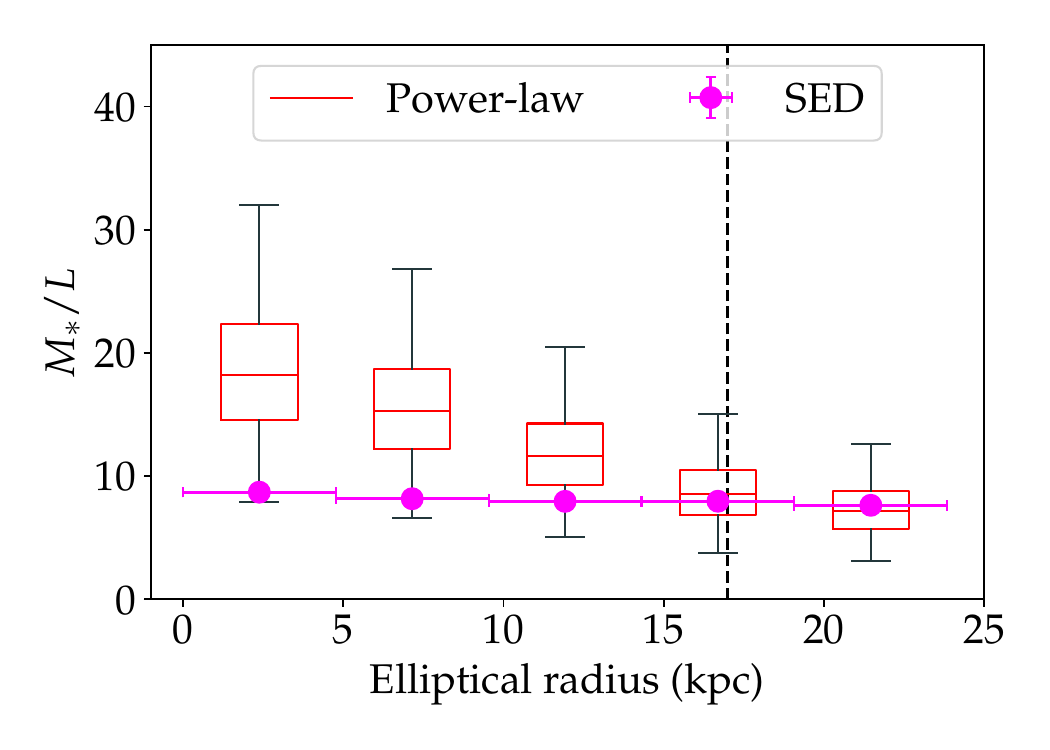}
    \end{minipage}
    \begin{minipage}{0.33\linewidth}
    \centering
    \includegraphics[width=\linewidth]{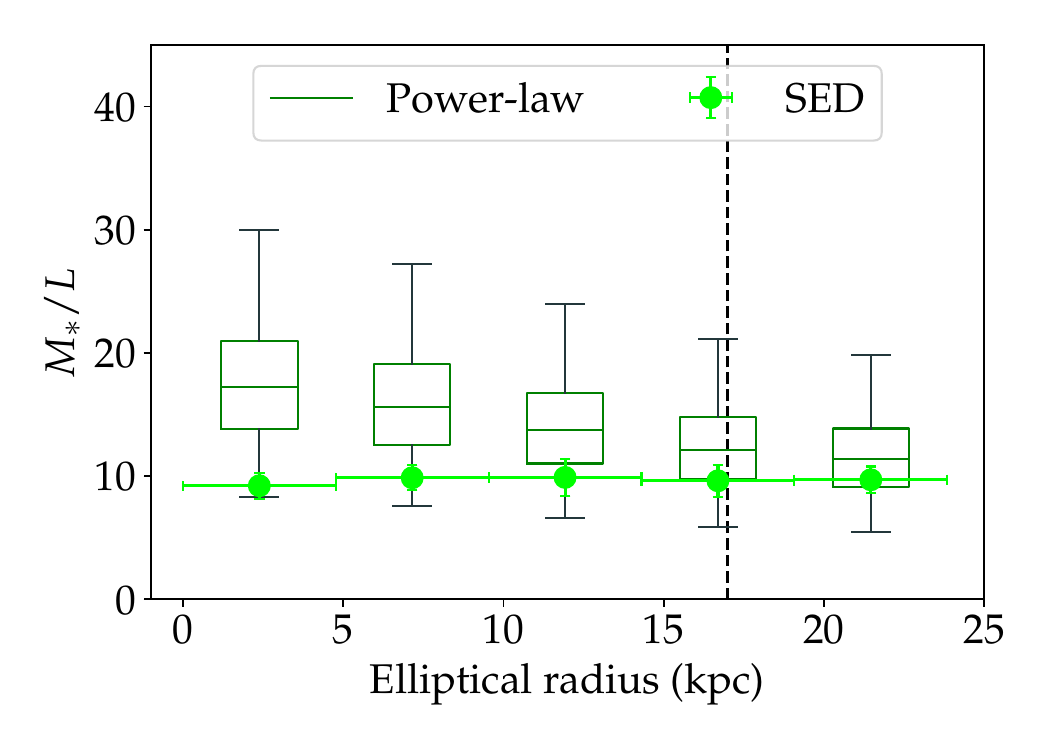}
    \end{minipage}

    \caption{Stellar-mass-to-light ratio $M_*/L$ as a function of the elliptical radius defined in B25a (section~5.2). In each plot, we highlight the BCG size using its half-light radius, represented by the dashed black line. The $1\sigma\, \rm CI$ and $3\sigma\, \rm CI$ of the posterior distribution of $M_*/L$ from the mass model are represented by the boxes and the associated error bars, respectively. The median is highlighted by the horizontal bars within each box. For the SED estimation of $M_*/L$, the error bars represent the $1\sigma\, \rm CI$. Each column represents a model with a ``BCG - ML 2'' parametrisation that used different SED models for the cluster member components. They are ordered from \textit{left to right}: \textsc{Bagpipes} Delayed SFH, \textsc{Bagpipes} double power-law SFH and \textsc{LePhare}. The colour scheme of the boxes represents the associated SED models. Each row represents a different test: fiducial fit (\textit{Top row}), gNFW fit (\textit{Middle row}) and power-law fit (\textit{Bottom row}).}
    \label{fig:DM-degeneracy-check}
    
\end{figure*}

Fig.~\ref{fig:DM-degeneracy-check} presents the results for each ``BCG - ML 2'' model for the three parametrisation where each column refers to a different ``BCG - ML 2'' model and each row a different DM parametrisation. When comparing the first and second rows of the figure, adopting a gNFW DM profile reduces the discrepancy between the SED and mass-model stellar mass-to-light ratios, yet the results still indicate an underestimation of the stellar mass by the SED fitting. The power-law profile in the last row shows the largest modification, with the required stellar mass almost halved. In particular, the SED estimation is now consistent with the $3\sigma\, \rm CI$ of the mass model estimation. However, the power-law profile is still not enough to recover the SED estimation within the $1\sigma\, \rm CI$.

Comparing the uncertainty from the top row of Fig.~\ref{fig:DM-degeneracy-check} and Fig.~\ref{fig:ML_profile}, it is likely that a proper model with a power-law or gNFW profile will have smaller $3\sigma\, \rm CI$ and $1\sigma\, \rm CI$. The agreement within $3\sigma\, \rm CI$ is unlikely to hold for most models, as the SED estimation is at the limit of that interval.

It appears that the DM profile is likely the largest source of misestimation of the stellar mass from the BCG \& ICL component. Although using a steeper mass profile, such as a gNFW or power-law, is not enough to make both estimates agree. As presented in Fig.~\ref{fig:alpha_profile}, the overestimation of the BCG stellar mass is at the $3\sigma$ limit of the discrepancies observed by \citet{Lu2024}. Hence, a steeper DM profile still tends to favour a discrepancy, but shows better agreement with \citet{Lu2024} results.

These three tests regarding our estimate of $\Upsilon^{\rm BCG}_{\rm *, lt}$ highlight that the main biases are the degeneracy between the DM and stellar mass of the BCG, as the two other tests did not show significant influence on the final results. Such biases were also found in the kinematic studies of ETG, although they have been mitigated by constraining the mass distribution using a 2D mapping of stellar kinematics \citep{Cappellari2012}. Future work should therefore tackle this issue by incorporating 2D kinematic constraints, which appear most promising given that baryons are more centrally concentrated than DM, thereby helping break the degeneracy between the stellar and DM mass components.


\subsection{Stellar-to-subhalo mass relation}
\label{sect:SsHMR_discuss}

Thanks to our new cluster member modelling, we can estimate the stellar-to-subhalo mass relations (SsHMR) with the double power-law defined in Sect.~\ref{sect:cluster_member_definition} for satellite galaxies in cluster cores. In the context of the SsHMR in cluster, satellite galaxies refer to gravitationally bound subhaloes of a cluster-size halo, which in our case translate to the cluster members without the central galaxy/BCG. Such relations are measured in cosmological simulations, and they should vary during the infall of galaxies within their host cluster \citep{niemiec2019}: the increase of stellar mass stops as galaxies are quenched by the cluster environment, while the outside-in stripping of DM by tidal forces of the cluster decreases sub-halo masses. In addition, self-interacting DM could cause an additional ‘evaporation’, thus further decreasing sub-halo masses during their infall \citep{Bhattacharyya2022, Sirks2022}. As galaxies that have spent a longer time in their host cluster tend to be located closer to the core, precisely measuring the radial evolution of the SsHMR could allow us to quantify the impact of tidal stripping, as well as possible evaporation due to DM self-interactions. Stacked galaxy-galaxy weak lensing analyses have placed some constraints on the SsHMR \citep{Li2016, sifon2015, niemiec2017, sifon2018, Wang2024}, but these measurements do not reach the cluster core, where stripping effects should be highest. To make the fairest comparisons, the types of galaxies (i.e. early or late type galaxies) and their distances to the cluster core should be considered \citep{niemiec2022}. In particular, \citet{niemiec2022} reported a double power-law fit of the SsHMR for satellite galaxies in the IllustrisTNG \citep{Pillepich2018,Weinberger2017} simulation. They reported the fit for the whole sample of satellite galaxies, as well as for different subsets based on galaxy type, redshift bin, or distance to the cluster core. Hence, to match the sample of galaxies considered from IllustrisTNG to our observational measurements, we combine different SsHMR to match the properties of our sample. We sample $1000$ SsHMRs, where $19$ per cent of the sample follows the SsHMR for satellite late-type galaxies. This proportion matches the proportion of late-type galaxies in our cluster member sample as determined by visual inspection. We classify cluster members as late-type galaxies if they exhibit complex morphologies with possible spiral arms or colour difference from the bulk of the population. The remaining cluster members are modelled assuming the SsHMR derived from satellite early-type galaxies with radii smaller than the maximum cluster-centric distance of our cluster member selection. Hence, the radial dependencies of the SsHMR are only included for early-type galaxies, as this dependence has not been probed by \citet{niemiec2022} for late-type galaxies.

\begin{figure*}
    \begin{minipage}{.49\linewidth}
    \centering
    \includegraphics[width=\linewidth]{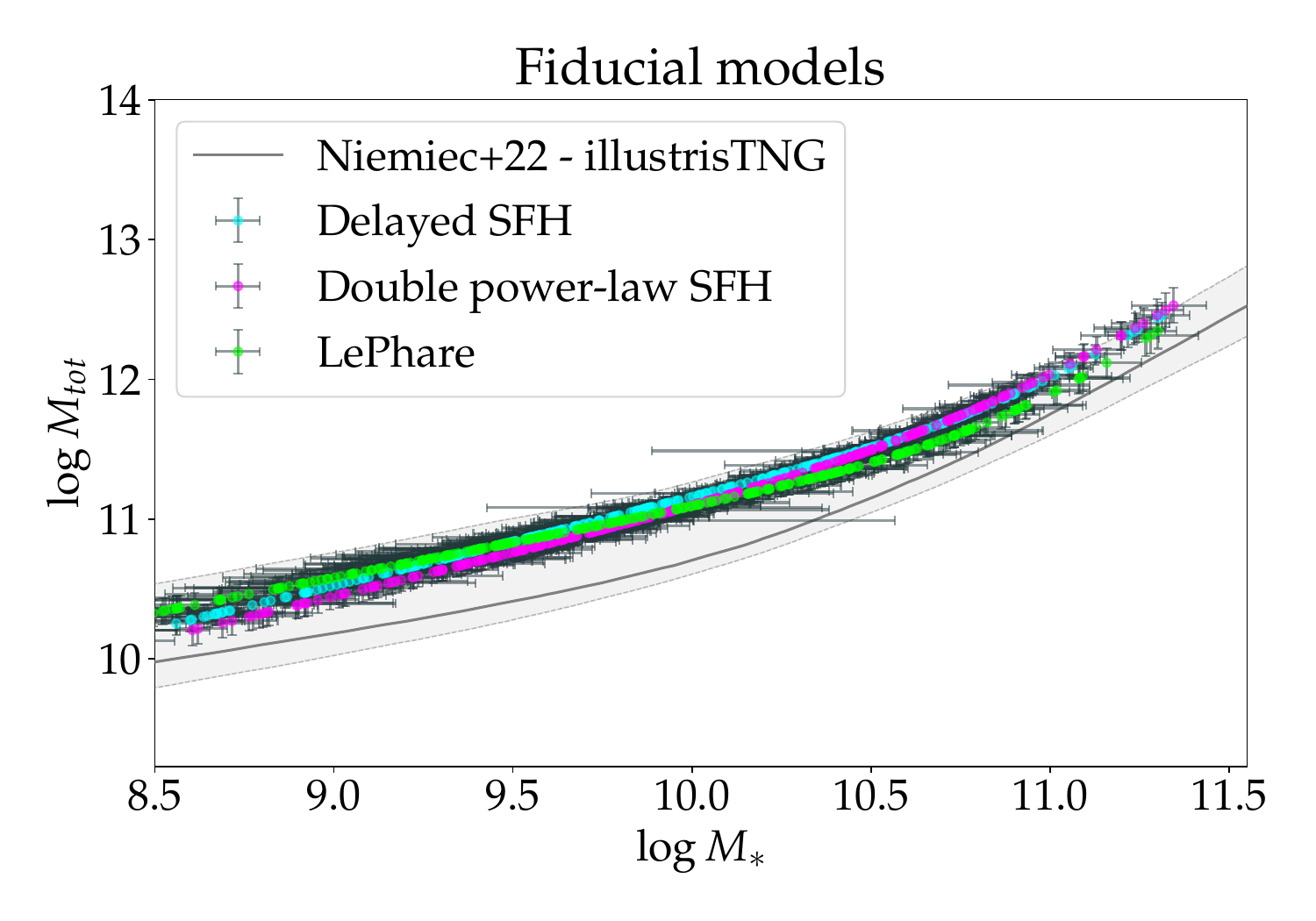}
    \end{minipage}
    \begin{minipage}{.49\linewidth}
    \centering
    \includegraphics[width=\linewidth]{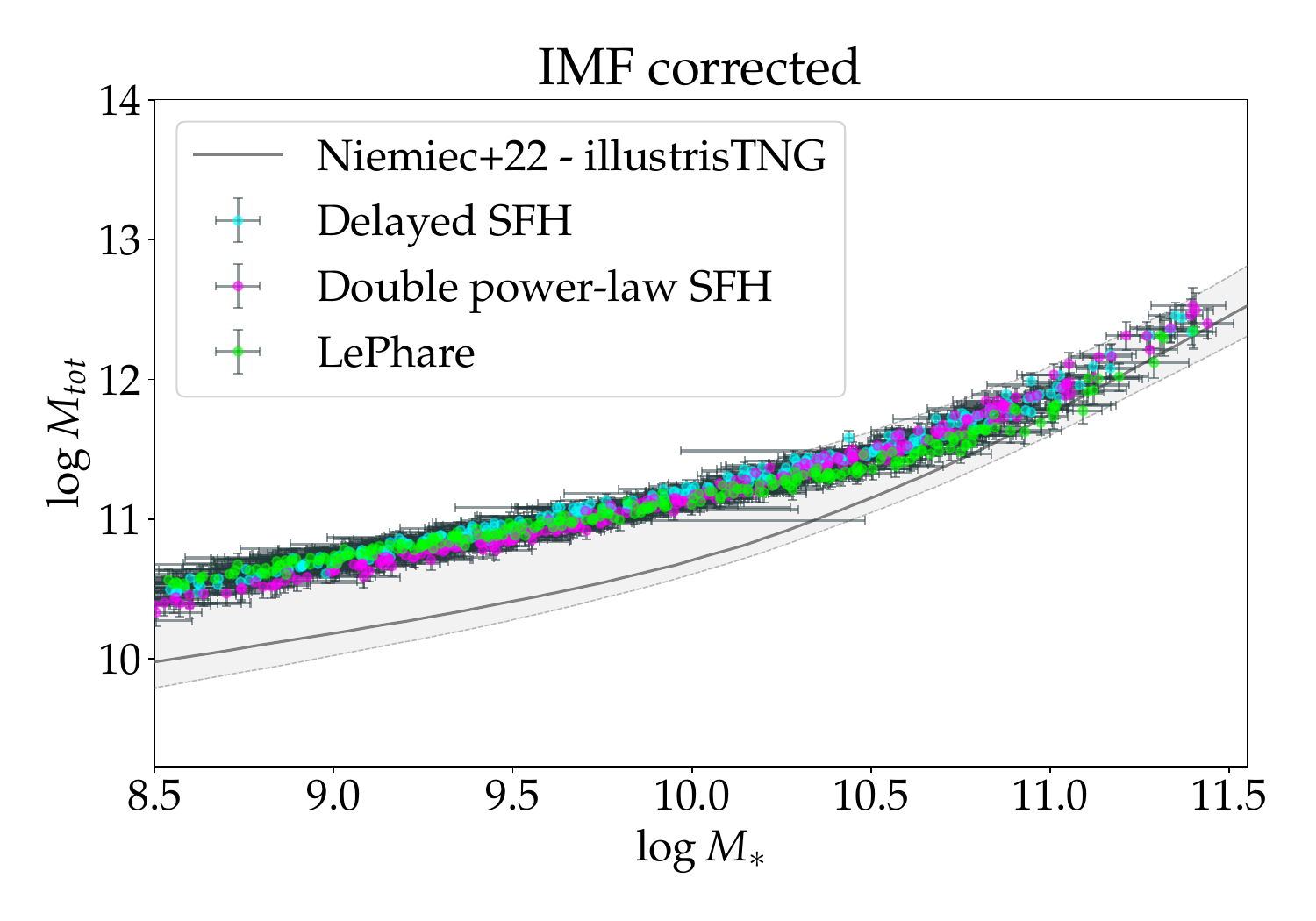}
    \end{minipage}
    \caption{Stellar-to-subhalo mass relation measured in this work compared to the results from the illustrisTNG simulations as reported in \citet{niemiec2022}. We represent our estimate through each cluster member. The positions of the colour points represent the median of their stellar masses, as determined by the SED fitting, or the total mass from the model posterior distribution. The error bars show the $1\sigma\, \rm CI$ for those distributions. Each colour represents different SED models with the Delayed SFH, double power-law SFH and \textsc{LePhare} models shown in cyan, magenta and lime, respectively. The results from the IllustrisTNG simulation are represented by the grey area, which shows the $1\sigma\, \rm CI$, while the plain lines show the median. The \textit{left panel} presents stellar-to-subhalo mass relations as defined in the mass models, while the \textit{right panel} presents the same relation with the stellar masses being corrected for the IMF variation as reported in \citet{Lu2024}. The IMF correction is based on the model-predicted velocity dispersion, $\sigma_e$, for the best-fitting model.}
    \label{fig:SsHMR}
\end{figure*}

The left panel of Fig.~\ref{fig:SsHMR} compares the SsHMR from IllustrisTNG with our measurement for each model with the ``BCG - ML 2'' parametrisation. The $1\sigma\, \rm CI$ of the SsHMR of our three models and IllustrisTNG overlap over the entire stellar mass range, from $10^{8.5}$ to $10^{11.5}$~$M_\odot$. We notice that our results are systematically above the median SsHMR measured from the simulation. It may highlight discrepancies between our model and the simulation SsHMR on subsamples, such as selections based on galaxy type, e.g., only early- or late-type galaxies. As in the current modelling, we do not distinguish between the two galaxy types; we leave any further analyses for future work.

Each SED model used to estimate the cluster member $M_*$ yields slightly different SsHMRs, which agree at most for $M_*\approx10^{10}$~$M_\odot$ and diverge as $M_*$ increases or decreases. Hence, we observe that the double power-law SFH model provides higher $M_*$ for the most massive cluster members and lower $M_*$ for the least massive. The \textsc{LePhare} SED model shows an opposite pattern, while the delayed SFH model is closer to the double power-law one to a lesser extent. Even if their $1\sigma\, \rm CI$ overlap with IllustrisTNG, they present different slopes. It may indicate some discrepancies in the details of the different galaxy populations within AS1063. Increasing the mass range of $M_*$ above $10^{11.5}$~$M_\odot$ would likely lead to a disagreement between our results and the simulation, particularly when considering the double power-law SFH model.

\subsection{Cluster member IMF variation}
\label{sect:IMF_CL}
In our SsHMR estimate, we use a fixed estimation of the cluster member $M_*$ as the posterior median from the SED fit. These SED models assume a Milky-Way like IMF, which does not agree with the kinematic modelling of early-type galaxies \citep{Posacki2015,Lu2024}. We use the best-fitting relation from \citet{Lu2024} for the expected mismatch between SED and kinematic estimate of $M_*$ to correct our SED estimate of $M_*$. We estimate $\sigma_e$ for each cluster member based on their mass and light distribution as we do for the cluster member likelihood (see Sect.~\ref{sect:cl_likelihoods}) and assume the best-fitting mass model. We apply this correcting factor to each cluster member in our sample without distinguishing between early- and late-type galaxies. 

The right panel of Fig.~\ref{fig:SsHMR} presents the SsHMR with the correcting factor applied to $M_*$ estimates with the SsHMR from IllustrisTNG. Similarly to the uncorrected SsHMR, the $1\sigma\, \rm CI$ of the observation and simulation overlap. We also observe that the SsHMR followed in AS1063 is systematically higher than the median SsHMR from IllustrisTNG. Although at the massive end, there is a better agreement with the trend between observations and simulations, as there is no tendency for discrepancies at $M_*\approx10^{10.75}$~$M_\odot$ or above. 

We try to optimise new mass models with the correct $M_*$, although the obtained best-fit model $\sigma_e$ are not in agreement with the ones used to estimate the correcting factor for $M_*$. These differences in correction factors exceed $50$ per cent for some cluster members. Hence, we reject these models. The correcting formula from \citet{Lu2024} as a polynomial of degree $1$ could be added to our modelling of cluster members with a split between early and late-type galaxies. It could be a new way of estimating IMF variations within the cluster galaxy population.

Our new modelling of cluster members appears to be a robust method for estimating SsHMR, as the overall trend aligns with large-scale simulation results. The estimates of the total masses of cluster members seem consistent across SED models, as the slope difference between them is expected from the difference in stellar masses. As highlighted by the slight differences between our results and illustrisTNG, anomalies may be found between observations and simulations, which can highlight differences in how the cluster environment strips DM in real cases.

\section{Conclusion}

In this series of two papers, we present the first mass modelling method that allows us to fully disentangle DM at the galaxy- and cluster-scale, but also each baryonic component. To achieve such details, we rely on multiple mass probes to specifically constrain the baryonic components with X-ray surface brightness or multi-band photometry. We combine total mass constraints from the BCG \& ICL stellar kinematics and strong lensing to constrain the mass profile from $\approx5$ to $250$~${\rm kpc}$, and use cluster member velocity dispersions to measure their total masses. We detail how we gather these sets of constraints in B25a.

Here, we focus on the mass modelling methodology and results. We test different parameterisations for the BCG \& ICL components, which favour a varying $\Upsilon^{\rm BCG}_{\rm *}$ with radius. In our case, a simple model with two coefficients, one for the BCG and one for the ICL, is favoured. In addition to providing a comprehensive mass model of AS1063, with each cluster component disentangled, we reduce the uncertainties on the total mass profile by $20-30$ per cent, depending on the considered radii. It allows us to measure the DM only profile with an error of less than $5$ per cent over the entire considered region. We obtain these results while presenting one of the most complex parametric mass models.

We then discuss the robustness of the estimates of the stellar masses contained in the BCG \& ICL component, as well as cluster members. We found that the stellar mass of the ICL agrees at $3\sigma$ with a Milky-Way like IMF as fitted through three different SED models. We find that the BCG requires a larger $\Upsilon^{\rm BCG}_{\rm *}$ that can be explained by the IMF variation observed in early-type galaxies, although AS1063 BCG is in the tails of the early-type galaxies distribution modelled by \citet{Lu2024}. We look for over- or under-estimates of $\Upsilon^{\rm BCG}_{\rm *}$ by adding a SMBH in the BCG, accounting for deviations from a Gaussian velocity distribution and looking at its degeneracy with the DM mass profile. We found that the most significant bias is the degeneracy with the DM profile, although a gNFW or power-law profile still yields $\Upsilon^{\rm BCG}_{\rm *}$ larger than the SED estimation. However, with those steeper DM profiles, the discrepancies are in better agreement with \citet{Lu2024} results. Besides this degeneracy, we found that only a SMBH with a mass above the $M_\bullet-\sigma_e$ relation, we find no significant influence on $\Upsilon^{\rm BCG}_{\rm *}$.

Regarding cluster members, we analyse our estimates of their total and stellar masses by comparing the SsHMR included in the model with results from the large-scale simulation IllustrisTNG \citep{niemiec2022}. We find a $1\sigma\, \rm CI$ agreement between the observational and simulated relations, although with some discrepancy at the higher mass end ($>10^{10.75}$~$M_\odot$) for the delayed and double power-law SFH models. When accounting for the IMF variations with the \citet{Lu2024} correction formula, we find a slightly better agreement for the massive end, as no data points are outside the $1\sigma\, \rm CI$ of the IllustrisTNG result. More detailed comparisons in future work will enable us to refine the comparisons and potentially identify discrepancies between observations and simulations for subsets of the galaxy population.

In this series, we extend the B24 framework to obtain a complete picture of the cluster mass distribution. This method can be further improved via a better treatment of the constraints or modelling hypotheses. With the development of hardware accelerators, the next logical step appears to be considering the full light information provided by multiple images, similar to what is done routinely with galaxy-scale lenses \citep[e.g. ][]{Nightingale2021}. At cluster-scale, it is already possible to reconstruct a multiple image system at fixed mass model \citep[e.g. ][]{Sharma2021}, although the combination of both the reconstruction and mass model optimisation has been out of reach. \citet{Acebron2024} is paving the way in that direction as they used a hybrid approach, where they combined the reconstruction of a single lensing system with the usual point-like approximation for the other systems. 

Beyond the methodological considerations, applying our method to a sample of clusters, ideally relaxed, will be instrumental in constraining DM properties by allowing for the precise estimation of the DM only profile. Crucially, our approach is the first step toward reproducing the treatment of \citet{Robertson2021} for observations aimed at measuring the self-interacting DM cross-sections, as we have an accurate census of the baryonic distribution. This method can also be leveraged to develop similar analyses for alternative DM models \citep[e.g. fuzzy DM profile from ][]{Chan2022} but also modified gravity, allowing us to set constraints beyond the single scope of the self-interacting DM paradigm.

The broader application of this method depends on the availability of comprehensive multi-wavelength datasets, which currently exist for only a limited number of clusters, such as the Hubble Frontier Fields. This work advances cluster mass modelling to new methodological and observational frontiers, establishing the observational framework required for the precise characterisation of the DM distribution.

\section*{Acknowledgements}
The authors are grateful to the referee for carefully reading the manuscript and for valuable suggestions and comments, which helped improve and clarify it. The authors thank Massimo Meneghetti, Raphaël Gavazzi, Michaela Hirschmann, and Mireia Montes for their helpful discussions at various stages of the project. The computations were performed at the University of Geneva using Yggdrasil and Boabab HPC services.
BB acknowledges the Swiss National Science Foundation (SNSF) for supporting this work.
ML acknowledges the Centre National de la Recherche Scientifique (CNRS) and the Centre National des Etudes Spatiale (CNES) for support.
MJ is supported by the United Kingdom Research and Innovation (UKRI) Future Leaders Fellowship `Using Cosmic Beasts to uncover the Nature of Dark Matter' (grant number MR/X006069/1).

\section*{Data Availability}
The mass models with a ``BCG - ML 2'' parametrisation are released
publicly with the paper, and are available at the Mikulski Archive for Space Telescopes as High Level Science Products via \href{https://doi.org/10.17909/t9-w6tj-wp63}{https://doi.org/10.17909/t9-w6tj-wp63}.



\bibliographystyle{mnras}
\bibliography{References} 

@ARTICLE{Cappellari2012,
       author = {{Cappellari}, Michele and {McDermid}, Richard M. and {Alatalo}, Katherine and {Blitz}, Leo and {Bois}, Maxime and {Bournaud}, Fr{\'e}d{\'e}ric and {Bureau}, M. and {Crocker}, Alison F. and {Davies}, Roger L. and {Davis}, Timothy A. and {de Zeeuw}, P.~T. and {Duc}, Pierre-Alain and {Emsellem}, Eric and {Khochfar}, Sadegh and {Krajnovi{\'c}}, Davor and {Kuntschner}, Harald and {Lablanche}, Pierre-Yves and {Morganti}, Raffaella and {Naab}, Thorsten and {Oosterloo}, Tom and {Sarzi}, Marc and {Scott}, Nicholas and {Serra}, Paolo and {Weijmans}, Anne-Marie and {Young}, Lisa M.},
        title = "{Systematic variation of the stellar initial mass function in early-type galaxies}",
      journal = {\nat},
         year = 2012,
        month = apr,
       volume = {484},
       number = {7395},
        pages = {485-488},
          doi = {10.1038/nature10972},
archivePrefix = {arXiv},
       eprint = {1202.3308},
 primaryClass = {astro-ph.CO},
       adsurl = {https://ui.adsabs.harvard.edu/abs/2012Natur.484..485C}
}

@ARTICLE{Suyu2010,
       author = {{Suyu}, S.~H. and {Halkola}, A.},
        title = "{The halos of satellite galaxies: the companion of the massive elliptical lens SL2S J08544-0121}",
      journal = {\aap},
         year = 2010,
        month = dec,
       volume = {524},
          eid = {A94},
        pages = {A94},
          doi = {10.1051/0004-6361/201015481},
archivePrefix = {arXiv},
       eprint = {1007.4815},
 primaryClass = {astro-ph.CO},
       adsurl = {https://ui.adsabs.harvard.edu/abs/2010A&A...524A..94S}
}

@ARTICLE{Suyu2012,
       author = {{Suyu}, S.~H. and {Hensel}, S.~W. and {McKean}, J.~P. and {Fassnacht}, C.~D. and {Treu}, T. and {Halkola}, A. and {Norbury}, M. and {Jackson}, N. and {Schneider}, P. and {Thompson}, D. and {Auger}, M.~W. and {Koopmans}, L.~V.~E. and {Matthews}, K.},
        title = "{Disentangling Baryons and Dark Matter in the Spiral Gravitational Lens B1933+503}",
      journal = {\apj},
         year = 2012,
        month = may,
       volume = {750},
       number = {1},
          eid = {10},
        pages = {10},
          doi = {10.1088/0004-637X/750/1/10},
archivePrefix = {arXiv},
       eprint = {1110.2536},
 primaryClass = {astro-ph.CO},
       adsurl = {https://ui.adsabs.harvard.edu/abs/2012ApJ...750...10S}
}

@ARTICLE{Atek2025,
       author = {{Atek}, Hakim and {Chisholm}, John and {Kokorev}, Vasily and {Endsley}, Ryan and {Pan}, Richard and {Furtak}, Lukas and {Chemerynska}, Iryna and {Richard}, Johan and {Claeyssens}, Ad{\'e}la{\"\i}de and {Oesch}, Pascal and {Fujimoto}, Seiji and {Naidu}, Rohan and {Korber}, Damien and {Schaerer}, Daniel and {Blaizot}, Jeremy and {Rosdahl}, Joki and {Adamo}, Angela and {Asada}, Yoshihisa and {Basu}, Arghyadeep and {Beauchesne}, Benjamin and {Berg}, Danielle and {Bezanson}, Rachel and {Bouwens}, Rychard and {Brammer}, Gabriel and {Dessauges-Zavadsky}, Miroslava and {Ellien}, Ama{\"e}l and {Ezziati}, Meriam and {Fei}, Qinyue and {Goovaerts}, Ilias and {Heurtier}, Sylvain and {Hsiao}, Tiger Yu-Yang and {Jecmen}, Michelle and {Khullar}, Gourav and {Kneib}, Jean-Paul and {Labb{\'e}}, Ivo and {Leclercq}, Floriane and {Marques-Chaves}, Rui and {Mason}, Charlotte and {McQuinn}, Kristen B.~W. and {Mu{\~n}oz}, Julian B. and {Natarajan}, Priyamvada and {Saldana-Lopez}, Alberto and {Stephenson}, Mabel G. and {Trebitsch}, Maxime and {Volonteri}, Marta and {Weibel}, Andrea and {Zitrin}, Adi},
        title = "{JWST's GLIMPSE: an overview of the deepest probe of early galaxy formation and cosmic reionization}",
      journal = {arXiv e-prints},
         year = 2025,
        month = nov,
          eid = {arXiv:2511.07542},
        pages = {arXiv:2511.07542},
          doi = {10.48550/arXiv.2511.07542},
archivePrefix = {arXiv},
       eprint = {2511.07542},
 primaryClass = {astro-ph.GA},
       adsurl = {https://ui.adsabs.harvard.edu/abs/2025arXiv251107542A}
}

@ARTICLE{Diego2026,
       author = {{Diego}, J.~M. and {Palencia}, J.~M. and {Goolsby}, C. and {Conselice}, C.~J. and {Lagattuta}, D.~J. and {Mahler}, G. and {Richard}, J. and {Sharon}, K. and {Williams}, L.~L.~R.},
        title = "{Hedorah, the first yellow supergiant Kaiju star candidate at $z\approx3$ revealed by behind AS1063}",
      journal = {arXiv e-prints},
         year = 2026,
        month = jan,
          eid = {arXiv:2601.11704},
        pages = {arXiv:2601.11704},
          doi = {10.48550/arXiv.2601.11704},
archivePrefix = {arXiv},
       eprint = {2601.11704},
 primaryClass = {astro-ph.GA},
       adsurl = {https://ui.adsabs.harvard.edu/abs/2026arXiv260111704D}
}

@ARTICLE{Beauchesne2025a,
       author = {{Beauchesne}, Benjamin and {Cl{\'e}ment}, Benjamin and {Limousin}, Marceau and {Alcalde Pampliega}, Bel{\'e}n and {Jauzac}, Mathilde and {Niemiec}, Anna and {Richard}, Johan and {Mahler}, Guillaume and {Diego}, Jose M. and {Hibon}, Pascale and {Koekemoer}, Anton M. and {Connor}, Thomas and {Kneib}, Jean-Paul and {Faisst}, Andreas L.},
        title = "{A comprehensive separation of dark matter and baryonic mass components in galaxy clusters I: Mass constraints from Abell S1063}",
      journal = {arXiv e-prints},
         year = 2025,
        month = sep,
          eid = {arXiv:2509.07762},
        pages = {arXiv:2509.07762},
          doi = {10.48550/arXiv.2509.07762},
archivePrefix = {arXiv},
       eprint = {2509.07762},
 primaryClass = {astro-ph.GA},
       adsurl = {https://ui.adsabs.harvard.edu/abs/2025arXiv250907762B}
}

@ARTICLE{Balestra2013,
       author = {{Balestra}, I. and {Vanzella}, E. and {Rosati}, P. and {Monna}, A. and {Grillo}, C. and {Nonino}, M. and {Mercurio}, A. and {Biviano}, A. and {Bradley}, L. and {Coe}, D. and {Fritz}, A. and {Postman}, M. and {Seitz}, S. and {Scodeggio}, M. and {Tozzi}, P. and {Zheng}, W. and {Ziegler}, B. and {Zitrin}, A. and {Annunziatella}, M. and {Bartelmann}, M. and {Benitez}, N. and {Broadhurst}, T. and {Bouwens}, R. and {Czoske}, O. and {Donahue}, M. and {Ford}, H. and {Girardi}, M. and {Infante}, L. and {Jouvel}, S. and {Kelson}, D. and {Koekemoer}, A. and {Kuchner}, U. and {Lemze}, D. and {Lombardi}, M. and {Maier}, C. and {Medezinski}, E. and {Melchior}, P. and {Meneghetti}, M. and {Merten}, J. and {Molino}, A. and {Moustakas}, L. and {Presotto}, V. and {Smit}, R. and {Umetsu}, K.},
        title = "{CLASH-VLT: spectroscopic confirmation of a z = 6.11 quintuply lensed galaxy in the Frontier Fields cluster RXC J2248.7-4431}",
      journal = {\aap},
         year = 2013,
        month = nov,
       volume = {559},
          eid = {L9},
        pages = {L9},
          doi = {10.1051/0004-6361/201322620},
archivePrefix = {arXiv},
       eprint = {1309.1593},
 primaryClass = {astro-ph.CO},
       adsurl = {https://ui.adsabs.harvard.edu/abs/2013A&A...559L...9B}
}

@ARTICLE{Cerny2025,
       author = {{Cerny}, Catherine and {Jauzac}, Mathilde and {Lagattuta}, David and {Niemiec}, Anna and {Mahler}, Guillaume and {Edge}, Alastair and {Massey}, Richard},
        title = "{The Kaleidoscope Survey: Strong Gravitational Lensing in Galaxy Clusters with Radial Arcs}",
      journal = {arXiv e-prints},
         year = 2025,
        month = jun,
          eid = {arXiv:2506.21531},
        pages = {arXiv:2506.21531},
          doi = {10.48550/arXiv.2506.21531},
archivePrefix = {arXiv},
       eprint = {2506.21531},
 primaryClass = {astro-ph.CO},
       adsurl = {https://ui.adsabs.harvard.edu/abs/2025arXiv250621531C}
}

@ARTICLE{Chan2022,
       author = {{Chan}, Hei Yin Jowett and {Ferreira}, Elisa G.~M. and {May}, Simon and {Hayashi}, Kohei and {Chiba}, Masashi},
        title = "{The diversity of core-halo structure in the fuzzy dark matter model}",
      journal = {\mnras},
         year = 2022,
        month = mar,
       volume = {511},
       number = {1},
        pages = {943-952},
          doi = {10.1093/mnras/stac063},
archivePrefix = {arXiv},
       eprint = {2110.11882},
 primaryClass = {astro-ph.CO},
       adsurl = {https://ui.adsabs.harvard.edu/abs/2022MNRAS.511..943C}
}

@ARTICLE{Sharma2021,
       author = {{Sharma}, Soniya and {Richard}, Johan and {Yuan}, Tiantian and {Patr{\'\i}cio}, Vera and {Kewley}, Lisa and {Rigby}, Jane R. and {Gupta}, Anshu and {Leethochawalit}, Nicha},
        title = "{Resolving star-forming clumps in a z {\ensuremath{\sim}} 2 lensed galaxy: a pixelated Bayesian approach}",
      journal = {\mnras},
         year = 2021,
        month = jul,
       volume = {505},
       number = {1},
        pages = {L1-L5},
          doi = {10.1093/mnrasl/slab040},
archivePrefix = {arXiv},
       eprint = {2104.07812},
 primaryClass = {astro-ph.GA},
       adsurl = {https://ui.adsabs.harvard.edu/abs/2021MNRAS.505L...1S}
}

@ARTICLE{Caminha2016,
       author = {{Caminha}, G.~B. and {Grillo}, C. and {Rosati}, P. and {Balestra}, I. and {Karman}, W. and {Lombardi}, M. and {Mercurio}, A. and {Nonino}, M. and {Tozzi}, P. and {Zitrin}, A. and {Biviano}, A. and {Girardi}, M. and {Koekemoer}, A.~M. and {Melchior}, P. and {Meneghetti}, M. and {Munari}, E. and {Suyu}, S.~H. and {Umetsu}, K. and {Annunziatella}, M. and {Borgani}, S. and {Broadhurst}, T. and {Caputi}, K.~I. and {Coe}, D. and {Delgado-Correal}, C. and {Ettori}, S. and {Fritz}, A. and {Frye}, B. and {Gobat}, R. and {Maier}, C. and {Monna}, A. and {Postman}, M. and {Sartoris}, B. and {Seitz}, S. and {Vanzella}, E. and {Ziegler}, B.},
        title = "{CLASH-VLT: A highly precise strong lensing model of the galaxy cluster RXC J2248.7-4431 (Abell S1063) and prospects for cosmography}",
      journal = {\aap},
         year = 2016,
        month = mar,
       volume = {587},
          eid = {A80},
        pages = {A80},
          doi = {10.1051/0004-6361/201527670},
archivePrefix = {arXiv},
       eprint = {1512.04555},
 primaryClass = {astro-ph.CO},
       adsurl = {https://ui.adsabs.harvard.edu/abs/2016A&A...587A..80C}
}

@ARTICLE{Diego2016,
       author = {{Diego}, Jose M. and {Broadhurst}, Tom and {Wong}, Jess and {Silk}, Joseph and {Lim}, Jeremy and {Zheng}, Wei and {Lam}, Daniel and {Ford}, Holland},
        title = "{A free-form mass model of the Hubble Frontier Fields cluster AS1063 (RXC J2248.7-4431) with over one hundred constraints}",
      journal = {\mnras},
         year = 2016,
        month = jul,
       volume = {459},
       number = {4},
        pages = {3447-3459},
          doi = {10.1093/mnras/stw865},
archivePrefix = {arXiv},
       eprint = {1512.07916},
 primaryClass = {astro-ph.CO},
       adsurl = {https://ui.adsabs.harvard.edu/abs/2016MNRAS.459.3447D}
}

@ARTICLE{Monna2014,
       author = {{Monna}, A. and {Seitz}, S. and {Greisel}, N. and {Eichner}, T. and {Drory}, N. and {Postman}, M. and {Zitrin}, A. and {Coe}, D. and {Halkola}, A. and {Suyu}, S.~H. and {Grillo}, C. and {Rosati}, P. and {Lemze}, D. and {Balestra}, I. and {Snigula}, J. and {Bradley}, L. and {Umetsu}, K. and {Koekemoer}, A. and {Kuchner}, U. and {Moustakas}, L. and {Bartelmann}, M. and {Ben{\'\i}tez}, N. and {Bouwens}, R. and {Broadhurst}, T. and {Donahue}, M. and {Ford}, H. and {Host}, O. and {Infante}, L. and {Jimenez-Teja}, Y. and {Jouvel}, S. and {Kelson}, D. and {Lahav}, O. and {Medezinski}, E. and {Melchior}, P. and {Meneghetti}, M. and {Merten}, J. and {Molino}, A. and {Moustakas}, J. and {Nonino}, M. and {Zheng}, W.},
        title = "{CLASH: z {\ensuremath{\sim}} 6 young galaxy candidate quintuply lensed by the frontier field cluster RXC J2248.7-4431}",
      journal = {\mnras},
         year = 2014,
        month = feb,
       volume = {438},
       number = {2},
        pages = {1417-1434},
          doi = {10.1093/mnras/stt2284},
archivePrefix = {arXiv},
       eprint = {1308.6280},
 primaryClass = {astro-ph.CO},
       adsurl = {https://ui.adsabs.harvard.edu/abs/2014MNRAS.438.1417M}
}

@ARTICLE{Johnson2014,
       author = {{Johnson}, Traci L. and {Sharon}, Keren and {Bayliss}, Matthew B. and {Gladders}, Michael D. and {Coe}, Dan and {Ebeling}, Harald},
        title = "{Lens Models and Magnification Maps of the Six Hubble Frontier Fields Clusters}",
      journal = {\apj},
         year = 2014,
        month = dec,
       volume = {797},
       number = {1},
          eid = {48},
        pages = {48},
          doi = {10.1088/0004-637X/797/1/48},
archivePrefix = {arXiv},
       eprint = {1405.0222},
 primaryClass = {astro-ph.CO},
       adsurl = {https://ui.adsabs.harvard.edu/abs/2014ApJ...797...48J}
}

@ARTICLE{Richard2014,
       author = {{Richard}, Johan and {Jauzac}, Mathilde and {Limousin}, Marceau and {Jullo}, Eric and {Cl{\'e}ment}, Benjamin and {Ebeling}, Harald and {Kneib}, Jean-Paul and {Atek}, Hakim and {Natarajan}, Priya and {Egami}, Eiichi and {Livermore}, Rachael and {Bower}, Richard},
        title = "{Mass and magnification maps for the Hubble Space Telescope Frontier Fields clusters: implications for high-redshift studies}",
      journal = {\mnras},
         year = 2014,
        month = oct,
       volume = {444},
       number = {1},
        pages = {268-289},
          doi = {10.1093/mnras/stu1395},
archivePrefix = {arXiv},
       eprint = {1405.3303},
 primaryClass = {astro-ph.CO},
       adsurl = {https://ui.adsabs.harvard.edu/abs/2014MNRAS.444..268R}
}

@ARTICLE{Sirks2022,
       author = {{Sirks}, Ellen L. and {Oman}, Kyle A. and {Robertson}, Andrew and {Massey}, Richard and {Frenk}, Carlos},
        title = "{The effects of self-interacting dark matter on the stripping of galaxies that fall into clusters}",
      journal = {\mnras},
         year = 2022,
        month = apr,
       volume = {511},
       number = {4},
        pages = {5927-5935},
          doi = {10.1093/mnras/stac406},
archivePrefix = {arXiv},
       eprint = {2109.03257},
 primaryClass = {astro-ph.CO},
       adsurl = {https://ui.adsabs.harvard.edu/abs/2022MNRAS.511.5927S}
}

@ARTICLE{Bhattacharyya2022,
       author = {{Bhattacharyya}, Joy and {Adhikari}, Susmita and {Banerjee}, Arka and {More}, Surhud and {Kumar}, Amit and {Nadler}, Ethan O. and {Chatterjee}, Suchetana},
        title = "{The Signatures of Self-interacting Dark Matter and Subhalo Disruption on Cluster Substructure}",
      journal = {\apj},
         year = 2022,
        month = jun,
       volume = {932},
       number = {1},
          eid = {30},
        pages = {30},
          doi = {10.3847/1538-4357/ac68e9},
archivePrefix = {arXiv},
       eprint = {2106.08292},
 primaryClass = {astro-ph.CO},
       adsurl = {https://ui.adsabs.harvard.edu/abs/2022ApJ...932...30B}
}

@ARTICLE{niemiec2019,
       author = {{Niemiec}, Anna and {Jullo}, Eric and {Giocoli}, Carlo and {Limousin}, Marceau and {Jauzac}, Mathilde},
        title = "{Dark matter stripping in galaxy clusters: a look at the stellar-to-halo mass relation in the Illustris simulation}",
      journal = {\mnras},
         year = 2019,
        month = jul,
       volume = {487},
       number = {1},
        pages = {653-666},
          doi = {10.1093/mnras/stz1318},
archivePrefix = {arXiv},
       eprint = {1811.04996},
 primaryClass = {astro-ph.GA},
       adsurl = {https://ui.adsabs.harvard.edu/abs/2019MNRAS.487..653N}
}

@ARTICLE{Mehrgan2024,
       author = {{Mehrgan}, Kianusch and {Thomas}, Jens and {Saglia}, Roberto and {Parikh}, Taniya and {Neureiter}, Bianca and {Erwin}, Peter and {Bender}, Ralf},
        title = "{Dynamical Stellar Mass-to-light Ratio Gradients: Evidence for Very Centrally Concentrated IMF Variations in ETGs?}",
      journal = {\apj},
         year = 2024,
        month = jan,
       volume = {961},
       number = {1},
          eid = {127},
        pages = {127},
          doi = {10.3847/1538-4357/acfe09},
archivePrefix = {arXiv},
       eprint = {2309.15911},
 primaryClass = {astro-ph.GA},
       adsurl = {https://ui.adsabs.harvard.edu/abs/2024ApJ...961..127M}
}

@article{Kumar2019, doi = {10.21105/joss.01143}, url = {https://doi.org/10.21105/joss.01143}, year = {2019}, publisher = {The Open Journal}, volume = {4}, number = {33}, pages = {1143}, author = {Ravin Kumar and Colin Carroll and Ari Hartikainen and Osvaldo Martin}, title = {ArviZ a unified library for exploratory analysis of Bayesian models in Python}, journal = {Journal of Open Source Software} }

@ARTICLE{Spergel2000,
       author = {{Spergel}, David N. and {Steinhardt}, Paul J.},
        title = "{Observational Evidence for Self-Interacting Cold Dark Matter}",
      journal = {\prl},
         year = 2000,
        month = apr,
       volume = {84},
       number = {17},
        pages = {3760-3763},
          doi = {10.1103/PhysRevLett.84.3760},
archivePrefix = {arXiv},
       eprint = {astro-ph/9909386},
 primaryClass = {astro-ph},
       adsurl = {https://ui.adsabs.harvard.edu/abs/2000PhRvL..84.3760S}
}

@ARTICLE{Lu2024,
       author = {{Lu}, Shengdong and {Zhu}, Kai and {Cappellari}, Michele and {Li}, Ran and {Mao}, Shude and {Xu}, Dandan},
        title = "{MaNGA DynPop - V. The dark-matter fraction versus stellar velocity dispersion relation and stellar initial mass function variations in galaxies: dynamical models and full spectrum fitting of integral-field spectroscopy}",
      journal = {\mnras},
         year = 2024,
        month = jun,
       volume = {530},
       number = {4},
        pages = {4474-4492},
          doi = {10.1093/mnras/stae1116},
archivePrefix = {arXiv},
       eprint = {2309.12395},
 primaryClass = {astro-ph.GA},
       adsurl = {https://ui.adsabs.harvard.edu/abs/2024MNRAS.530.4474L}
}

@article{Spearman1904,
 ISSN = {00029556},
 URL = {http://www.jstor.org/stable/1412159},
 author = {C. Spearman},
 journal = {The American Journal of Psychology},
 number = {1},
 pages = {72--101},
 publisher = {University of Illinois Press},
 title = {The Proof and Measurement of Association between Two Things},
 urldate = {2024-12-20},
 volume = {15},
 year = {1904}
}

@string{june = {June}}

@ARTICLE{Dutton2011,
       author = {{Dutton}, Aaron A. and {Brewer}, Brendon J. and {Marshall}, Philip J. and {Auger}, Matthew W. and {Treu}, Tommaso and {Koo}, David C. and {Bolton}, Adam S. and {Holden}, Bradford P. and {Koopmans}, Leon V.~E.},
        title = "{The SWELLS survey - II. Breaking the disc-halo degeneracy in the spiral galaxy gravitational lens SDSS J2141-0001}",
      journal = {\mnras},
         year = 2011,
        month = nov,
       volume = {417},
       number = {3},
        pages = {1621-1642},
          doi = {10.1111/j.1365-2966.2011.18706.x},
archivePrefix = {arXiv},
       eprint = {1101.1622},
 primaryClass = {astro-ph.CO},
       adsurl = {https://ui.adsabs.harvard.edu/abs/2011MNRAS.417.1621D}
}

@ARTICLE{Buchner2014,
       author = {{Buchner}, J. and {Georgakakis}, A. and {Nandra}, K. and {Hsu}, L. and {Rangel}, C. and {Brightman}, M. and {Merloni}, A. and {Salvato}, M. and {Donley}, J. and {Kocevski}, D.},
        title = "{X-ray spectral modelling of the AGN obscuring region in the CDFS: Bayesian model selection and catalogue}",
      journal = {\aap},
         year = 2014,
        month = apr,
       volume = {564},
          eid = {A125},
        pages = {A125},
          doi = {10.1051/0004-6361/201322971},
archivePrefix = {arXiv},
       eprint = {1402.0004},
 primaryClass = {astro-ph.HE},
       adsurl = {https://ui.adsabs.harvard.edu/abs/2014A&A...564A.125B}
}

@article{scikit-learn,
  title={Scikit-learn: Machine Learning in {P}ython},
  author={Pedregosa, F. and Varoquaux, G. and Gramfort, A. and Michel, V.
          and Thirion, B. and Grisel, O. and Blondel, M. and Prettenhofer, P.
          and Weiss, R. and Dubourg, V. and Vanderplas, J. and Passos, A. and
          Cournapeau, D. and Brucher, M. and Perrot, M. and Duchesnay, E.},
  journal={Journal of Machine Learning Research},
  volume={12},
  pages={2825--2830},
  year={2011}
}

@ARTICLE{sifon2018,
       author = {{Sif{\'o}n}, Crist{\'o}bal and {Herbonnet}, Ricardo and {Hoekstra}, Henk and {van der Burg}, Remco F.~J. and {Viola}, Massimo},
        title = "{The galaxy-subhalo connection in low-redshift galaxy clusters from weak gravitational lensing}",
      journal = {\mnras},
         year = 2018,
        month = jul,
       volume = {478},
       number = {1},
        pages = {1244-1264},
          doi = {10.1093/mnras/sty1161},
archivePrefix = {arXiv},
       eprint = {1706.06125},
 primaryClass = {astro-ph.GA},
       adsurl = {https://ui.adsabs.harvard.edu/abs/2018MNRAS.478.1244S}
}

@ARTICLE{sifon2015,
       author = {{Sif{\'o}n}, Crist{\'o}bal and {Cacciato}, Marcello and {Hoekstra}, Henk and {Brouwer}, Margot and {van Uitert}, Edo and {Viola}, Massimo and {Baldry}, Ivan and {Brough}, Sarah and {Brown}, Michael J.~I. and {Choi}, Ami and {Driver}, Simon P. and {Erben}, Thomas and {Grado}, Aniello and {Heymans}, Catherine and {Hildebrandt}, Hendrik and {Joachimi}, Benjamin and {de Jong}, Jelte T.~A. and {Kuijken}, Konrad and {McFarland}, John and {Miller}, Lance and {Nakajima}, Reiko and {Napolitano}, Nicola and {Norberg}, Peder and {Robotham}, Aaron S.~G. and {Schneider}, Peter and {Verdoes Kleijn}, Gijs},
        title = "{The masses of satellites in GAMA galaxy groups from 100 square degrees of KiDS weak lensing data}",
      journal = {\mnras},
         year = 2015,
        month = dec,
       volume = {454},
       number = {4},
        pages = {3938-3951},
          doi = {10.1093/mnras/stv2051},
archivePrefix = {arXiv},
       eprint = {1507.00737},
 primaryClass = {astro-ph.CO},
       adsurl = {https://ui.adsabs.harvard.edu/abs/2015MNRAS.454.3938S}
}

@ARTICLE{niemiec2017,
       author = {{Niemiec}, Anna and {Jullo}, Eric and {Limousin}, Marceau and {Giocoli}, Carlo and {Erben}, Thomas and {Hildebrant}, Hendrik and {Kneib}, Jean-Paul and {Leauthaud}, Alexie and {Makler}, Martin and {Moraes}, Bruno and {Pereira}, Maria E.~S. and {Shan}, Huanyuan and {Rozo}, Eduardo and {Rykoff}, Eli and {Van Waerbeke}, Ludovic},
        title = "{Stellar-to-halo mass relation of cluster galaxies}",
      journal = {\mnras},
         year = 2017,
        month = oct,
       volume = {471},
       number = {1},
        pages = {1153-1166},
          doi = {10.1093/mnras/stx1667},
archivePrefix = {arXiv},
       eprint = {1703.03348},
 primaryClass = {astro-ph.CO},
       adsurl = {https://ui.adsabs.harvard.edu/abs/2017MNRAS.471.1153N}
}

@ARTICLE{Li2016,
       author = {{Li}, Ran and {Shan}, Huanyuan and {Kneib}, Jean-Paul and {Mo}, Houjun and {Rozo}, Eduardo and {Leauthaud}, Alexie and {Moustakas}, John and {Xie}, Lizhi and {Erben}, Thomas and {Van Waerbeke}, Ludovic and {Makler}, Martin and {Rykoff}, Eli and {Moraes}, Bruno},
        title = "{Measuring subhalo mass in redMaPPer clusters with CFHT Stripe 82 Survey}",
      journal = {\mnras},
         year = 2016,
        month = may,
       volume = {458},
       number = {3},
        pages = {2573-2583},
          doi = {10.1093/mnras/stw494},
archivePrefix = {arXiv},
       eprint = {1507.01464},
 primaryClass = {astro-ph.CO},
       adsurl = {https://ui.adsabs.harvard.edu/abs/2016MNRAS.458.2573L}
}

@ARTICLE{Wang2024,
       author = {{Wang}, Chunxiang and {Li}, Ran and {Shan}, Huanyuan and {Xu}, Weiwei and {Yao}, Ji and {Jing}, Yingjie and {Gao}, Liang and {Li}, Nan and {Xie}, Yushan and {Zhu}, Kai and {Yang}, Hang and {Chen}, Qingze},
        title = "{Assessing mass-loss and stellar-to-halo mass ratio of satellite galaxies: a galaxy-galaxy lensing approach utilizing DECaLS DR8 data}",
      journal = {\mnras},
         year = 2024,
        month = feb,
       volume = {528},
       number = {2},
        pages = {2728-2741},
          doi = {10.1093/mnras/stae121},
archivePrefix = {arXiv},
       eprint = {2305.13694},
 primaryClass = {astro-ph.GA},
       adsurl = {https://ui.adsabs.harvard.edu/abs/2024MNRAS.528.2728W}
}

@article{Acebron2024,
 adsurl = {https://ui.adsabs.harvard.edu/abs/2024arXiv241001883A},
 archiveprefix = {arXiv},
 author = {Ana {Acebron} and Claudio {Grillo} and Sherry H. {Suyu} and Giuseppe {Angora} and Pietro {Bergamini} and Gabriel B. {Caminha} and Sebastian {Ertl} and Amata {Mercurio} and Mario {Nonino} and Piero {Rosati} and Han {Wang} and Andrea {Bolamperti} and Massimo {Meneghetti} and Stefan {Schuldt} and Eros {Vanzella}},
 eid = {arXiv:2410.01883},
 eprint = {2410.01883},
 journal = {arXiv e-prints},
 month = {October},
 pages = {arXiv:2410.01883},
 primaryclass = {astro-ph.GA},
 title = {{The next step in galaxy cluster strong lensing: modeling the surface brightness of multiply-imaged sources}},
 year = {2024}
}

@article{Bonyadi2017,
    author = {Bonyadi, Mohammad Reza and Michalewicz, Zbigniew},
    title = "{Particle Swarm Optimization for Single Objective Continuous Space Problems: A Review}",
    journal = {Evolutionary Computation},
    volume = {25},
    number = {1},
    pages = {1-54},
    year = {2017},
    month = {03},
    issn = {1063-6560},
    doi = {10.1162/EVCO_r_00180},
    url = {https://doi.org/10.1162/EVCO\_r\_00180},
    eprint = {https://direct.mit.edu/evco/article-pdf/25/1/1/1535702/evco\_r\_00180.pdf},
}

@article{Hol2006,
  title={On Resampling Algorithms for Particle Filters},
  author={Jeroen D. Hol and Thomas B. Schon and Fredrik K. Gustafsson},
  journal={2006 IEEE Nonlinear Statistical Signal Processing Workshop},
  year={2006},
  pages={79-82},
  url={https://api.semanticscholar.org/CorpusID:215951528}
}

@inbook{Akaike1973,
 address = {New York, NY},
 author = {Akaike, Hirotogu},
 booktitle = {Selected Papers of Hirotugu Akaike},
 doi = {10.1007/978-1-4612-1694-0_15},
 editor = {Parzen, Emanuel
and Tanabe, Kunio
and Kitagawa, Genshiro},
 isbn = {978-1-4612-1694-0},
 pages = {199--213},
 publisher = {Springer New York},
 title = {Information Theory and an Extension of the Maximum Likelihood Principle},
 url = {https://doi.org/10.1007/978-1-4612-1694-0_15},
 year = {1998}
}

@article{Beauchesne2021,
 adsurl = {https://ui.adsabs.harvard.edu/abs/2021MNRAS.506.2002B},
 archiveprefix = {arXiv},
 author = {{Beauchesne}, Benjamin and {Cl{\'e}ment}, Benjamin and {Richard}, Johan and {Kneib}, Jean-Paul},
 doi = {10.1093/mnras/stab1684},
 eprint = {2106.05029},
 journal = {\mnras},
 month = {September},
 number = {2},
 pages = {2002-2019},
 primaryclass = {astro-ph.CO},
 title = {{Improving parametric mass modelling of lensing clusters through a perturbative approach}},
 volume = {506},
 year = {2021}
}

@article{Beauchesne2024,
 adsurl = {https://ui.adsabs.harvard.edu/abs/2024MNRAS.527.3246B},
 archiveprefix = {arXiv},
 author = {{Beauchesne}, Benjamin and {Cl{\'e}ment}, Benjamin and {Hibon}, Pascale and {Limousin}, Marceau and {Eckert}, Dominique and {Kneib}, Jean-Paul and {Richard}, Johan and {Natarajan}, Priyamvada and {Jauzac}, Mathilde and {Montes}, Mireia and {Mahler}, Guillaume and {Claeyssens}, Ad{\'e}la{\"\i}de and {Jeanneau}, Alexandre and {Koekemoer}, Anton M. and {Lagattuta}, David and {Pagul}, Amanda and {S{\'a}nchez}, Javier},
 doi = {10.1093/mnras/stad3308},
 eprint = {2301.10907},
 journal = {\mnras},
 month = {January},
 number = {2},
 pages = {3246-3275},
 primaryclass = {astro-ph.CO},
 title = {{A new step forward in realistic cluster lens mass modelling: analysis of Hubble Frontier Field Cluster Abell S1063 from joint lensing, X-ray, and galaxy kinematics data}},
 volume = {527},
 year = {2024}
}

@article{Bergamini2019,
 adsurl = {https://ui.adsabs.harvard.edu/abs/2019A&A...631A.130B},
 archiveprefix = {arXiv},
 author = {{Bergamini}, P. and {Rosati}, P. and {Mercurio}, A. and {Grillo}, C. and {Caminha}, G.~B. and {Meneghetti}, M. and {Agnello}, A. and {Biviano}, A. and {Calura}, F. and {Giocoli}, C. and {Lombardi}, M. and {Rodighiero}, G. and {Vanzella}, E.},
 doi = {10.1051/0004-6361/201935974},
 eid = {A130},
 eprint = {1905.13236},
 journal = {\aap},
 month = {November},
 pages = {A130},
 primaryclass = {astro-ph.GA},
 title = {{Enhanced cluster lensing models with measured galaxy kinematics}},
 volume = {631},
 year = {2019}
}

@article{Bonamigo2017,
 adsurl = {https://ui.adsabs.harvard.edu/abs/2017ApJ...842..132B},
 archiveprefix = {arXiv},
 author = {{Bonamigo}, M. and {Grillo}, C. and {Ettori}, S. and {Caminha}, G.~B. and {Rosati}, P. and {Mercurio}, A. and {Annunziatella}, M. and {Balestra}, I. and {Lombardi}, M.},
 doi = {10.3847/1538-4357/aa75cc},
 eid = {132},
 eprint = {1705.10322},
 journal = {\apj},
 month = {June},
 number = {2},
 pages = {132},
 primaryclass = {astro-ph.GA},
 title = {{Joining X-Ray to Lensing: An Accurate Combined Analysis of MACS J0416.1-2403}},
 volume = {842},
 year = {2017}
}

@article{Bonamigo2018,
 adsurl = {https://ui.adsabs.harvard.edu/abs/2018ApJ...864...98B},
 archiveprefix = {arXiv},
 author = {{Bonamigo}, M. and {Grillo}, C. and {Ettori}, S. and {Caminha}, G.~B. and {Rosati}, P. and {Mercurio}, A. and {Munari}, E. and {Annunziatella}, M. and {Balestra}, I. and {Lombardi}, M.},
 doi = {10.3847/1538-4357/aad4a7},
 eid = {98},
 eprint = {1807.10286},
 journal = {\apj},
 month = {September},
 number = {1},
 pages = {98},
 primaryclass = {astro-ph.GA},
 title = {{Dissection of the Collisional and Collisionless Mass Components in a Mini Sample of CLASH and HFF Massive Galaxy Clusters at z {\ensuremath{\approx}} 0.4}},
 volume = {864},
 year = {2018}
}

@article{Bovy2015,
 adsurl = {https://ui.adsabs.harvard.edu/abs/2015ApJS..216...29B},
 archiveprefix = {arXiv},
 author = {{Bovy}, Jo},
 doi = {10.1088/0067-0049/216/2/29},
 eid = {29},
 eprint = {1412.3451},
 journal = {\apjs},
 month = {February},
 number = {2},
 pages = {29},
 primaryclass = {astro-ph.GA},
 title = {{galpy: A python Library for Galactic Dynamics}},
 volume = {216},
 year = {2015}
}

@article{Buchner2019,
 adsurl = {https://ui.adsabs.harvard.edu/abs/2019PASP..131j8005B},
 archiveprefix = {arXiv},
 author = {{Buchner}, Johannes},
 doi = {10.1088/1538-3873/aae7fc},
 eprint = {1707.04476},
 journal = {\pasp},
 month = {October},
 number = {1004},
 pages = {108005},
 primaryclass = {stat.CO},
 title = {{Collaborative Nested Sampling: Big Data versus Complex Physical Models}},
 volume = {131},
 year = {2019}
}

@article{Buchner2021,
 adsurl = {https://ui.adsabs.harvard.edu/abs/2021JOSS....6.3001B},
 archiveprefix = {arXiv},
 author = {{Buchner}, Johannes},
 doi = {10.21105/joss.03001},
 eid = {3001},
 eprint = {2101.09604},
 journal = {The Journal of Open Source Software},
 month = {April},
 number = {60},
 pages = {3001},
 primaryclass = {stat.CO},
 title = {{UltraNest - a robust, general purpose Bayesian inference engine}},
 volume = {6},
 year = {2021}
}

@article{Cappellari2008,
 adsurl = {https://ui.adsabs.harvard.edu/abs/2008MNRAS.390...71C},
 archiveprefix = {arXiv},
 author = {{Cappellari}, Michele},
 doi = {10.1111/j.1365-2966.2008.13754.x},
 eprint = {0806.0042},
 journal = {\mnras},
 month = {October},
 number = {1},
 pages = {71-86},
 primaryclass = {astro-ph},
 title = {{Measuring the inclination and mass-to-light ratio of axisymmetric galaxies via anisotropic Jeans models of stellar kinematics}},
 volume = {390},
 year = {2008}
}

@ARTICLE{Gavazzi2005,
       author = {{Gavazzi}, R.},
        title = "{Projection effects in cluster mass estimates: the case of MS2137-23}",
      journal = {\aap},
         year = 2005,
        month = dec,
       volume = {443},
       number = {3},
        pages = {793-804},
          doi = {10.1051/0004-6361:20053166},
archivePrefix = {arXiv},
       eprint = {astro-ph/0503696},
 primaryClass = {astro-ph},
       adsurl = {https://ui.adsabs.harvard.edu/abs/2005A&A...443..793G}
}

@article{Cappellari2020,
 adsurl = {https://ui.adsabs.harvard.edu/abs/2020MNRAS.494.4819C},
 archiveprefix = {arXiv},
 author = {{Cappellari}, Michele},
 doi = {10.1093/mnras/staa959},
 eprint = {1907.09894},
 journal = {\mnras},
 month = {June},
 number = {4},
 pages = {4819-4837},
 primaryclass = {astro-ph.GA},
 title = {{Efficient solution of the anisotropic spherically aligned axisymmetric Jeans equations of stellar hydrodynamics for galactic dynamics}},
 volume = {494},
 year = {2020}
}

@article{Collett2018,
 adsurl = {https://ui.adsabs.harvard.edu/abs/2018Sci...360.1342C},
 archiveprefix = {arXiv},
 author = {{Collett}, Thomas E. and {Oldham}, Lindsay J. and {Smith}, Russell J. and {Auger}, Matthew W. and {Westfall}, Kyle B. and {Bacon}, David and {Nichol}, Robert C. and {Masters}, Karen L. and {Koyama}, Kazuya and {van den Bosch}, Remco},
 doi = {10.1126/science.aao2469},
 eprint = {1806.08300},
 journal = {Science},
 month = {June},
 number = {6395},
 pages = {1342-1346},
 primaryclass = {astro-ph.CO},
 title = {{A precise extragalactic test of General Relativity}},
 volume = {360},
 year = {2018}
}

@article{diego2005,
 adsurl = {https://ui.adsabs.harvard.edu/abs/2005MNRAS.360..477D},
 archiveprefix = {arXiv},
 author = {{Diego}, J.~M. and {Protopapas}, P. and {Sandvik}, H.~B. and
{Tegmark}, M.},
 doi = {10.1111/j.1365-2966.2005.09021.x},
 eprint = {astro-ph/0408418},
 journal = {\mnras},
 month = {June},
 number = {2},
 pages = {477-491},
 primaryclass = {astro-ph},
 title = {{Non-parametric inversion of strong lensing systems}},
 volume = {360},
 year = {2005}
}

@article{diego2007,
 adsurl = {https://ui.adsabs.harvard.edu/abs/2007MNRAS.375..958D},
 archiveprefix = {arXiv},
 author = {{Diego}, J.~M. and {Tegmark}, M. and {Protopapas}, P. and {Sand
vik}, H.~B.},
 doi = {10.1111/j.1365-2966.2007.11380.x},
 eprint = {astro-ph/0509103},
 journal = {\mnras},
 month = {March},
 number = {3},
 pages = {958-970},
 primaryclass = {astro-ph},
 title = {{Combined reconstruction of weak and strong lensing data with WSLAP}},
 volume = {375},
 year = {2007}
}

@article{Eliasdottir2007,
 adsurl = {https://ui.adsabs.harvard.edu/abs/2007arXiv0710.5636E},
 archiveprefix = {arXiv},
 author = {{El{\'\i}asd{\'o}ttir}, {\'A}rd{\'\i}s and {Limousin}, Marceau and {Richard}, Johan and {Hjorth}, Jens and {Kneib}, Jean-Paul and {Natarajan}, Priya and {Pedersen}, Kristian and {Jullo}, Eric and {Paraficz}, Danuta},
 doi = {10.48550/arXiv.0710.5636},
 eid = {arXiv:0710.5636},
 eprint = {0710.5636},
 journal = {arXiv e-prints},
 month = {October},
 pages = {arXiv:0710.5636},
 primaryclass = {astro-ph},
 title = {{Where is the matter in the Merging Cluster Abell 2218?}},
 year = {2007}
}

@article{Faber&Jackson1976,
 adsurl = {https://ui.adsabs.harvard.edu/abs/1976ApJ...204..668F},
 author = {{Faber}, S.~M. and {Jackson}, R.~E.},
 doi = {10.1086/154215},
 journal = {\apj},
 month = {March},
 pages = {668-683},
 title = {{Velocity dispersions and mass-to-light ratios for elliptical galaxies.}},
 volume = {204},
 year = {1976}
}

@article{Granata2022,
 adsurl = {https://ui.adsabs.harvard.edu/abs/2022A&A...659A..24G},
 archiveprefix = {arXiv},
 author = {{Granata}, G. and {Mercurio}, A. and {Grillo}, C. and {Tortorelli}, L. and {Bergamini}, P. and {Meneghetti}, M. and {Rosati}, P. and {Caminha}, G.~B. and {Nonino}, M.},
 doi = {10.1051/0004-6361/202141817},
 eid = {A24},
 eprint = {2107.09079},
 journal = {\aap},
 month = {March},
 pages = {A24},
 primaryclass = {astro-ph.GA},
 title = {{Improved strong lensing modelling of galaxy clusters using the Fundamental Plane: Detailed mapping of the baryonic and dark matter mass distribution of Abell S1063}},
 volume = {659},
 year = {2022}
}

@article{Hu2000,
 adsurl = {https://ui.adsabs.harvard.edu/abs/2000PhRvL..85.1158H},
 archiveprefix = {arXiv},
 author = {{Hu}, Wayne and {Barkana}, Rennan and {Gruzinov}, Andrei},
 doi = {10.1103/PhysRevLett.85.1158},
 eprint = {astro-ph/0003365},
 journal = {\prl},
 month = {August},
 number = {6},
 pages = {1158-1161},
 primaryclass = {astro-ph},
 title = {{Fuzzy Cold Dark Matter: The Wave Properties of Ultralight Particles}},
 volume = {85},
 year = {2000}
}

@article{Hyde2009,
 adsurl = {https://ui.adsabs.harvard.edu/abs/2009MNRAS.396.1171H},
 archiveprefix = {arXiv},
 author = {{Hyde}, Joseph B. and {Bernardi}, Mariangela},
 doi = {10.1111/j.1365-2966.2009.14783.x},
 eprint = {0810.4924},
 journal = {\mnras},
 month = {June},
 number = {2},
 pages = {1171-1185},
 primaryclass = {astro-ph},
 title = {{The luminosity and stellar mass Fundamental Plane of early-type galaxies}},
 volume = {396},
 year = {2009}
}

@article{jullo2007,
 adsurl = {https://ui.adsabs.harvard.edu/abs/2007NJPh....9..447J},
 archiveprefix = {arXiv},
 author = {{Jullo}, E. and {Kneib}, J. -P. and {Limousin}, M. and
{El{\'\i}asd{\'o}ttir}, {\'A}. and {Marshall}, P.~J. and {Verdugo}, T.},
 doi = {10.1088/1367-2630/9/12/447},
 eprint = {0706.0048},
 journal = {New Journal of Physics},
 month = {December},
 number = {12},
 pages = {447},
 primaryclass = {astro-ph},
 title = {{A Bayesian approach to strong lensing modelling of galaxy clusters}},
 volume = {9},
 year = {2007}
}

@article{Kneib1993,
 adsurl = {https://ui.adsabs.harvard.edu/abs/1993A&A...273..367K},
 author = {{Kneib}, J. -P. and {Mellier}, Y. and {Fort}, B. and {Mathez}, G.},
 journal = {\aap},
 month = {June},
 pages = {367},
 title = {{The distribution of dark matter in distant cluster-lenses : modelling modelling A 370.}},
 volume = {273},
 year = {1993}
}

@article{Limousin2022,
 adsurl = {https://ui.adsabs.harvard.edu/abs/2022A&A...664A..90L},
 archiveprefix = {arXiv},
 author = {{Limousin}, Marceau and {Beauchesne}, Benjamin and {Jullo}, Eric},
 doi = {10.1051/0004-6361/202243278},
 eid = {A90},
 eprint = {2202.02992},
 journal = {\aap},
 month = {August},
 pages = {A90},
 primaryclass = {astro-ph.CO},
 title = {{Dark matter in galaxy clusters: Parametric strong-lensing approach}},
 volume = {664},
 year = {2022}
}

@article{lotz2017,
 adsurl = {https://ui.adsabs.harvard.edu/abs/2017ApJ...837...97L},
 archiveprefix = {arXiv},
 author = {{Lotz}, J.~M. and {Koekemoer}, A. and {Coe}, D. and {Grogin}, N. and
{Capak}, P. and {Mack}, J. and {Anderson}, J. and {Avila}, R. and
{Barker}, E.~A. and {Borncamp}, D. and {Brammer}, G. and {Durbin}, M. and
{Gunning}, H. and {Hilbert}, B. and {Jenkner}, H. and {Khandrika}, H. and
{Levay}, Z. and {Lucas}, R.~A. and {MacKenty}, J. and {Ogaz}, S. and
{Porterfield}, B. and {Reid}, N. and {Robberto}, M. and {Royle}, P. and
{Smith}, L.~J. and {Storrie-Lombardi}, L.~J. and {Sunnquist}, B. and
{Surace}, J. and {Taylor}, D.~C. and {Williams}, R. and {Bullock}, J. and
{Dickinson}, M. and {Finkelstein}, S. and {Natarajan}, P. and
{Richard}, J. and {Robertson}, B. and {Tumlinson}, J. and {Zitrin}, A. and
{Flanagan}, K. and {Sembach}, K. and {Soifer}, B.~T. and {Mountain}, M.},
 doi = {10.3847/1538-4357/837/1/97},
 eid = {97},
 eprint = {1605.06567},
 journal = {\apj},
 month = {March},
 number = {1},
 pages = {97},
 primaryclass = {astro-ph.GA},
 title = {{The Frontier Fields: Survey Design and Initial Results}},
 volume = {837},
 year = {2017}
}

@article{Mahler2023,
 adsurl = {https://ui.adsabs.harvard.edu/abs/2023ApJ...945...49M},
 archiveprefix = {arXiv},
 author = {{Mahler}, Guillaume and {Jauzac}, Mathilde and {Richard}, Johan and {Beauchesne}, Benjamin and {Ebeling}, Harald and {Lagattuta}, David and {Natarajan}, Priyamvada and {Sharon}, Keren and {Atek}, Hakim and {Claeyssens}, Ad{\'e}la{\"\i}de and {Cl{\'e}ment}, Benjamin and {Eckert}, Dominique and {Edge}, Alastair and {Kneib}, Jean-Paul and {Niemiec}, Anna},
 doi = {10.3847/1538-4357/acaea9},
 eid = {49},
 eprint = {2207.07101},
 journal = {\apj},
 month = {March},
 number = {1},
 pages = {49},
 primaryclass = {astro-ph.GA},
 title = {{Precision Modeling of JWST's First Cluster Lens SMACS J0723.3-7327}},
 volume = {945},
 year = {2023}
}

@article{Mercurio2021,
 adsurl = {https://ui.adsabs.harvard.edu/abs/2021A&A...656A.147M},
 archiveprefix = {arXiv},
 author = {{Mercurio}, A. and {Rosati}, P. and {Biviano}, A. and {Annunziatella}, M. and {Girardi}, M. and {Sartoris}, B. and {Nonino}, M. and {Brescia}, M. and {Riccio}, G. and {Grillo}, C. and {Balestra}, I. and {Caminha}, G.~B. and {De Lucia}, G. and {Gobat}, R. and {Seitz}, S. and {Tozzi}, P. and {Scodeggio}, M. and {Vanzella}, E. and {Angora}, G. and {Bergamini}, P. and {Borgani}, S. and {Demarco}, R. and {Meneghetti}, M. and {Strazzullo}, V. and {Tortorelli}, L. and {Umetsu}, K. and {Fritz}, A. and {Gruen}, D. and {Kelson}, D. and {Lombardi}, M. and {Maier}, C. and {Postman}, M. and {Rodighiero}, G. and {Ziegler}, B.},
 doi = {10.1051/0004-6361/202142168},
 eid = {A147},
 eprint = {2109.03305},
 journal = {\aap},
 month = {December},
 pages = {A147},
 primaryclass = {astro-ph.GA},
 title = {{CLASH-VLT: Abell S1063. Cluster assembly history and spectroscopic catalogue}},
 volume = {656},
 year = {2021}
}

@article{Montes2019,
 adsurl = {https://ui.adsabs.harvard.edu/abs/2019MNRAS.482.2838M},
 archiveprefix = {arXiv},
 author = {{Montes}, Mireia and {Trujillo}, Ignacio},
 doi = {10.1093/mnras/sty2858},
 eprint = {1807.11488},
 journal = {\mnras},
 month = {January},
 number = {2},
 pages = {2838-2851},
 primaryclass = {astro-ph.GA},
 title = {{Intracluster light: a luminous tracer for dark matter in clusters of galaxies}},
 volume = {482},
 year = {2019}
}

@article{Moster2013,
 adsurl = {https://ui.adsabs.harvard.edu/abs/2013MNRAS.428.3121M},
 archiveprefix = {arXiv},
 author = {{Moster}, Benjamin P. and {Naab}, Thorsten and {White}, Simon D.~M.},
 doi = {10.1093/mnras/sts261},
 eprint = {1205.5807},
 journal = {\mnras},
 month = {February},
 number = {4},
 pages = {3121-3138},
 primaryclass = {astro-ph.CO},
 title = {{Galactic star formation and accretion histories from matching galaxies to dark matter haloes}},
 volume = {428},
 year = {2013}
}

@article{Newman2013b,
 adsurl = {https://ui.adsabs.harvard.edu/abs/2013ApJ...765...25N},
 archiveprefix = {arXiv},
 author = {{Newman}, Andrew B. and {Treu}, Tommaso and {Ellis}, Richard S. and {Sand}, David J.},
 doi = {10.1088/0004-637X/765/1/25},
 eid = {25},
 eprint = {1209.1392},
 journal = {\apj},
 month = {March},
 number = {1},
 pages = {25},
 primaryclass = {astro-ph.CO},
 title = {{The Density Profiles of Massive, Relaxed Galaxy Clusters. II. Separating Luminous and Dark Matter in Cluster Cores}},
 volume = {765},
 year = {2013}
}

@article{niemiec2022,
 adsurl = {https://ui.adsabs.harvard.edu/abs/2022MNRAS.512.6021N},
 archiveprefix = {arXiv},
 author = {{Niemiec}, Anna and {Giocoli}, Carlo and {Cohen}, Ethan and {Jauzac}, Mathilde and {Jullo}, Eric and {Limousin}, Marceau},
 doi = {10.1093/mnras/stac832},
 eprint = {2201.07817},
 journal = {\mnras},
 month = {June},
 number = {4},
 pages = {6021-6037},
 primaryclass = {astro-ph.CO},
 title = {{Scatter in the satellite galaxy SHMR: fitting functions, scaling relations, and physical processes from the IllustrisTNG simulation}},
 volume = {512},
 year = {2022}
}

@article{Nightingale2021,
 adsurl = {https://ui.adsabs.harvard.edu/abs/2021JOSS....6.2825N},
 archiveprefix = {arXiv},
 author = {{Nightingale}, James. and {Hayes}, Richard and {Kelly}, Ashley and {Amvrosiadis}, Aristeidis and {Etherington}, Amy and {He}, Qiuhan and {Li}, Nan and {Cao}, XiaoYue and {Frawley}, Jonathan and {Cole}, Shaun and {Enia}, Andrea and {Frenk}, Carlos and {Harvey}, David and {Li}, Ran and {Massey}, Richard and {Negrello}, Mattia and {Robertson}, Andrew},
 doi = {10.21105/joss.02825},
 eid = {2825},
 eprint = {2106.01384},
 journal = {The Journal of Open Source Software},
 month = {February},
 number = {58},
 pages = {2825},
 primaryclass = {astro-ph.IM},
 title = {{PyAutoLens: Open-Source Strong Gravitational Lensing}},
 volume = {6},
 year = {2021}
}

@article{Oldham2018a,
 adsurl = {https://ui.adsabs.harvard.edu/abs/2018MNRAS.474.4169O},
 archiveprefix = {arXiv},
 author = {{Oldham}, Lindsay and {Auger}, Matthew},
 doi = {10.1093/mnras/stx2969},
 eprint = {1711.05813},
 journal = {\mnras},
 month = {March},
 number = {3},
 pages = {4169-4185},
 primaryclass = {astro-ph.GA},
 title = {{Galaxy structure from multiple tracers - III. Radial variations in M87's IMF}},
 volume = {474},
 year = {2018}
}

@article{Pillepich2018,
 adsurl = {https://ui.adsabs.harvard.edu/abs/2018MNRAS.473.4077P},
 archiveprefix = {arXiv},
 author = {{Pillepich}, Annalisa and {Springel}, Volker and {Nelson}, Dylan and {Genel}, Shy and {Naiman}, Jill and {Pakmor}, R{\"u}diger and {Hernquist}, Lars and {Torrey}, Paul and {Vogelsberger}, Mark and {Weinberger}, Rainer and {Marinacci}, Federico},
 doi = {10.1093/mnras/stx2656},
 eprint = {1703.02970},
 journal = {\mnras},
 month = {January},
 number = {3},
 pages = {4077-4106},
 primaryclass = {astro-ph.GA},
 title = {{Simulating galaxy formation with the IllustrisTNG model}},
 volume = {473},
 year = {2018}
}

@article{Posacki2015,
 adsurl = {https://ui.adsabs.harvard.edu/abs/2015MNRAS.446..493P},
 archiveprefix = {arXiv},
 author = {{Posacki}, Silvia and {Cappellari}, Michele and {Treu}, Tommaso and {Pellegrini}, Silvia and {Ciotti}, Luca},
 doi = {10.1093/mnras/stu2098},
 eprint = {1407.5633},
 journal = {\mnras},
 month = {January},
 number = {1},
 pages = {493-509},
 primaryclass = {astro-ph.GA},
 title = {{The stellar initial mass function of early-type galaxies from low to high stellar velocity dispersion: homogeneous analysis of ATLAS$^{3D}$ and Sloan Lens ACS galaxies}},
 volume = {446},
 year = {2015}
}

@article{Postman2012,
 adsurl = {https://ui.adsabs.harvard.edu/abs/2012ApJS..199...25P},
 archiveprefix = {arXiv},
 author = {{Postman}, Marc and {Coe}, Dan and {Ben{\'\i}tez}, Narciso and
{Bradley}, Larry and {Broadhurst}, Tom and {Donahue}, Megan and
{Ford}, Holland and {Graur}, Or and {Graves}, Genevieve and
{Jouvel}, Stephanie and {Koekemoer}, Anton and {Lemze}, Doron and
{Medezinski}, Elinor and {Molino}, Alberto and {Moustakas}, Leonidas and
{Ogaz}, Sara and {Riess}, Adam and {Rodney}, Steve and {Rosati}, Piero and
{Umetsu}, Keiichi and {Zheng}, Wei and {Zitrin}, Adi and
{Bartelmann}, Matthias and {Bouwens}, Rychard and {Czakon}, Nicole and
{Golwala}, Sunil and {Host}, Ole and {Infante}, Leopoldo and
{Jha}, Saurabh and {Jimenez-Teja}, Yolanda and {Kelson}, Daniel and
{Lahav}, Ofer and {Lazkoz}, Ruth and {Maoz}, Dani and
{McCully}, Curtis and {Melchior}, Peter and {Meneghetti}, Massimo and
{Merten}, Julian and {Moustakas}, John and {Nonino}, Mario and
{Patel}, Brandon and {Reg{\"o}s}, Enik{\"o} and {Sayers}, Jack and
{Seitz}, Stella and {Van der Wel}, Arjen},
 doi = {10.1088/0067-0049/199/2/25},
 eid = {25},
 eprint = {1106.3328},
 journal = {\apjs},
 month = {April},
 number = {2},
 pages = {25},
 primaryclass = {astro-ph.CO},
 title = {{The Cluster Lensing and Supernova Survey with Hubble: An Overview}},
 volume = {199},
 year = {2012}
}

@article{Robertson2021,
 adsurl = {https://ui.adsabs.harvard.edu/abs/2021MNRAS.501.4610R},
 archiveprefix = {arXiv},
 author = {{Robertson}, Andrew and {Massey}, Richard and {Eke}, Vincent and {Schaye}, Joop and {Theuns}, Tom},
 doi = {10.1093/mnras/staa3954},
 eprint = {2009.07844},
 journal = {\mnras},
 month = {March},
 number = {3},
 pages = {4610-4634},
 primaryclass = {astro-ph.CO},
 title = {{The surprising accuracy of isothermal Jeans modelling of self-interacting dark matter density profiles}},
 volume = {501},
 year = {2021}
}

@article{Salpeter1955,
 adsurl = {https://ui.adsabs.harvard.edu/abs/1955ApJ...121..161S},
 author = {{Salpeter}, Edwin E.},
 doi = {10.1086/145971},
 journal = {\apj},
 month = {January},
 pages = {161},
 title = {{The Luminosity Function and Stellar Evolution.}},
 volume = {121},
 year = {1955}
}

@article{Schwarz1978,
 author = {Schwarz, Gideon},
 doi = {10.1214/aos/1176344136},
 fjournal = {Annals of Statistics},
 journal = {Ann. Statist.},
 month = {03},
 number = {2},
 pages = {461--464},
 publisher = {The Institute of Mathematical Statistics},
 title = {Estimating the Dimension of a Model},
 url = {https://doi.org/10.1214/aos/1176344136},
 volume = {6},
 year = {1978}
}

@article{Shajib2021,
 adsurl = {https://ui.adsabs.harvard.edu/abs/2021MNRAS.503.2380S},
 archiveprefix = {arXiv},
 author = {{Shajib}, Anowar J. and {Treu}, Tommaso and {Birrer}, Simon and {Sonnenfeld}, Alessandro},
 doi = {10.1093/mnras/stab536},
 eprint = {2008.11724},
 journal = {\mnras},
 month = {May},
 number = {2},
 pages = {2380-2405},
 primaryclass = {astro-ph.GA},
 title = {{Dark matter haloes of massive elliptical galaxies at z {\ensuremath{\sim}} 0.2 are well described by the Navarro-Frenk-White profile}},
 volume = {503},
 year = {2021}
}

@article{simon2024,
 adsurl = {https://ui.adsabs.harvard.edu/abs/2024MNRAS.527.2341S},
 archiveprefix = {arXiv},
 author = {{Simon}, David A. and {Cappellari}, Michele and {Hartke}, Johanna},
 doi = {10.1093/mnras/stad3309},
 eprint = {2303.18229},
 journal = {\mnras},
 month = {January},
 number = {2},
 pages = {2341-2361},
 primaryclass = {astro-ph.GA},
 title = {{Supermassive black hole mass in the massive elliptical galaxy M87 from integral-field stellar dynamics using OASIS and MUSE with adaptive optics: assessing systematic uncertainties}},
 volume = {527},
 year = {2024}
}

@article{Smith2020,
 adsurl = {https://ui.adsabs.harvard.edu/abs/2020ARA&A..58..577S},
 author = {{Smith}, Russell J.},
 doi = {10.1146/annurev-astro-032620-020217},
 journal = {\araa},
 month = {August},
 pages = {577-615},
 title = {{Evidence for Initial Mass Function Variation in Massive Early-Type Galaxies}},
 volume = {58},
 year = {2020}
}

@article{steinhardt2020,
 adsurl = {https://ui.adsabs.harvard.edu/abs/2020ApJS..247...64S},
 archiveprefix = {arXiv},
 author = {{Steinhardt}, Charles L. and {Jauzac}, Mathilde and {Acebron}, Ana and {Atek}, Hakim and {Capak}, Peter and {Davidzon}, Iary and {Eckert}, Dominique and {Harvey}, David and {Koekemoer}, Anton M. and {Lagos}, Claudia D.~P. and {Mahler}, Guillaume and {Montes}, Mireia and {Niemiec}, Anna and {Nonino}, Mario and {Oesch}, P.~A. and {Richard}, Johan and {Rodney}, Steven A. and {Schaller}, Matthieu and {Sharon}, Keren and {Strolger}, Louis-Gregory and {Allingham}, Joseph and {Amara}, Adam and {Bah{\'e}}, Yannick and {B{\oe}hm}, C{\'e}line and {Bose}, Sownak and {Bouwens}, Rychard J. and {Bradley}, Larry D. and {Brammer}, Gabriel and {Broadhurst}, Tom and {Ca{\~n}as}, Rodrigo and {Cen}, Renyue and {Cl{\'e}ment}, Benjamin and {Clowe}, Douglas and {Coe}, Dan and {Connor}, Thomas and {Darvish}, Behnam and {Diego}, Jose M. and {Ebeling}, Harald and {Edge}, A.~C. and {Egami}, Eiichi and {Ettori}, Stefano and {Faisst}, Andreas L. and {Frye}, Brenda and {Furtak}, Lukas J. and {G{\'o}mez-Guijarro}, C. and {Remolina Gonz{\'a}lez}, J.~D. and {Gonzalez}, Anthony and {Graur}, Or and {Gruen}, Daniel and {Harvey}, David and {Hensley}, Hagan and {Hovis-Afflerbach}, Beryl and {Jablonka}, Pascale and {Jha}, Saurabh W. and {Jullo}, Eric and {Kneib}, Jean-Paul and {Kokorev}, Vasily and {Lagattuta}, David J. and {Limousin}, Marceau and {von der Linden}, Anja and {Linzer}, Nora B. and {Lopez}, Adrian and {Magdis}, Georgios E. and {Massey}, Richard and {Masters}, Daniel C. and {Maturi}, Matteo and {McCully}, Curtis and {McGee}, Sean L. and {Meneghetti}, Massimo and {Mobasher}, Bahram and {Moustakas}, Leonidas A. and {Murphy}, Eric J. and {Natarajan}, Priyamvada and {Neyrinck}, Mark and {O'Connor}, Kyle and {Oguri}, Masamune and {Pagul}, Amanda and {Rhodes}, Jason and {Rich}, R. Michael and {Robertson}, Andrew and {Sereno}, Mauro and {Shan}, Huanyuan and {Smith}, Graham P. and {Sneppen}, Albert and {Squires}, Gordon K. and {Tam}, Sut-Ieng and {Tchernin}, C{\'e}line and {Toft}, Sune and {Umetsu}, Keiichi and {Weaver}, John R. and {van Weeren}, R.~J. and {Williams}, Liliya L.~R. and {Wilson}, Tom J. and {Yan}, Lin and {Zitrin}, Adi},
 doi = {10.3847/1538-4365/ab75ed},
 eid = {64},
 eprint = {2001.09999},
 journal = {\apjs},
 month = {April},
 number = {2},
 pages = {64},
 primaryclass = {astro-ph.GA},
 title = {{The BUFFALO HST Survey}},
 volume = {247},
 year = {2020}
}

@article{tortorelli2018,
 adsurl = {https://ui.adsabs.harvard.edu/abs/2018MNRAS.477..648T},
 archiveprefix = {arXiv},
 author = {{Tortorelli}, Luca and {Mercurio}, Amata and {Paolillo}, Maurizio and {Rosati}, Piero and {Gargiulo}, Adriana and {Gobat}, Raphael and {Balestra}, Italo and {Caminha}, G.~B. and {Annunziatella}, Marianna and {Grillo}, Claudio and {Lombardi}, Marco and {Nonino}, Mario and {Rettura}, Alessandro and {Sartoris}, Barbara and {Strazzullo}, Veronica},
 doi = {10.1093/mnras/sty617},
 eprint = {1803.02375},
 journal = {\mnras},
 month = {June},
 number = {1},
 pages = {648-668},
 primaryclass = {astro-ph.GA},
 title = {{The Kormendy relation of galaxies in the Frontier Fields clusters: Abell S1063 and MACS J1149.5+2223}},
 volume = {477},
 year = {2018}
}

@article{vandenBosch2016,
 adsurl = {https://ui.adsabs.harvard.edu/abs/2016ApJ...831..134V},
 archiveprefix = {arXiv},
 author = {{van den Bosch}, Remco C.~E.},
 doi = {10.3847/0004-637X/831/2/134},
 eid = {134},
 eprint = {1606.01246},
 journal = {\apj},
 month = {November},
 number = {2},
 pages = {134},
 primaryclass = {astro-ph.GA},
 title = {{Unification of the fundamental plane and Super Massive Black Hole Masses}},
 volume = {831},
 year = {2016}
}

@article{Watanabe2010,
 adsurl = {https://ui.adsabs.harvard.edu/abs/2010arXiv1004.2316W},
 archiveprefix = {arXiv},
 author = {{Watanabe}, Sumio},
 eid = {arXiv:1004.2316},
 eprint = {1004.2316},
 journal = {arXiv e-prints},
 month = {April},
 pages = {arXiv:1004.2316},
 primaryclass = {cs.LG},
 title = {{Asymptotic Equivalence of Bayes Cross Validation and Widely Applicable Information Criterion in Singular Learning Theory}},
 year = {2010}
}

@article{Weinberger2017,
 adsurl = {https://ui.adsabs.harvard.edu/abs/2017MNRAS.465.3291W},
 archiveprefix = {arXiv},
 author = {{Weinberger}, Rainer and {Springel}, Volker and {Hernquist}, Lars and {Pillepich}, Annalisa and {Marinacci}, Federico and {Pakmor}, R{\"u}diger and {Nelson}, Dylan and {Genel}, Shy and {Vogelsberger}, Mark and {Naiman}, Jill and {Torrey}, Paul},
 doi = {10.1093/mnras/stw2944},
 eprint = {1607.03486},
 journal = {\mnras},
 month = {March},
 number = {3},
 pages = {3291-3308},
 primaryclass = {astro-ph.GA},
 title = {{Simulating galaxy formation with black hole driven thermal and kinetic feedback}},
 volume = {465},
 year = {2017}
}

@article{Wright2017,
 adsurl = {https://ui.adsabs.harvard.edu/abs/2017MNRAS.470..283W},
 archiveprefix = {arXiv},
 author = {{Wright}, A.~H. and {Robotham}, A.~S.~G. and {Driver}, S.~P. and {Alpaslan}, M. and {Andrews}, S.~K. and {Baldry}, I.~K. and {Bland-Hawthorn}, J. and {Brough}, S. and {Brown}, M.~J.~I. and {Colless}, M. and {da Cunha}, E. and {Davies}, L.~J.~M. and {Graham}, Alister W. and {Holwerda}, B.~W. and {Hopkins}, A.~M. and {Kafle}, P.~R. and {Kelvin}, L.~S. and {Loveday}, J. and {Maddox}, S.~J. and {Meyer}, M.~J. and {Moffett}, A.~J. and {Norberg}, P. and {Phillipps}, S. and {Rowlands}, K. and {Taylor}, E.~N. and {Wang}, L. and {Wilkins}, S.~M.},
 doi = {10.1093/mnras/stx1149},
 eprint = {1705.04074},
 journal = {\mnras},
 month = {September},
 number = {1},
 pages = {283-302},
 primaryclass = {astro-ph.GA},
 title = {{Galaxy And Mass Assembly (GAMA): the galaxy stellar mass function to z = 0.1 from the r-band selected equatorial regions}},
 volume = {470},
 year = {2017}
}

@article{Zhao1996,
 adsurl = {https://ui.adsabs.harvard.edu/abs/1996MNRAS.278..488Z},
 archiveprefix = {arXiv},
 author = {{Zhao}, Hongsheng},
 doi = {10.1093/mnras/278.2.488},
 eprint = {astro-ph/9509122},
 journal = {\mnras},
 month = {January},
 number = {2},
 pages = {488-496},
 primaryclass = {astro-ph},
 title = {{Analytical models for galactic nuclei}},
 volume = {278},
 year = {1996}
}




\appendix
\section{Mass model parameters and posteriors}
\label{app:model_parameters}
\begin{figure*}
    \centering
    \includegraphics[width=0.9\linewidth]{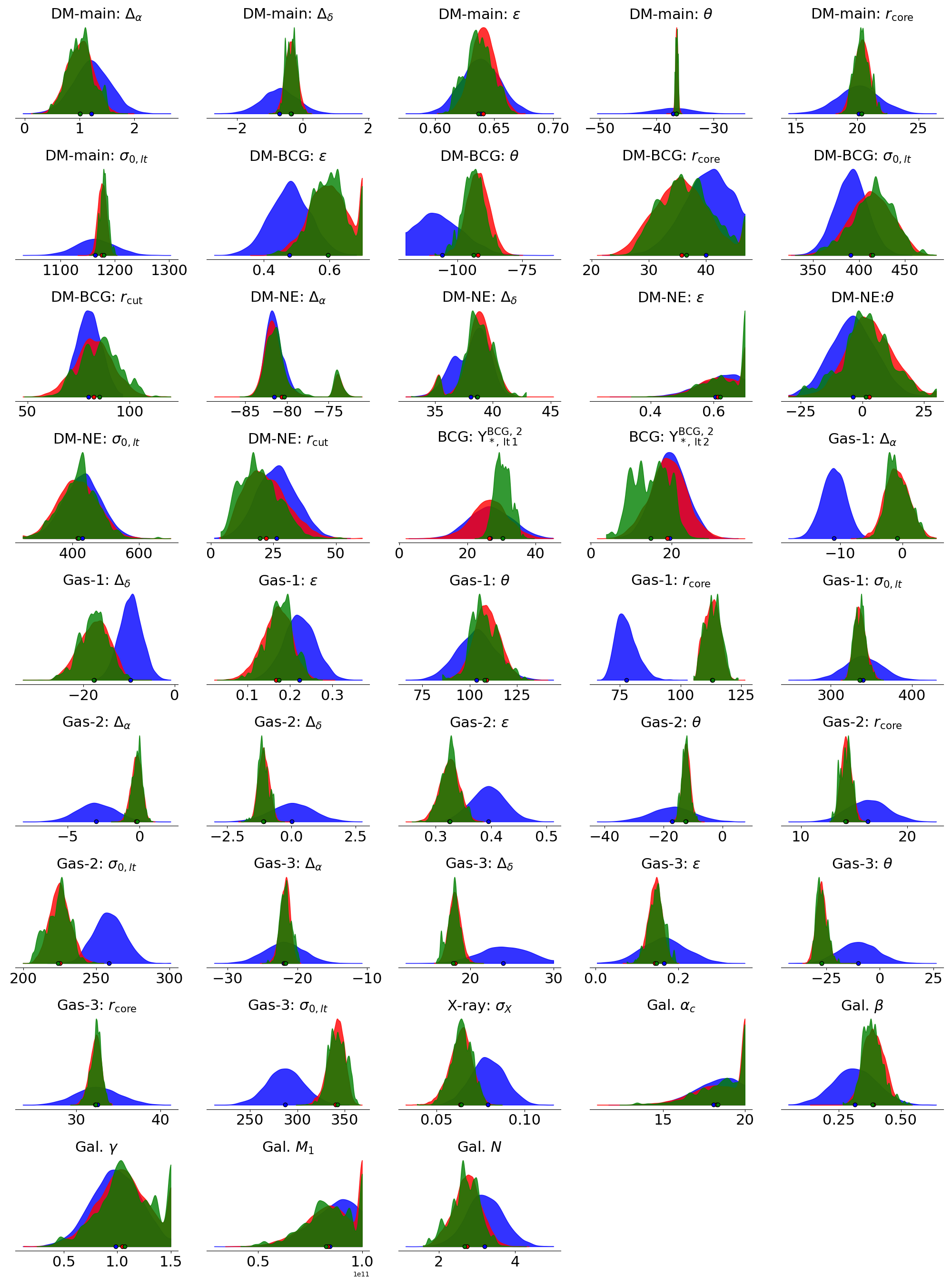}
    \caption{1D posterior density of each model parameter. The model considered here has a ``BCG - ML 2'' parametrisation and the delayed SFH model. Each colour represents the distribution at a different step of our optimisation process as detailed in Sect.~\ref{sect:optimisation-process}. The blue distribution represents the initial approximation based on a multivariate normal distribution. The red and green posteriors show the biased and final approximate posteriors or the distributions before and after the importance sampling step. dPIEs parameters are presented with the relative x-coordinate $\Delta_\alpha$ (${\rm arcsec}$), the relative y-coordinate $\Delta_\delta$ (${\rm arcsec}$), ellipticity $\epsilon$, position angle $\theta$ (degree), core radius $r_{\rm core}$ (${\rm arcsec}$), cut radius $r_{\rm cut}$ (${\rm arcsec}$) and central velocity dispersion $\sigma_{0,lt}$ (${\rm km/s}$). For the BCG \& ICL component, only its normalisation ($\Upsilon^{\rm BCG,2}_{\rm *,lt,1/2}$ ) is presented. The cluster member parameters are shown by the spatial scale, $\alpha_c$, the power-law slopes, $\delta$ and $\gamma$, the turn-off mass, $M_1$ ($M_\odot$), and the normalisation, $N$.}
    \label{fig:Posterior_delayed}
\end{figure*}

\begin{figure*}
    \centering
    \includegraphics[width=0.9\linewidth]{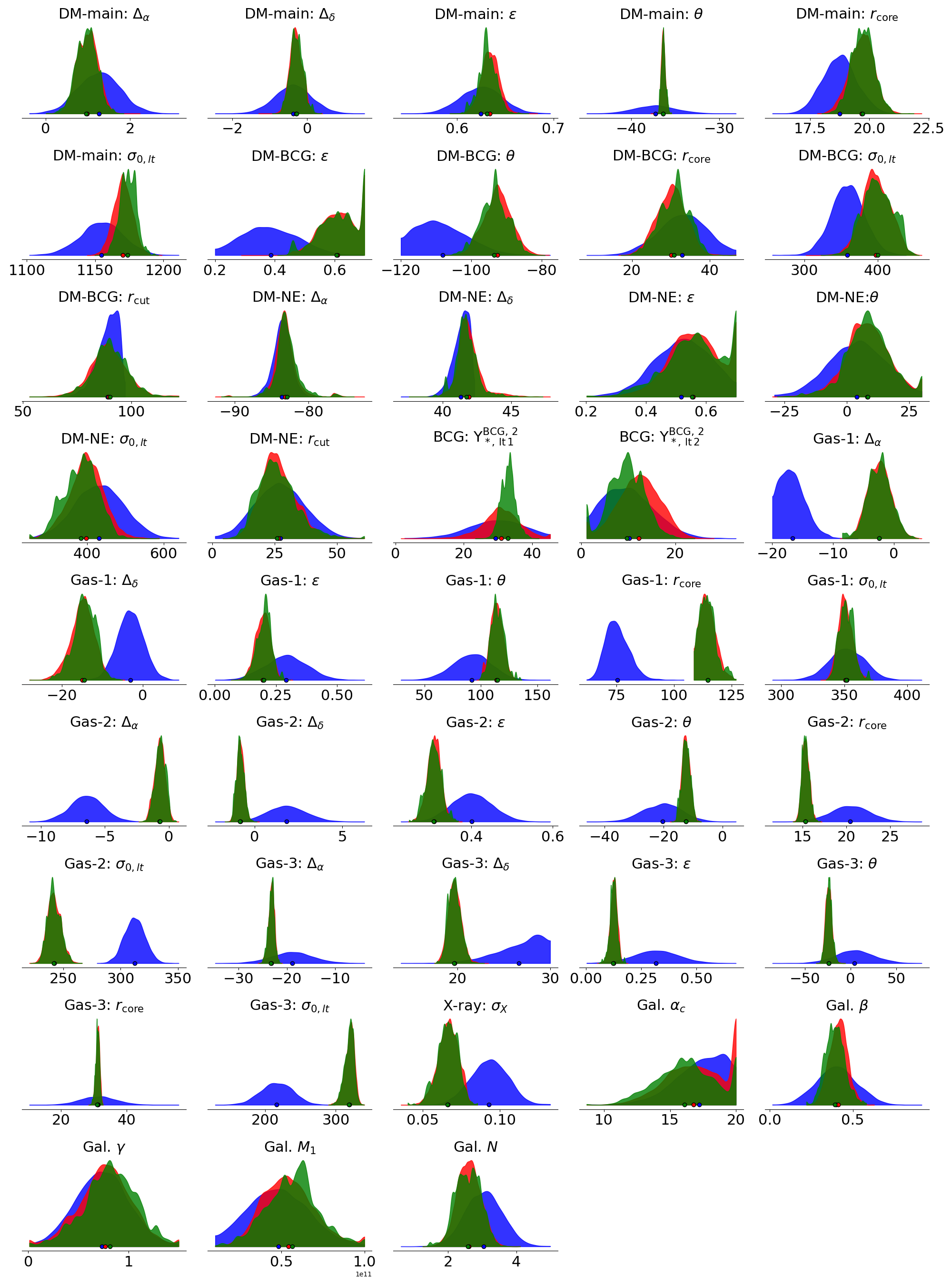}
    \caption{Same as Fig.~\ref{fig:Posterior_delayed} for the model with the double power-law SFH.}
    \label{fig:Posterior_dbplaw}
\end{figure*}

\begin{figure*}
    \centering
    \includegraphics[width=0.9\linewidth]{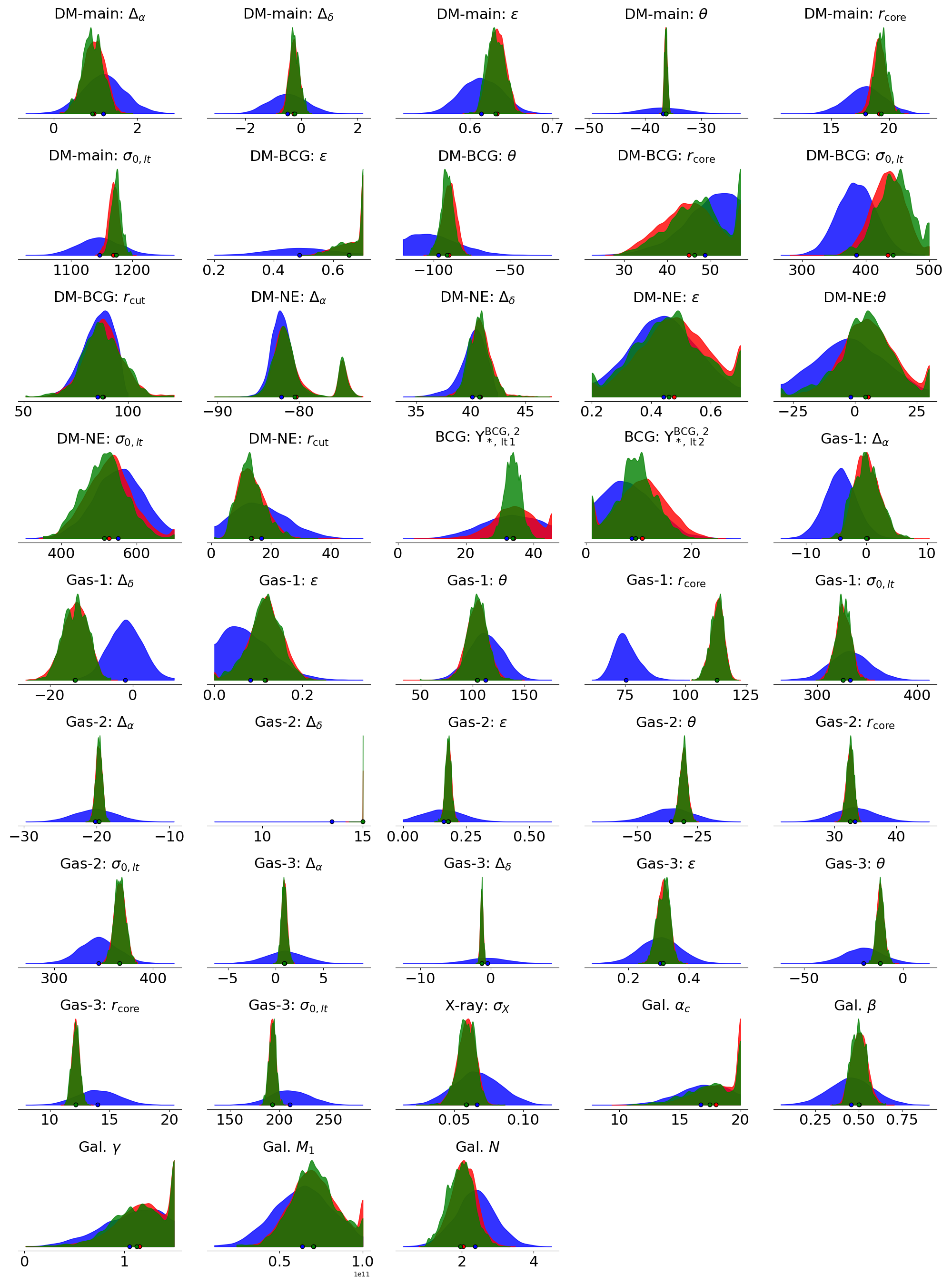}
    \caption{Same as Fig.~\ref{fig:Posterior_delayed} for the model with \textsc{LePhare} SED model.}
    \label{fig:Posterior_lephare}
\end{figure*}

In this appendix, we present the model posteriors for the three models with a ``BCG - ML 2'' parametrisation. The 1D posterior distributions of each model parameter are presented in Fig.~\ref{fig:Posterior_delayed}, Fig.~\ref{fig:Posterior_dbplaw} and Fig.~\ref{fig:Posterior_lephare}. It shows the models with a delayed SFH, a double power-law SFH and the \textsc{LePhare} SED model, respectively. In our optimisation procedure detailed in Sect.~\ref{sect:optimisation-process}, we use three different distributions that are presented in different colours. To speed up the inference process, we use a first approximation of the posterior based on a multivariate normal distribution, shown in blue. We use the 1D posterior of this multivariate to transform our parameter space to increase the volume represented by the model posteriors with respect to the priors. The posteriors obtained by the nested sampling runs are presented in red. In this run, parameters are not constrained by the BCG \& ICL kinematics. The last distribution presented in green is the final approximation of the posterior obtained after the importance sampling step to incorporate the BCG \& ICL constraints. 


We are presenting tabulated values of each parameter with their $1\sigma$ CI and best-fitting values in Table~\ref{Tab:param_delayed}, Table~\ref{Tab:param_dbplaw} and Table~\ref{Tab:param_lePhare}. These are associated with the models having a delayed SFH, a double power-law SFH or \textsc{LePhare} SED model, respectively.

\begin{table*}
\centering
 
\begin{tabular}{ccccc}
    \hline\\
    ID & $\Delta\alpha$&$\Delta\delta$&$\phi$ &$\epsilon$\\
    & $\rm{kpc}$&$\rm{kpc}$&$\rm{degree}$&\\
    \\
    \hline\hline\\
    DM-main&$5.0^{+1.0}_{-1.1}(5.7)$&$-1.7^{+0.8}_{-0.8}(-2.5)$&$-36.4^{+0.2}_{-0.2}(-36.3)$&$0.636^{+0.012}_{-0.011}(0.631)$\\
    DM-BCG&$[0.0]$&$[0.0]$&$-93.7^{+4.2}_{-3.9}(-91.6)$&$0.597^{+0.048}_{-0.052}(0.67)$\\
    DM-NE&$-399.8^{+14.5}_{-5.5}(-399.3)$&$190.5^{+6.2}_{-5.6}(201.0)$&$1.3^{+9.4}_{-8.8}(20.4)$&$0.624^{+0.075}_{-0.072}(0.643)$\\
    Gas-1&$-5.0^{+10.7}_{-7.0}(-2.0)$&$-86.2^{+14.3}_{-16.8}(-100.9)$&$107.2^{+7.9}_{-6.1}(96.7)$&$0.175^{+0.027}_{-0.032}(0.163)$\\
    Gas-2&$-0.8^{+1.4}_{-1.8}(-0.7)$&$-5.5^{+1.2}_{-0.8}(-5.0)$&$-12.4^{+1.3}_{-1.4}(-12.9)$&$0.326^{+0.014}_{-0.017}(0.332)$\\
    Gas-3&$-106.9^{+4.4}_{-3.6}(-112.3)$&$88.6^{+3.1}_{-4.3}(94.2)$&$-27.6^{+3.1}_{-2.1}(-27.5)$&$0.147^{+0.018}_{-0.018}(0.128)$\\

    \\
    \hline
    \end{tabular}

\vspace*{0.8cm}
\begin{tabular}{cccc}
    \hline\\
    ID &$r_{\rm core}$&$r_{\rm cut}$&$\sigma_{0\rm,lt}$\\
    & $\rm{kpc}$&$\rm{kpc}$&$\rm{km/s}$\\
    \\
    \hline\hline\\
    DM-main&$99.9^{+3.2}_{-2.8}(94.1)$&$[3000]$&$1179^{+7}_{-7}(1167)$\\
    DM-BCG&$178.1^{+24.4}_{-22.6}(205.2)$&$420^{+44}_{-48}(419)$&$416^{+20}_{-25}(430)$\\
    DM-NE&$[2.5]$&$93^{+44}_{-37}(79)$&$419^{+52}_{-57}(513)$\\
    Gas-1&$557.7^{+16.5}_{-18.4}(537.0)$&$[1250]$&$335^{+6}_{-7}(330)$\\
    Gas-2&$69.9^{+2.6}_{-3.1}(74.7)$&$[1250]$&$224^{+7}_{-10}(235)$\\
    Gas-3&$159.3^{+2.7}_{-3.5}(158.3)$&$[1250]$&$342^{+10}_{-9}(333)$\\
    \\
    \hline
    \end{tabular}

\vspace*{0.5cm}
\begin{tabular}{ccc}
    \hline\\
    ID &$\Upsilon^{\rm BCG,\, 2}_{\rm *,\,lt\,1}$& $\Upsilon^{\rm BCG,\, 2}_{\rm *,\,lt\,2}$\\
    \\
    \hline\hline\\
    BCG&$30.22^{+2.52}_{-2.23}(32.7)$&$14.94^{+4.8}_{-5.03}(11.09)$\\
    \\
    \hline
\end{tabular}

\vspace*{0.5cm}
\begin{tabular}{cccccc}
    \hline\\
    ID&$\alpha_c$&$\delta$&$\gamma$&$M_1$&$N$\\
    &&&&$10^{10}\,M_\odot$&\\
    \\
    \hline\hline \\
    Gal.&$18.62^{+1.36}_{-1.75}(18.49)$&$0.38^{+0.04}_{-0.03}(0.35)$&$1.06^{+0.28}_{-0.22}(0.97)$&$8.31^{+1.09}_{-1.18}(8.1)$&$2.69^{+0.34}_{-0.38}(2.74)$\\
    \\
    \hline
\end{tabular}

\caption{Model parameters distributions of each mass component for the model with a ``BCG - ML 2'' parametrisation and the delayed SFH model. Starting from the top, the first two tables present the mass component modelled with dPIEs, the third table shows the MGE component, and the last one shows the cluster member population parameters. Each parameter is represented as $\text{median}^{84 \text{ per cent limit}}_{16 \text{ per cent limit}}(\text{best-fit})$ except the ones fixed a priori that are shown in brackets. $\Delta\alpha$ and $\Delta\delta$ are the coordinates of dPIE haloes relative to the BCG centre at $\alpha({\rm J2000})=342.183213$ and $\delta({\rm J2000})=-44.530897$. The $\rm lt$ label refers to the velocity dispersions of dPIEs as implemented in \textsc{Lenstool}.}
\label{Tab:param_delayed}
\end{table*}

\begin{table*}
\centering

\begin{tabular}{ccccc}
    \hline\\
    ID & $\Delta\alpha$&$\Delta\delta$&$\phi$ &$\epsilon$\\
    & $\rm{kpc}$&$\rm{kpc}$&$\rm{degree}$&\\
    \\
    \hline\hline\\
    DM-main&$4.7^{+1.1}_{-1.2}(3.8)$&$-1.4^{+0.8}_{-0.8}(-0.5)$&$-36.3^{+0.2}_{-0.2}(-36.3)$&$0.63^{+0.009}_{-0.008}(0.616)$\\
    DM-BCG&$[0.0]$&$[0.0]$&$-93.3^{+3.9}_{-4.0}(-95.8)$&$0.61^{+0.063}_{-0.066}(0.691)$\\
    DM-NE&$-408.9^{+6.7}_{-4.8}(-416.7)$&$205.0^{+3.5}_{-2.9}(209.3)$&$8.4^{+8.2}_{-8.1}(13.6)$&$0.559^{+0.096}_{-0.1}(0.572)$\\
    Gas-1&$-11.6^{+9.3}_{-9.6}(0.8)$&$-70.7^{+13.2}_{-13.3}(-83.6)$&$114.3^{+6.1}_{-5.4}(113.2)$&$0.203^{+0.024}_{-0.034}(0.192)$\\
    Gas-2&$-3.2^{+2.0}_{-2.0}(-2.9)$&$-4.0^{+1.1}_{-1.0}(-3.3)$&$-12.4^{+1.5}_{-1.3}(-12.2)$&$0.307^{+0.018}_{-0.018}(0.308)$\\
    Gas-3&$-114.0^{+2.5}_{-3.4}(-114.1)$&$96.5^{+3.4}_{-3.2}(94.1)$&$-24.0^{+3.2}_{-3.3}(-26.8)$&$0.123^{+0.014}_{-0.014}(0.146)$\\

    \\
    \hline
    \end{tabular}

\vspace*{0.8cm}
\begin{tabular}{cccc}
    \hline\\
    ID &$r_{\rm core}$&$r_{\rm cut}$&$\sigma_{0\rm,lt}$\\
    & $\rm{kpc}$&$\rm{kpc}$&$\rm{km/s}$\\
    \\
    \hline\hline\\
    DM-main&$97.0^{+2.2}_{-2.4}(93.6)$&$[3000]$&$1174^{+5}_{-5}(1172)$\\
    DM-BCG&$153.2^{+17.7}_{-20.9}(184.4)$&$442^{+37}_{-36}(447)$&$398^{+23}_{-19}(406)$\\
    DM-NE&$[2.5]$&$124^{+38}_{-34}(110)$&$387^{+40}_{-51}(421)$\\
    Gas-1&$562.9^{+16.8}_{-15.8}(582.1)$&$[1250]$&$351^{+5}_{-5}(355)$\\
    Gas-2&$75.3^{+2.1}_{-1.8}(72.5)$&$[1250]$&$241^{+6}_{-5}(234)$\\
    Gas-3&$152.9^{+2.7}_{-3.1}(155.1)$&$[1250]$&$319^{+5}_{-7}(327)$\\
    \\
    \hline
    \end{tabular}

\vspace*{0.5cm}
\begin{tabular}{ccc}
    \hline\\
    ID &$\Upsilon^{\rm BCG,\, 2}_{\rm *,\,lt\,1}$& $\Upsilon^{\rm BCG,\, 2}_{\rm *,\,lt\,2}$\\
    \\
    \hline\hline\\
    BCG&$32.99^{+1.91}_{-2.16}(33.38)$&$9.7^{+3.58}_{-3.64}(1.95)$\\
    \\
    \hline
\end{tabular}

\vspace*{0.5cm}
\begin{tabular}{cccccc}
    \hline\\
    ID&$\alpha_c$&$\delta$&$\gamma$&$M_1$&$N$\\
    &&&&$10^{10}\,M_\odot$&\\
    \\
    \hline\hline \\
    Gal.&$16.09^{+2.21}_{-2.21}(13.77)$&$0.39^{+0.05}_{-0.05}(0.37)$&$0.82^{+0.25}_{-0.25}(0.9)$&$5.72^{+1.33}_{-1.67}(6.44)$&$2.57^{+0.36}_{-0.38}(2.75)$\\
    \\
    \hline
\end{tabular}

\caption{Same as Table~\ref{Tab:param_delayed} for the model with the double power-law SFH.}
\label{Tab:param_dbplaw}
\end{table*}

\begin{table*}
\centering
 
\begin{tabular}{ccccc}
    \hline\\
    ID & $\Delta\alpha$&$\Delta\delta$&$\phi$ &$\epsilon$\\
    & $\rm{kpc}$&$\rm{kpc}$&$\rm{degree}$&\\
    \\
    \hline\hline\\
    DM-main&$4.4^{+1.2}_{-1.0}(4.5)$&$-1.2^{+0.9}_{-0.7}(-1.3)$&$-36.2^{+0.2}_{-0.2}(-36.0)$&$0.63^{+0.011}_{-0.01}(0.623)$\\
    DM-BCG$[0.0]$&$[0.0]$&$-91.2^{+3.9}_{-3.8}(-87.4)$&$0.661^{+0.038}_{-0.055}(0.634)$\\
    DM-NE&$-401.4^{+32.6}_{-7.0}(-399.4)$&$200.5^{+4.7}_{-4.6}(200.0)$&$4.4^{+10.2}_{-10.6}(-10.4)$&$0.457^{+0.124}_{-0.117}(0.6)$\\
    Gas-1&$-0.3^{+11.4}_{-11.0}(0.7)$&$-67.5^{+14.6}_{-15.5}(-55.6)$&$104.6^{+9.1}_{-9.3}(98.9)$&$0.115^{+0.035}_{-0.036}(0.046)$\\
    Gas-2&$-96.9^{+2.1}_{-2.1}(-98.3)$&$73.8^{+0.0}_{0.0}(73.8)$&$-30.7^{+1.4}_{-1.7}(-32.1)$&$0.179^{+0.011}_{-0.011}(0.196)$\\
    Gas-3&$4.3^{+1.4}_{-1.4}(3.1)$&$-6.3^{+0.9}_{-0.9}(-6.6)$&$-11.5^{+1.9}_{-1.9}(-9.1)$&$0.317^{+0.018}_{-0.024}(0.291)$\\

    \\
    \hline
    \end{tabular}

\vspace*{0.8cm}
\begin{tabular}{cccc}
    \hline\\
    ID &$r_{\rm core}$&$r_{\rm cut}$&$\sigma_{0\rm,lt}$\\
    & $\rm{kpc}$&$\rm{kpc}$&$\rm{km/s}$\\
    \\
    \hline\hline\\
    DM-main&$95.0^{+2.9}_{-2.6}(95.6)$&$[3000]$&$1173^{+7}_{-8}(1171)$\\
    DM-BCG&$229.3^{+32.4}_{-33.0}(231.0)$&$430^{+49}_{-42}(498)$&$444^{+25}_{-29}(432)$\\
    DM-NE&$[2.5]$&$62^{+24}_{-19}(63)$&$514^{+55}_{-61}(461)$\\
    Gas-1&$556.9^{+12.0}_{-13.9}(534.2)$&$[1250]$&$325^{+6}_{-7}(323)$\\
    Gas-2&$160.2^{+2.7}_{-2.9}(158.0)$&$[1250]$&$366^{+5}_{-5}(363)$\\
    Gas-3&$59.8^{+1.5}_{-1.6}(59.0)$&$[1250]$&$192^{+3}_{-3}(191)$\\
    \\
    \hline
    \end{tabular}

\vspace*{0.5cm}
\begin{tabular}{ccc}
    \hline\\
    ID &$\Upsilon^{\rm BCG,\, 2}_{\rm *,\,lt\,1}$& $\Upsilon^{\rm BCG,\, 2}_{\rm *,\,lt\,2}$\\
    \\
    \hline\hline\\
    BCG-1&$33.79^{+2.26}_{-2.3}(31.8)$&$9.41^{+3.74}_{-3.61}(8.73)$\\
    \\
    \hline
\end{tabular}

\vspace*{0.5cm}
\begin{tabular}{cccccc}
    \hline\\
    ID&$\alpha_c$&$\delta$&$\gamma$&$M_1$&$N$\\
    &&&&$10^{10}\,M_\odot$&\\
    \\
    \hline\hline \\
    Gal.&$17.68^{+2.1}_{-2.19}(19.99)$&$0.49^{+0.04}_{-0.04}(0.47)$&$1.15^{+0.26}_{-0.28}(1.37)$&$6.98^{+1.42}_{-1.25}(7.77)$&$1.96^{+0.33}_{-0.38}(2.57)$\\
    \\
    \hline
\end{tabular}

\caption{Same as Table~\ref{Tab:param_delayed} for the model with \textsc{LePhare} SED model.}
\label{Tab:param_lePhare}
\end{table*}


\bsp	
\label{lastpage}
\end{document}